\def\isweb{1}
\documentclass{pdg}

\reviewtitle{Machine Learning}
\reviewauthor{J.M. Duarte (UC San Diego), U. Seljak (UC Berkeley; LBNL) and K. Terao (SLAC; Stanford U.)}
\reviewlabel{ml}
\ischaptertrue

\usepackage{mathtools} 
\usepackage{xfrac} 
\usepackage{amsmath}

\newcommand*\dif{\mathop{}\!\mathrm{d}} 

\DeclareMathOperator*{\argmax}{arg\,max}
\DeclareMathOperator*{\argmin}{arg\,min}
\newcommand{\aka}{\textit{a.k.a.}}

\usepackage{makeidx}
\makeindex

\ifdefined\isbooklet
\externalcitedocument{ml-full}
\fi

\IfFileExists{crossref.aux}{\externaldocument{crossref}}{}

\begin{document}

\ifdefined\isbook
\setcounter{page}{711}
\fi
\ifdefined\isbooklet
\setcounter{page}{262}
\fi

\ifdefined\isbook
\rppthumb
\fi

\ifdefined\isbooklet
\fontsize{8pt}{9pt}\selectfont
\input{ml-booklet}
\else
\begin{bibunit}
%
%





%
\pdgtitle


\written{August 2025}

\setcounter{tocdepth}{2}
\tableofcontents

\section{Introduction}


This chapter gives an overview of the core concepts of machine learning\index{machine learning} (ML)---the use of algorithms that learn from data, identify patterns, and make predictions or decisions without being explicitly programmed---that are relevant to particle physics with some examples of applications to the energy, intensity, cosmic, and accelerator frontiers.
ML is an enormous field that has grown substantially in the last decade, largely driven by the emergence of so-called deep learning (DL)~\cite{2015Natur.521..436L, schmidhuber2015deep}.
ML has a long history in particle physics going back to the late 1980s and early 1990s; see Refs.~\cite{Radovic:2018dip,Guest:2018yhq,Carleo:2019ptp} for recent reviews. 
ML is a subset of artificial intelligence\index{artificial intelligence} (AI), which generally refers to computational systems that can perform tasks typically associated with human intelligence, such as learning, reasoning, problem-solving, perception, and decision-making.

Physicists are exploring and contributing to machine learning at an unprecedented rate, which poses a challenge for those who wish to have an up-to-date view of the field.
This motivated an effort to create \textit{A Living Review of Machine Learning for Particle and Nuclear Physics}~\cite{Feickert:2021ajf}, which can be accessed here: \url{https://iml-wg.github.io/HEPML-LivingReview/}.
At the time of writing, the Living Review included more than 1,800 references organized hierarchically by topic.
Although we make references to some of these papers, this chapter focuses on the methodology and does not attempt to give a comprehensive review of the applications.

Machine learning and artificial intelligence have a mathematical foundation that is closely tied to statistics (see Ch.~\crossref{stat}), the calculus of variations, approximation theory, and optimal control theory.
Nevertheless, there have been tremendous advances in recent years, driven by increased computational power, enormous datasets, and new insights,  that are impacting physics and society.

The topic can be organized along a few axes, which we use to structure this section.
First, there are different learning paradigms, for example, supervised learning\index{supervised learning}, unsupervised learning\index{unsupervised learning}, and reinforcement learning\index{RL, reinforcement learning}.
Within these paradigms, there are various tasks; for example, classification\index{classification} and regression\index{regression}---which have been the primary use of ML in particle physics---are examples of supervised learning.
In addition to the learning paradigm and tasks, there are various types of machine learning models that generically process some input and produce some output.
The types of models vary based on what they are modeling (\eg, so-called discriminative\index{discriminative model} vs. generative models\index{generative model}), as well as how they are implemented (\eg, neural networks\index{neural network}, decision trees\index{decision tree}, or kernel machines\index{kernel machine}).
Next, there are the issues around training\index{training} or learning within the context of a given task and model class, which connects to optimization\index{optimization} and regularization\index{regularization}. 
We will briefly discuss the various considerations that emerge in the application of machine learning methods to physics, such as the treatment of systematic uncertainty, the interpretability\index{interpretability} of the models, and the incorporation of symmetry. 


\subsection{A gentle introduction with a representative example}

We will use a specific, familiar example to introduce the various ingredients in context before factorizing and abstracting them.
Consider the task of \textit{classifying} energy deposits in a particle detector as coming from electrons or protons.
For this example, let the detector data consist of energy deposits in $d$ sensors so that the data can be represented as a \textit{feature vector}\index{feature vector} $x \in \mathbb{R}^d$.
Different components of $x$ may correspond to physical quantities with different units (\eg, units of energy, momentum, or position). 

Due to the complex interactions of particles in the detector, we do not have an explicit probability model for the high-dimensional data for the electron and proton scenarios, but we do have a simulator that allows us to generate Monte Carlo samples for each.
This allows us to assemble a \textit{training dataset}\index{training dataset} $\{x_i, y_i\}_{i=1,\dots,n}$, where $y$ is a \textit{label}\index{label} that identifies how the example was generated (\eg, $y=0$ for electrons and $y=1$ for protons).
We would like to find a function that accurately \textit{predicts} the label on new data.
Because we have feature-label pairs, this is a \textit{supervised learning} problem.
We can use a \textit{neural network} to provide a flexible family of functions $f_\phi: \mathbb{R}^d \to \mathbb{R}$, where $\phi$ denotes the internal parameters of the neural network (\ie, the weights and biases that we will discuss in Sec.~\ref{ML:sec:nn}).
The goal of \textit{training}\index{training} is to find the value of the parameters $\phi$ that provide the `best' predictions.
This is made concrete through a \textit{loss function}\index{loss} $\mathcal{L}(y, f_\phi(x))$.
Instead of the obvious zero-one loss, which is 0 if $f_\phi(x) = y$ and 1 if not, we use the squared-loss $\mathcal{L}_{\rm sq}(y, f_\phi(x)) = (y - f_\phi(x))^2$, which will be motivated in Sec.~\ref{ML:sec:classification}.
We can evaluate the average of the loss on the training set of size $n$, known as the \textit{empirical risk}\index{risk} or \textit{training loss} $\mathcal{R}_\textrm{emp}(f_\phi) = \sum_{i=1}^n \mathcal{L}(y_i, f_\phi(x_i))/n$.
\textit{Training}\index{training} refers to numerically minimizing the empirical risk.
We can numerically optimize the model through \textit{gradient descent}\index{gradient descent}, which iteratively adjusts the parameters of the network according to $\phi^{t+1} = \phi^t - \lambda \nabla_\phi \mathcal{R}_\textrm{emp}(f_\phi)$, where $\lambda$ is the \textit{learning rate}\index{learning rate}.

Once the optimization is complete and we obtain the solution $\hat\phi$, it is natural to assess the quality of the trained model $f_{\hat\phi}$ on an independent \textit{testing dataset}\index{testing dataset}\footnote{It is important to never use the testing dataset to make decisions about the model.
For this purpose, another independent dataset, usually called the \emph{validation dataset}\index{validation dataset}, should be used (see Sec.~\ref{ML:sec:generalization}).}.
The empirical risk evaluated on the testing set is often larger than on the training set, and large differences indicate \textit{overfitting}\index{overfitting}, meaning that the model does not generalize well to the unseen data.
The ability to accurately predict on unseen data is referred to as \textit{generalization}\index{generalization} and the empirical risk on the test data provides a measure of the \textit{generalization error}.
In order to reduce the generalization error one might explore different model choices (\eg, neural network architectures), additional regularization\index{regularization} terms in the loss function, different learning rates, optimization algorithms, or early stopping\index{early stopping} criterion in the optimization.

In order to produce a binary electron vs. proton decision from the continuous output of the neural network, one typically chooses a threshold (\ie, classify as proton if $f_{\hat\phi}(x)>c$).
The choice of the threshold $c$ is often referred to as a working point and it sets the tradeoff between electron and proton efficiencies, fake rates, and purities.
A \textit{receiver operating characteristic curve}\index{ROC, receiver operating characteristic curve}, or ROC curve, is used to summarize the tradeoff between true positive rate (TPR) and false positive rate (FPR).
Importantly, the characterization of the efficiency and rejection power (or equivalently the ROC curve) requires labeled data.
In a particle physics context, it is recognized that the simulation is not perfect and the mismodeling is associated to the presence of systematic uncertainty.
The discrepancy between the distribution of the training dataset and the distribution of the data where the model will be applied is referred to as \textit{domain shift}\index{domain shift} or \textit{distribution shift}\index{distribution shift}.
While mismodeling in the training dataset might lead to a suboptimal classifier, the real source of systematic uncertainty comes from the mismatch between the data used to characterize the performance of the classifier and the unlabeled data that the classifier is applied to.
This motivates the use of data-driven methods to calibrate the resulting model. 

This example provides a vertical slice through the various aspects of supervised machine learning in particle physics.
Now we factorize and abstract the various ingredients in order to provide a more general treatment with a broader scope. 





\section{Supervised learning}\label{ML:sec:supervised}

Supervised learning\index{supervised learning} generally refers to the class of problems where the training dataset\index{training dataset} are presented as input-output pairs $\{(x_i, y_i)\}_{i=1,\dots,n}$, where $x_i \in \mathcal{X}$ are the input features and $y_i \in \mathcal{Y}$ are the corresponding target labels.
Furthermore, it is typically the case that $x_i$ and $y_i$ are independent and identically distributed (i.i.d.) according to the data distribution $p(x, y)$, $(x_i,y_i) \overset{\text{i.i.d.}}{\sim} p(x,y)$, though $p(x,y)$ is usually not known explicitly.


The goal of supervised learning is find a function $f:\mathcal{X} \to {Y}$ that `best' captures the relationship between the input features and the corresponding target labels, similar to parameter estimation described in Sec.~\crossref{stat:sec:paramest} of the Statistics chapter.
Section~\ref{ML:sec:loss_risk_Remp} discusses how we quantify which function is best.

\subsection{Loss, risk, empirical risk}\label{ML:sec:loss_risk_Remp}

The term \textit{learning} in machine learning generally refers to optimization of some objective, which can be thought of as minimizing \textit{risk}\index{risk}.
The risk brings together three main ingredients.
The first is the \textit{model family} $\mathcal{F}$ (where $f\in \mathcal{F}$ is the quantity that we vary during optimization), the second is the \textit{loss function} $\mathcal{L}$, and the third is a data distribution $p(x, y)$. 
The \textit{risk}\index{risk} for a model $f\in \mathcal{F}$ is defined as its expected loss
\begin{equation}
    \label{ML:eq:risk}
    \mathcal{R}[f] \coloneqq \mathbb{E}_{p(x, y)}[\mathcal{L}(y, f(x))] \equiv \int \mathcal{L}(y, f(x))\, p(x, y) \dif x \dif y \;,
\end{equation}
where $\mathbb{E}_p[\cdot]$ refers to the expectation with respect to the distribution $p$.
Written this way, the risk is a functional, and the idealized goal for machine learning is to solve the optimization problem
\begin{equation}
    \label{ML:eq:fstar}
    f^{*}=\argmin _{{f\in {\mathcal{F}}}}\mathcal{R}[f] \; , 
\end{equation}
where $\mathcal{F}$ would include all possible functions. 


One of the defining characteristics of machine learning in practice is that one does not know the data distribution $p(x, y)$, but does have access to samples from that distribution, \ie $\{x_i, y_i\}_{i=1,\dots,n}$ with $(x_i, y_i) \overset{\text{i.i.d.}}{\sim} p(x, y)$.
This leads to the corresponding \textit{empirical risk}\index{risk}
\begin{equation}
    \label{ml:eq:Remp}
        \mathcal{R}_\textrm{emp}[f] \coloneqq \mathbb{E}_{\hat{p}(x, y)}[\mathcal{L}(y, f(x))] \equiv \frac{1}{n} \sum_{i=1}^n  \mathcal{L}(y_i, f(x_i)) \;,
\end{equation}
where $\hat{p}(x, y) = \frac{1}{n} \sum_{i=1}^n \delta(x - x_i)\delta(y - y_i)$ is referred to as the empirical distribution of the dataset $\{(x_i, y_i)\}_{i=1,\dots,n}$. 
The \textit{empirical risk minimization}\index{empirical risk}\index{risk minimization} principle is a core idea in statistical learning theory~\cite{vapnik2013nature}, which approximates $f^*$ with its empirical analogue
\begin{equation}
    \label{ML:eq:fhat}
    \hat{f}=\argmin _{{f\in {\hat{\mathcal{F}}}}}\mathcal{R}_\textrm{emp}[f] \; ,
\end{equation}
where $\hat{\mathcal{F}}$ is the set of all possible functions
parametrized by the model parameters $\phi$. In an idealized 
infinite parameter limit machine learning functions, 
such as neural networks\index{neural network}, are often universal 
approximators\index{universal approximation}, meaning they cover all functions and  $\hat{\mathcal{F}}=\mathcal{F}$.
For finite size models, this may not be a valid approximation.
Expressivity of the network characterizes this universality property and is a function of the network architecture and its parameters such as 
width and depth of neural network layers. 
If the expressivity is too small 
it leads to 
underfitting\index{underfitting}.
However, an equally important consideration 
is the risk of overfitting\index{overfitting} if we
optimize Eq.~\ref{ML:eq:fhat} for too 
long or use an unrestricted model class (see Sec.~\ref{ML:sec:regularization}).


While the loss function may quantify some well-motivated notion of risk, it is also common to design loss functions so that $f^*$ has some desired property.
In Secs.~\ref{ML:sec:regression}--\ref{ML:sec:regularization}, we will consider several such loss functions where one can show that the corresponding $f^*$ has the desired property even if the form of the loss is not obvious. 
Furthermore, there are often multiple loss functions that can lead to the same $f^*$.
Thus, one can think of machine learning as solving Eq.~\ref{ML:eq:fhat} with a sufficiently flexible model, powerful optimization algorithms, and practical considerations to break the degeneracy between different loss functions that lead to the same $f^*$.
As we shall see, commonly used loss functions can also be mathematically derived from a probabilistic approach.

\subsection{Regression}\label{ML:sec:regression}

The goal of regression\index{regression} is to predict a label $y\in \mathcal{Y}$ given an input feature vector $x \in \mathcal{X}$.
Typically, the label is a real-valued scalar, but $\mathcal{X}$ can be $\mathbb{R}^d$ or some more structured target (\eg, an image, sequence, graph, quantile, or distribution).
When $\mathcal{Y}$ is discrete, the task is usually referred to as classification (see Sec.~\ref{ML:sec:classification}); however, the two are closely related and \textit{logistic regression}\index{logistic regression} is an example where the model predicts a continuous probability associated to the possible label values. 
In elementary statistical language, the target label $y$ is often called a dependent variable, while the feature $x$ is called the independent variable. In classical statistics, one often assumes a model for the data such as 
\begin{equation}
    \label{ML:eq:classical_regression}
    y_i = f_\phi(x_i) + e_i \;,
\end{equation}
where $e_i$ is an additive error term that is often assumed to be independent of $x$ and normally distributed.
This leads to classic approaches like least-squares (see Sec.~\crossref{stat:sec:ls}), and when the model $f_\phi$ is linear in $\phi$ (not in $x$!) linear regression\index{linear regression}, which has a closed-form solution.
However, we can relax these assumptions and consider the general case of an arbitrary joint distribution $p(x,y)$, which can be written as $p(y|x)p(x)$ without loss of generality (see Sec.~\crossref{prob:sec:probGeneral}).
Consider the \textit{squared error} as a loss function, which corresponds to the mean-squared error (MSE)\index{MSE, mean-squared error}\index{L2 loss} empirical risk:
\begin{equation}
    \label{ML:eq:squared-error}
    \mathcal{L}_\textrm{MSE}(y, f(x)) = (y - f(x))^2 \;.
\end{equation}
One might expect that the squared error would only be appropriate in the case that the conditional distribution $p(y|x)$ is normally distributed, but one can use the calculus of variations to show that in general 
\begin{equation}
    \label{ML:eq:fstar_MSE}
    f^*_\textrm{MSE}(x) = \mathbb{E}_{p(y|x)} [y] \;,
\end{equation}
that is the optimal regressor for the MSE is the conditional expectation of $y$ given $x$. 

One issue with the squared-error as a loss function is that it is very sensitive to outliers.
Alternatively, one can use the absolute error $|y - f(x)|$\index{MAE, mean-absolute error}\index{L1 loss} as a loss function\footnote{The absolute error and squared error are often denoted as L1 and L2 errors, respectively, in reference to the corresponding norms.}.
However, the discontinuous derivative of the absolute (L1) error leads to challenges in optimization.
As a result there are various other loss functions, such as the Huber loss\index{Huber loss}, that aim to be both robust and more amenable to optimization. 

Note that this framing of regression yields a function $f(x)$ that only provides a point estimate for $y$. An alternative approach to regression is to model the full conditional distribution $p(y|x)$.
One such example is Gaussian process regression\index{GP, Gaussian process}, which is discussed in Sec.~\ref{ML:sec:GP}.
In that probabilistic approach, one can still obtain a point estimator, such as the conditional expectation or the maximum 
a posteriori (MAP) estimator
\begin{equation}
    f^*(x)=\argmax_y p(y|x) \; ,
\end{equation}
and one can also derive uncertainty estimates on the predicted value $y$ (see Sec.~\ref{ML:sec:uncert} for more details).
In this setting, the prior distribution on the model family is closely related to the concept of regularization, which we touch on in Secs.~\ref{ML:sec:regularization} and~\ref{ML:sec:GP}.

When one directly models $p(y|x)$, or goes further to model the joint distribution $p(x,y) = p(y|x)p(x)$, then one can use maximum likelihood for the loss function.
In that approach, the problem is really one of density estimation, which is a type of unsupervised learning that we discuss in Sec.~\ref{ML:sec:density_estimation}. 
These two approaches are a classic examples of two different approaches to modeling.
Regression with $f^*_\textrm{MSE}(x)$ is the prototypical example of \textit{discriminative} modeling\index{discriminative model}, while modeling the joint distribution is a prototypical example of \textit{generative} modeling\index{generative model}.
Generally, discriminative approaches with supervised learning outperform generative approaches when there is sufficient data, but generative approaches can be beneficial in data-starved settings~\cite{NgJ01}.

\subsection{Classification}\label{ML:sec:classification}

The goal of classification\index{classification} is to predict one of a finite number of class labels $y\in \mathcal{Y}$ given an input feature vector $x \in \mathbb{X}$.
It is similar to regression in this way, but the focus is on discrete target space $\mathcal{Y}$.
An important special case is when the label can only take on one of two values (\eg, ``signal'' or ``background''), which is referred to as binary classification\index{binary classification} and is equivalent to simple hypothesis testing in statistics.
It is common for a classifier to be the composition of two functions: $f(g(x))$.
The first function $g:\mathcal{X} \to \mathbb{R}^{|\mathcal{Y}|}$ predicts continuous probabilities for each class (\ie, $g_c(x) \approx p(y=c|x)$).
The second function $f:\mathbb{R}^{|\mathcal{Y}|}\to \mathcal{Y}$ then chooses the discrete label $y\in\mathcal{Y}$, such as $f(g(x)) = \argmax_{c} g_{c}(x) \approx \argmax_y p(y|x)$.
This is the case for both classical methods like logistic regression and modern, deep learning approaches to classification; therefore, we will use the term probabilistic classifier for $g(x)$ or just classifier when it is clear in context.

An intuitive loss function for classification is the zero-one loss, which simply counts the number of mis-classifications:
\begin{equation}
    \label{ML:eq:zero-one}
       \mathcal{L}_\textrm{0/1}(y, f(x))= 
\begin{cases}
    0 ,& \text{if } f(x) = y\\
    1,              & \text{otherwise} \; .
\end{cases}
\end{equation}
The zero-one loss can also be written as $\mathcal{L}_\textrm{0/1}(y, f(x)) = \mathbf{1}(y \ne f(x))$, where $\mathbf{1}(\cdot)$ is the indicator function.
The zero-one loss is non-differentiable, so it does not pair well with gradient-based optimization. 

For binary classification, one can use $y=\{0,1\}$ as numerical values for the class labels and the \textit{binary cross-entropy} loss function\index{cross-entropy loss}
\begin{equation}
\label{ML:eq:binary-cross-entropy}
\mathcal{L}_\textrm{bxe}(y, g(x)) = - 
\left [ y \log(g(x)) + (1-y)\log (1- g(x))\right ]\,.
\end{equation}
The resulting model will approximate $f^*_\textrm{bxe}(x)$, which takes on the form 
\begin{equation}
    \label{ML:eq:fstar_MSE_binary}
    f^*_\textrm{bxe}(x) = \mathbb{E}_{p(y|x)} [y]  = p(y=1|x) = \frac{p(x|y=1)p(y=1)}{p(x|y=0)p(y=0)+p(x|y=1)p(y=1)} \;.
\end{equation}
That is the binary cross-entropy loss for binary classification\index{classification} leads to the Bayesian posterior probability that the label $y=1$ given the feature vector $x$ (see Bayes theorem in Sec.~\crossref{prob:sec:probGeneral}).

Equation~\ref{ML:eq:fstar_MSE_binary} highlights an important feature of supervised learning relevant for particle physics: the joint distribution $p(x,y)$ of the training dataset implies a prior distribution $p(y)$ on the labels or classes.
This prior distribution reflects the frequency in the training dataset, not necessarily in the real data.
When applying the resulting model to a different dataset with the same conditional distribution (data likelihood) $p(x|y)$ for the features and a different prior $p^\prime(y)$ for the labels, the probabilistic interpretation of the result will not be properly calibrated, meaning $g(x) \not\approx p(y|x)$.
A common choice for binary classification is to use a balanced training dataset with $p(y=0)=p(y=1)=\frac{1}{2}$, while in many cases the true $p^\prime(y=1)$ in the experimental data might be very small (\ie, low signal-to-background), unknown, or zero (\ie, a hypothetical particle that does not exist).

If $p^\prime(y)$ and $p(y)$ are known then Bayes theorem can be used to re-calibrate the posterior $p(y|x)$ from one prior to another.
One example of such re-calibration is the correspondence of binary classification to simple hypothesis tests in frequentist statistics discussed in Sec.~\crossref{stat:sec:hyptest} of the Statistics chapter.
In that setting, the Neyman-Pearson lemma states that the optimal classifier is given by the likelihood ratio
\begin{equation}
    \label{ML:eq:fstar_neyman-pearson}
    f^*_\textrm{NP}(x) = \frac{p(x|y=1)}{p(x|y=0)} \;,
\end{equation}
which does not depend on the prior probabilities $p^\prime(y=0)$ or $p^\prime(y=1)$ as in Eq.~\ref{ML:eq:fstar_MSE_binary}, or, equivalently, assumes equal priors $p^\prime(y=0)=p^\prime(y=1)$.
Bayes theorem can be used to show that the two functions, $f^*_\textrm{NP}(x)$ and $f^*_\textrm{bxe}(x)$, are related by a one-to-one, monotonic transformation
\begin{equation}
    \label{ML:eq:lrt}
    f^*_\textrm{NP}(x) = \frac{p(y=0)}{p(y=1)}\frac{f^*_\textrm{bxe}(x)}{1-f^*_\textrm{bxe}(x)} \;,
\end{equation}
which is referred to as the \textit{likelihood-ratio trick}\index{likelihood-ratio trick} and plays an important role in simulation-based inference\index{simulation-based inference}\index{inference} (see Sec.~\ref{ML:sec:SBI}).

A standard way to evaluate the performance of a classifier is to evaluate the true positive rate (TPR)---the proportion of $y=1$ samples that are correctly identified based on a fixed threshold $g(x) > c$---as a function of the false positive rate (FPR)---the proportion of $y=0$ samples that are misidentified based on the same fixed threshold.
Plotting these values generates a graph known as the receiver operating characteristic (ROC)\index{ROC, receiver operating characteristic} curve.
Importantly, the monotonic transformation of Eq.~\ref{ML:eq:lrt} does not impact the tradeoff between FPR and TPR, therefore the ROC curves for $f^*_\textrm{NP}(x)$ and $f^*_\textrm{bxe}(x)$ are identical and do not depend on the prior probabilities $p(y)$.
This property has been leveraged in \textit{weakly supervised}\index{weakly supervised learning} approaches~\cite{Metodiev:2017vrx} to train a classifier in data without access to labels as long as one has two datasets with different $p(y=1)/p(y=0)$ ratios and the same conditional distribution $p(x|y)$ of the features given the labels. 

A generalization of Eq.~\ref{ML:eq:binary-cross-entropy} that applies to multiple classes, is the \textit{categorical cross-entropy} loss\index{categorical cross-entropy loss}
\begin{equation}
    \label{ML:eq:cross-entropy}
    \mathcal{L}_\textrm{xe}(y, f(x))= - \sum_{c\in |\mathcal{Y}|} \mathbf{1}(y=c) \log(f_c(x)) \; ,
\end{equation}
where $f:\mathcal{X} \to \mathbb{R}^{|\mathcal{Y}|}$ and the indicator function picks out the term in the sum for the corresponding class label $y$.
This loss can be derived by maximizing the posterior of Eq.~\ref{ML:eq:pxy} using a discrete set of class labels $y$, which identifies $f_c(x)=\tilde{f}(y=c|x)=p(y=c|x)$ and thus assumes the constraint $\sum_c f_c(x)=1$ and $f_c(x)\ge 0$ (see Sec.~\ref{ML:sec:softmax} for an activation function that enforces this).
The function $\tilde{f}(y|x)$ can be interpreted as a conditional distribution, \ie, an approximation to the true posterior $p(y|x)$.
The risk associated to the cross entropy loss function is 
\begin{equation}
    \label{ML:eq:xe_risk}
    \mathcal{R}_\textrm{xe}[f] = \mathbb{E}_{p(x,y)}\left[ -\sum_{c\in |\mathcal{Y}|} \mathbf{1}(y=c) \log f_c(x) \right] = - \sum_{c\in |\mathcal{Y}|} p(y=c) \mathbb{E}_{p(x|y)}[ \log \tilde{f}(y=c|x)]
    \;.
\end{equation}
This is equivalent to $\mathcal{R}_\textrm{xe}[f] = \mathbb{E}_{p(x)}[H[p(y|x), \tilde{f}(y|x)]]$, where 
\begin{equation}
\label{ML:eq:cross}
H[p,f]\equiv \mathbb{E}_p[-\log f] = - \int p(x)\log (f(x))\dif x
\end{equation} is the cross entropy between the two distributions.
One can use a Lagrange multiplier to enforce the normalization constraint and the calculus of variations to show that
\begin{equation}
    \label{ML:eq:fstar_xe}
    f^*_{\textrm{xe},c}(x) = p(y=c|x)\;,
\end{equation}
which is equivalent to the solution in Eq.~\ref{ML:eq:fstar_MSE_binary} in the binary case.

This approach is closely related to the loss functions that are used for density estimation, the forward Kullback-Leibler (KL) divergence\index{KL, Kullback-Leibler divergence}, and the maximum likelihood estimation. 
Minimizing cross entropy $H[p,f_\phi]$ to $\phi$ is equivalent to minimizing the forward KL divergence
\begin{equation}\label{ML:eq:KL}
    \text{KL}(p\|f_\phi)\coloneqq \mathbb{E}_p[ \log p(x))- \log f_\phi] = H[p,f_\phi] - H[p] \; , 
\end{equation}
where $H[p] \coloneqq \int p(x) \log p(x) dx$ is the entropy and independent of $f_\phi$.
The KL divergence $\text{KL}[p\|f] \ge 0$, and equal if and only if $p=f$.

Unlike in the binary classification\index{binary classification} case, the multi-class classifier\index{multi-class classification} is sensitive to the priors $p(y)$ used in training.
This leads to complications as often the class proportions are unknown.
For example, one might be interested in classifying a signal when multiple backgrounds are present and the relative proportion of those different background components is uncertain.
Ideally one would like the class proportions for the background components used in training to match those in the data, which presents an additional training challenge if those proportions are heavily unbalanced.

\subsection{Generalization and model complexity}
\label{ML:sec:generalization}
With a sufficiently flexible model, it is possible to fit the training dataset\index{training dataset} very well, though the model might not \textit{generalize}\index{generalization} well to unseen data, a phenomenon known as \textit{overfitting}\index{overfitting}.
More concretely, for a nonnegative loss function one might have $\mathcal{R}_\textrm{emp}[\hat{f}] \to 0$, while the true risk $\mathcal{R}[\hat{f}]$ might be large.
Conversely, \textit{underfitting} occurs when a model is unable to capture the relationship between the inputs and labels accurately, resulting in large empirical and true risks.
While it is generally not possible to evaluate $\mathcal{R}[\hat{f}]$ exactly because we do not know $p(u)$, we can use an independent dataset (also called validation dataset\index{validation dataset}) to obtain an unbiased estimate of it.
This \textit{cross-validation}\index{cross-validation} method motivates the test-train-validation split of the data. 

Intuitively, a model with many parameters has more flexibility and is more prone to overfitting\index{overfitting}.
However, some highly over-parameterized models (that have large subspaces of their parameters where $\mathcal{R}_\textrm{emp}[\hat{f_\phi}] \to 0$) generalize well~\cite{zhang2021understanding-2, nakkiran2019deep}.
Often this is achieved through \textit{regularization}\index{generalization}, both explicit and implicit (Sec.~\ref{ML:sec:regularization}).

Two main sources of error prevent models from generalizing beyond their training dataset.
One is \textit{bias} arising from erroneous assumptions in the model and the other is \textit{variance} arising from sensitivity to statistical fluctuations in the training dataset.
The \textit{bias-variance decomposition}\index{bias-variance tradeoff} is a way of analyzing a model's expected risk as a sum of bias and variance terms.
Concretely, if $\mathcal{L}$ is the squared loss, one can decompose the expected risk $\mathbb{E}_\mathcal{D}[\mathcal{R}_\textrm{emp}[\hat f_\phi]]$ over all possible training datasets $\mathcal{D}$ into three terms~\cite{hastie01statisticallearning,Mostafa2012},
\begin{align}
    \mathbb{E}_\mathcal{D} \left [\mathcal{R}[\hat f_\phi^\mathcal{D}]\right ] &= \mathbb{E}_\mathcal{D} \mathbb{E}_{p(x, y)} \left [ (y - \hat f^\mathcal{D}_\phi(x))^2\right ] \\
    &=\mathbb{E}_{p(x)}\left [ \underbrace{\mathbb{E}_{p(y|x)} \left[(y - \bar y)^2\right]}_{\text{noise}} + \underbrace{\mathbb{E}_\mathcal{D} \left [(\hat f_\phi(x) - \bar f(x))^2 \right ]}_{\text{variance}} + \underbrace{(\bar y - \bar f(x))^2}_{\text{bias}} \right ]\,,
    \label{ML:eq:bias_variance}
\end{align}
where $\bar f(x) \equiv \mathbb{E}_\mathcal{D}\left [ \hat f_\phi^\mathcal{D}(x)\right]$ is the ``average'' prediction of the model over different possible training datasets.
In this expression, the first term is the inherent ``noise'' in the dataset, \ie, the variance of $y$ around its mean, which is zero if $y$ is deterministically related to $x$.
The second term is the variance of the model around its average when considering different training datasets, and the third term is the squared bias, \ie, the difference between the average prediction and the true conditional mean.

Classically, there is a correspondence between overfitting and underfitting and the concepts of bias and variance discussed in Sec.~\crossref{stat:sec:paramest} on parameter estimation.
Overfitting implies high variance: the model class is too complex and retraining yields vastly different models.
Variance tends to increase with model complexity and decrease with more training data.
Underfitting implies high bias: the model class is too simple and has a large error rate.
Thus, there exists a tradeoff between bias and variance, shown schematically in Fig.~\ref{ML:fig:bias_variance} (left).

However, in modern machine learning, very high-capacity models such as neural networks can be trained to exactly fit the data, and yet obtain high accuracy on test data~\cite{Belkin_2019}, as shown in Fig.~\ref{ML:fig:bias_variance} (right).
This phenomenon is known as ``double descent.''
The apparent contradiction may be addressed by considering the regularizing (Sec.~\ref{ML:sec:regularization}) effects of neural network training, specifically stochastic gradient descent (Sec.~\ref{ML:sec:sgd}).

\begin{pdgxfigure}
    \centering
  \includegraphics[width=0.34\textwidth]{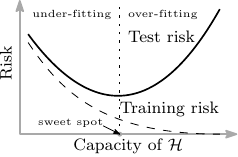}
  \includegraphics[width=0.6\textwidth]{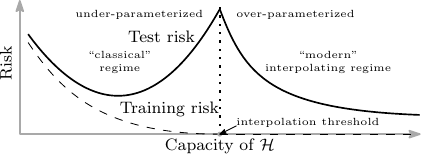}
    \caption{Curves for training risk (dashed line) and test risk (solid line) from Belkin et al. in Proceedings of the National Academy of Sciences, 2019.
    The classical U-shaped risk curve arising from the bias-variance trade-off (left) and the double descent risk curve (right), which incorporates the U-shaped risk curve (\ie, the ``classical'' regime) together with the observed behavior from using high capacity function classes (\ie the ``modern'' regime).}
    \label{ML:fig:bias_variance}
\end{pdgxfigure}

\subsection{Regularization}
\label{ML:sec:regularization}
The trained model $\hat{f}$, or equivalently, the parameters of the trained model $\hat{\phi}$ can be thought of as point estimates of $f^*$.
The bias-variance tradeoff means that introducing a small bias can often lead to a significant reduction in variance.
This motivates the explicit addition of a \textit{regularization}\index{regularization} term to the loss function, which will introduce some bias $f^*_\textrm{reg} \ne f^*$.
A common form of regularization is to penalize by the L2 norm\index{L2 regularization} of the parameters (\ie $\lVert \phi \rVert^2$), which is referred to as \textit{L2 or Tikhonov regularization}.
This appears in the form of penalized maximum likelihood, and it is also commonly used in unfolding~\cite{Kuusela:2015xqa}.
Alternatively, one can penalize by the L1 norm\index{L1 regularization} $\lVert \phi\rVert$, which is known as \textit{L1 regularization}.
One can also interpret the regularization term as an explicit prior on the parameters, and the resulting model as the Bayesian maximum a posteriori (MAP)\index{MAP, maximum a posteriori} estimator.
When L1 or L2 regularization is paired with linear regression, it is known as \textit{LASSO regression} or \textit{ridge regression}, respectively.
In addition, L2 regularization paired with kernel machines gives rise to Gaussian process regression.

These two types of explicit regularization generally have solutions with different properties.
For example, L1 regularization naturally induces sparsity, \ie, a fraction of the parameters are nearly zero, whereas L2 regularization tends to keep all parameters nonzero but with lower magnitudes, as illustrated in Fig.~\ref{ML:fig:regularization}.
Because L1 regularization sets certain parameters to zero, it is often used as part of feature selection and model compression techniques, as discussed in Sec.~\ref{ML:sec:deployment}.

\begin{pdgxfigure}
    \centering
    \includegraphics[width=0.45\textwidth]{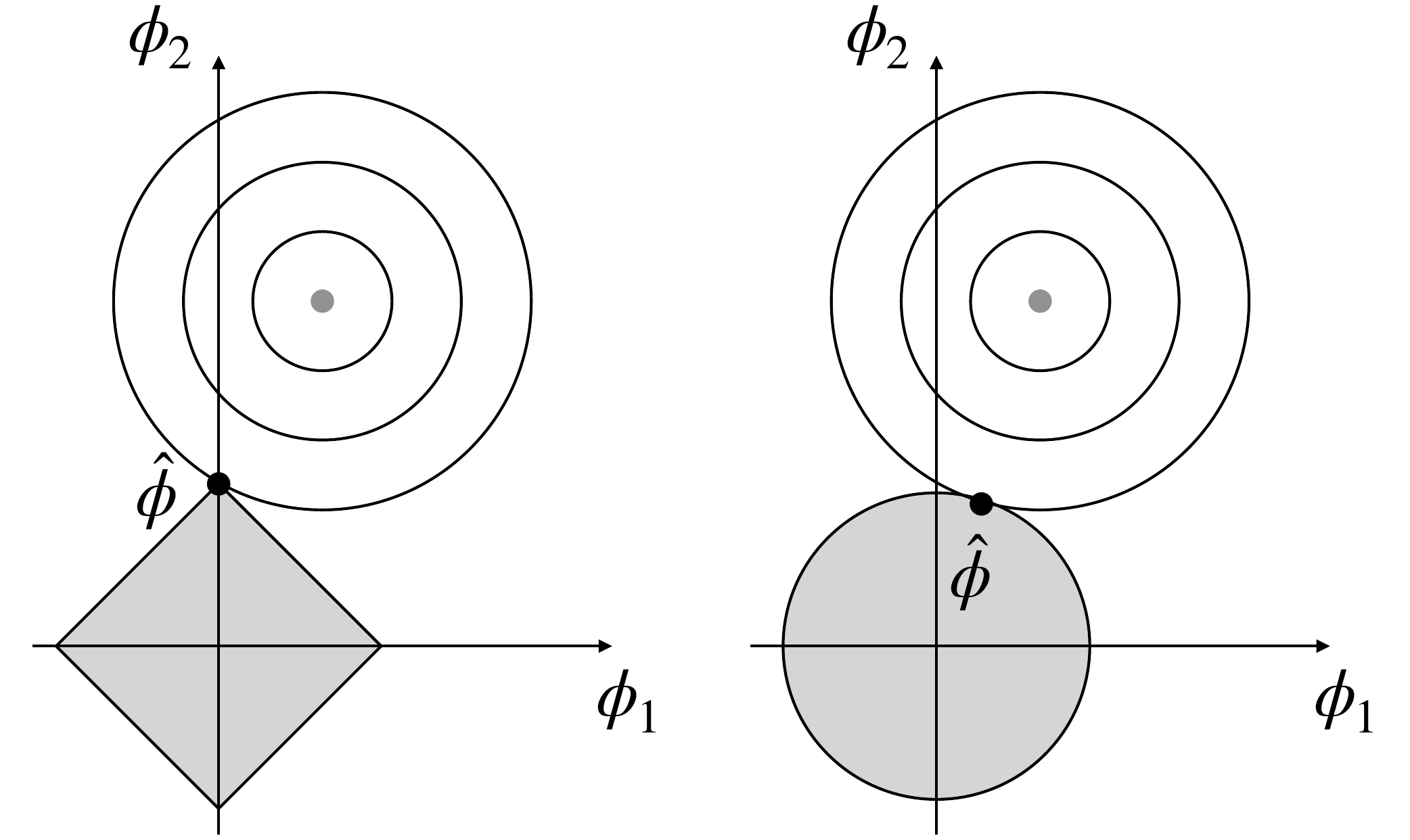}
    \caption{Depiction of L1 (left) and L2 (right) regularization constraint regions and the contours of an unregularized loss function.
        The intersection with the L1 constraint region gives an optimal value $\hat\phi$ that is sparse, \ie, $\phi_1=0$, while the L2 contraint region yields an optimal value $\hat\phi$ where both $\phi_1$ and $\phi_2$ are small, but nonzero.}
    \label{ML:fig:regularization}
\end{pdgxfigure}

Another form of regularization is to restrict the model class $\hat{\mathcal{F}}$.
For example, a neural network and a sequence of narrow step functions (delta functions) can both be shown to be 
universal approximators\index{universal approximation} in infinite parameter size limit, but on real world  examples the former generalizes much better than the latter.
Within the class of neural network\index{neural network} models, convolutional neural networks\index{CNN, convolutional neural network} are a subset of generic feedforward neural networks that approximately preserve translational symmetry (see Sec.~\ref{ML:sec:cnn} for more discussion).
These types of choices are often encoded in the architecture of a neural network and are broadly referred to as \textit{inductive bias}\index{inductive bias} in the model.

In addition to explicit regularization terms in the loss function or through restrictions to the model class, it is also possible to regularize implicitly. 
One implicit regularization is through early stopping~\cite{DBLP:journals/corr/RosascoTV14,Kuusela:2015xqa}\index{early stopping}, where we monitor the 
loss on the training dataset\index{training dataset} and the loss on held-out validation dataset\index{validation dataset}.
While the training loss continues to decrease with more gradient descent cycles\index{gradient descent}, the validation loss may not, and early stopping stops the training\index{training} when validation loss flattens out or begins to increase.
Another powerful form of regularization used in deep learning models is known as \textit{dropout}~\cite{dropout}\index{dropout}, which randomly removes some some parts of the model during training and can be thought of as implementing a type of model averaging~\cite{baldi2013understanding}.

The chosen numerical optimization procedure can also act as an implicit regularization.
In the case of highly over-parameterized models where there is a large degenerate parameter space that achieves zero loss, $\Phi_0 = \{\phi | \mathcal{R}_\textrm{emp}[f_\phi]\} = 0$, the dynamics of the optimization algorithm will break the degeneracy and favor some particular $\hat{\phi} \in \Phi_0$ as if an additional regularization term was included.
Despite zero loss and over-parametrization, the corresponding generalization error may be small, a phenomenon called \textit{benign overfitting}\index{overfitting}~\cite{Belkin18}. 
Different optimization algorithms will have different implicit regularization effects, and thus favor different parameter points in $\Phi_0$ that will have different generalization error~\cite{pmlr-v80-gunasekar18a}.
Understanding this interaction is a topic of contemporary research in machine learning~\cite{zdeborova2020understanding}. 

Some methods such as Gaussian process (GP) do not require
optimization, and instead use linear algebra to obtain the 
solution. Benign overfitting
is explicit for GP in that in the absence
of noise the solution goes through 
all the training data, yet it generalizes well if the kernel 
is well chosen. Infinitely 
wide neural networks have 
an explicit correspondence to 
Gaussian process~\cite{neal1994priors}.
When applied to deep 
networks this leads to the concept of neural tangent kernel~\cite{JacotHG18}. 

\section{Unsupervised learning}\label{ML:sec:unsupervised}
Unsupervised learning\index{unsupervised learning} generally refers to the class of problems that use unlabeled training dataset $\{x_i\}_{i=1,\dots,n}$, where $x_i \in \mathcal{X}$ are the input features. Furthermore, it is typically assumed that $(x_i) \overset{\text{i.i.d.}}{\sim} p(x)$, though $p(x)$ is usually not known explicitly. Finally, the loss function in unsupervised learning takes on the special form $\mathcal{L}(x, f(x))$.
This class of learning has many 
different applications, such as 
density estimation, anomaly detection, 
generative learning, representation learning and clustering, each 
with the corresponding set of methods. Some
of these tasks can be achieved with 
the same methods, \eg normalizing flows (see Sec.~\ref{ML:sec:flows}) can perform density estimation, 
generative sampling 
and anomaly detection.

\subsection{Representation learning\index{representation learning}, compression, and autoencoders}\label{ML:sec:representation}


A recurring topic in machine learning and statistics is how to represent the data.
Much of classical statistics involves constructing a low-dimensional summary statistic that extracts the relevant information from the data for a particular task (a sufficient statistic in the language of classical 
statistics).
There is a spectrum of representations with tradeoffs. At one end of this spectrum is lossless compression that allows one to encode the data into a smaller, intermediate representation that carries all the information since it can be decoded back into the original data.
At the other end of the spectrum is something like the likelihood ratio, which is a single scalar that carries the relevant information needed for hypothesis testing for a single hypothesis, but it discards all the other information that might be needed for other tasks, such as testing other hypotheses.
An intermediate point in this spectrum is the process of feature engineering\index{feature engineering}, which refers to the creation of new features $\mathcal{X}^\prime$ from the original features $\mathcal{X}$ in hopes that the downstream task will be easier with the new features.
For example, instead of working directly with the energy and momentum of particles, one might compute invariant masses or angles between particles.
This type of feature engineering generally improves performance for shallow neural networks and decision trees; however, with the rise of deep learning this is often no longer necessary and may limit performance compared to working with the original features~\cite{Guest:2018yhq}.
One can think of the intermediate layers of a neural network between the input and the output a representation of the data that is good for the task at hand, and by training all the layers of the network simultaneously (or ``end-to-end'') one can see the intermediate layers as a learned representations.
For a review, see Ref.~\cite{bengio2013representation}.

An example of a linear dimensionality reduction representation and data compression is principal component analysis (PCA)\index{PCA, principle component analysis} of data ${x}\in \mathbb{R}^d$ at fixed latent space dimensionality $k$ ($k<d$), which finds the orthogonal linear transformation, ${O}$,
\begin{equation}
    {O}: \mathbb{R}^k \to \mathbb{R}^d, {z} \mapsto {O}{z},\, {O}{O}^\intercal{=}I_d
\end{equation}
that maximizes the data variance in the latent space.
Maximizing the variance of the transformed data is equivalent to minimizing the average reconstruction error (the residual variance in data space), 
\begin{equation}
    \label{ML:eq:reconstruction_error}
    \mathcal{L}_\text{reco}(x, f(x)) = \lVert x - f(x) \rVert^2 \;.
\end{equation}
 A PCA can thus be interpreted as a linear, orthogonal model that is trained to minimize the $L_2$-distance between the input data and the reconstructed data given the fixed dimensionality $k$.
 In practice, the PCA problem can be solved analytically without the use of optimization algorithms or the loss function: the principal components are given by the eigenvectors of the data covariance matrix.

A suitable latent space dimensionality, $k$, is chosen by ordering the eigenvalues, $\lambda_i$, of the data covariance in descending order, and keeping only the first few eigenvectors that correspond to the largest eigenvalues.
The cut is often made at dimensionalities that capture around 90\% of the data variance.
For many data sets this results in $k\ll d$. The average reconstruction error that originates from the discarded eigenvalues is $\sigma_{\mathrm{reco}}^2{=}\sum_{i=k+1}^d \lambda_i$.


Another common type of representation learning and 
nonlinear dimensionality reduction is based on the \textit{autoencoder}\index{autoencoder} $f = g\circ e:\mathcal{X}\to \mathcal{X}$, where $e:\mathcal{X}\to \mathcal{Z}$ is referred to as the \textit{encoder}\index{encoder} and  $g:\mathcal{Z}\to \mathcal{X}$ is referred to as the
\textit{generator} or \textit{decoder}\index{decoder}. Typically the dimensionality of $\mathcal{Z}$ is much less than $\mathcal{X}$, and $z=e(x)$ can be thought of as a compressed representation of the input.
The intermediate space $\mathcal{Z}$ is sometimes referred to as the bottleneck\index{bottleneck} or the latent space of the autoencoder.
If the bottleneck is sufficiently large and the encoder and decoder are sufficiently flexible, then the function $f$ could just be the identity (\ie, lossless compression).
However, if the encoder and decoder are not sufficiently flexible or the dimensionality of the latent space is not large enough there will be some reconstruction error.
Therefore, the reconstruction error of Eq.~\ref{ML:eq:reconstruction_error} serves as a natural loss function of an autoencoder. 

Once trained, the encoder $e(x)$ can be used independently of the decoder to provide a generic low-dimensional representation of the data. The flexibility of this approach is attractive; however, there are no guarantees that this representation will be optimal for the other task. Indeed, the transition from pre-trained autoencoders to end-to-end learning is one of the important trends that characterized the onset of the deep learning era.

While achieving zero reconstruction error may seem good as it would imply lossless compression, it often performs poorly in practice.
First, the encoder may be overfit to the training dataset and not generalize well to held out data.
This can be addressed by adding a 
prior to the training, discussed in Sec. \ref{ML:sec:vae}. 
Second, it may not be robust to domain shift (see Sec.~\ref{ML:sec:domain_adaptation}).

\subsection{Clustering}\label{ML:sec:clustering}

The goal of clustering\index{clustering} is to group the data $\{x_i\}_{i=1,\dots,n}$ into $k$ groups, or \textit{clusters}, usually with $k \ll n$.
Intuitively, if two data points belong to the same cluster, then they should be similar in some sense.
Conversely, if two data points are very different, then they should be assigned to different clusters.
The notion of similarity usually is based on some heuristic, and there are a variety of algorithmic and probabilistic clustering algorithms.
In some cases $k$ is specified, while in others it is determined by the clustering algorithm.
There is also a distinction between flat clustering that directly partitions the data into $k$ clusters and hierarchical clustering where clusters are nested hierarchically as the name suggests.
In many cases, clustering uses some notion of distance $d(x_i, x_j)$, which may be the $L_p$ norm $\lVert x_i - x_j \rVert_p$. 

One of the most common clustering algorithms is known as $k$-means\index{k-means clustering}, where $k$ is specified by the user and results in sets $S = \{S_1, \dots, S_k\}$
that minimize the variance of each cluster.
Thus, the objective is
\begin{equation}
    \label{ML:eq:kmeans1}    
{\displaystyle {\underset {\mathbf {S} }{\operatorname {arg\,min} }}\sum _{i=1}^{k}\sum _{\mathbf {x} \in S_{i}}\left\|\mathbf {x} -{\boldsymbol {\mu }}_{i}\right\|^{2}={\underset {\mathbf {S} }{\operatorname {arg\,min} }}\sum _{i=1}^{k}|S_{i}|\operatorname {Var} S_{i}} = {\displaystyle {\underset {\mathbf {S} }{\operatorname {arg\,min} }}\sum _{i=1}^{k}\,{\frac {1}{2|S_{i}|}}\,\sum _{\mathbf {x} ,\mathbf {y} \in S_{i}}\left\|\mathbf {x} -\mathbf {y} \right\|^{2}}
\end{equation}
where $\mu_i$ is the mean of points in $S_i$. 
\label{ML:eq:kmeans2}
$k$-means can be interpreted as a Gaussian mixture density estimation of $p(x)$, where all 
the Gaussians are isotropic. It can be generalized to a Gaussian mixture model, where both the means and the covariance matrix are estimated. 

Among the other class of algorithms that determine $k$, density-based spatial clustering of applications with noise (DBSCAN) is one of the most frequently used.
DBSCAN clusters points based on a distance metric (\eg, Euclidean) defined for each application.
Two hyperparameters are $\epsilon$, the maximum distance threshold to determine whether a neighboring point belongs to the same cluster, and the minimum sample size for a group of close points to be identified as a valid cluster or noise.
While DBSCAN is robust against irregularly shaped clusters with a simple distance-based metric, single threshold parameter $\epsilon$ shared to distinguish all clusters can be challenging.
Hierarchical DBSCAN (HDBSCAN) generalizes to varying densities by building a hierarchy of density-based clusters across all $\epsilon$ via mutual-reachability distances, then extracts the most stable clusters from a condensed tree. 

Finally, neural networks are often used for clustering in particle physics.
One use case is to transform the data points into a latent space where clustering is performed using an unsupervised, traditional algorithm. 
For example, an input dataset may not follow an isotropic gaussian distribution which is assumed by $k$-means, but one can design a neural network to learn a transformation into the latent space where this assumption holds.
Another use case is to use neural network directly for clustering operation.
Examples include object detection~\cite{Acciarri_2017} and segmentation~\cite{Domine:2019zhm,Koh:2020snv} in computer vision (see Sec.~\ref{ML:sec:CNNApps}) as well as clustering of graph nodes via edge classification~\cite{Farrell:DLPS2017,Farrell:2018cjr,DeepLearnPhysics:2020hut,ExaTrkX:2021abe,Dezoort:2021kfk} (see Sec.~\ref{ML:sec:gnn}).


\subsection{Density estimation}\label{ML:sec:density_estimation}

The goal of density estimation\index{density estimation} is to estimate a distribution $p(x)$ based on samples $\{x_i\}_{i=1,\dots,n}$ with $x_i \overset{\text{i.i.d.}}{\sim} p(x)$. Conceptually, this is the same goal as when fitting a parameterized distribution $f(x;\theta)$ to data using the method of maximum likelihood as described in Sec. \crossref{stat:sec:ml} of the chapter on statistics. In practice, the difference in the machine learning context has to do with the flexibility of the model and the dimensionality of the data. A highly-flexible model, which can effectively approximate any distribution, is referred to as a non-parametric model (though, ironically, usually this means the model has many parameters). In contrast, typical maximum likelihood fits in particle physics are based on restricted families of distributions with relatively few parameters and the data is typically one- or two-dimensional, though occasionally five- or six-dimensional. 

Maximizing the likelihood function in Eq.~\crossref{stat:eq:likelihood}, $\mathcal{L}(\theta) = \prod_{i=1}^n f(x_i; \theta)$ is equivalent to minimizing the empirical risk:
\begin{equation}
    \label{ML:eq:max_likelihood} 
    \mathcal{R}_\text{emp,xe}[f_\phi] = - \frac{1}{n} \sum_{i=1}^n \log f_\phi(x) \;,
\end{equation}
where we adopt the notation used in this chapter.
The loss is simply $\mathcal{L}_\textrm{}(x,f_\phi(x)) = -\log f_\phi(x)$, and the corresponding risk is 
\begin{equation}
    \label{ML:eq:xe_risk_density} 
    \mathcal{R}_\text{xe}[f_\phi] = \mathbb{E}_{p(x)}[- \log f_\phi(x) ] 
    \;,
\end{equation}
which is the cross entropy $H[p, f_\phi]$.
For density estimation, the model is usually constructed to enforce $\int f_\phi(x) dx =1$ and $f_\phi(x) \ge 0$ so that it can be interpreted as a distribution.
With this constraint, one can show that $f^*_\text{xe}(x) = p(x)$. 
This is not the only form of training: 
flow matching and diffusion methods 
train on a different objective, 
discussed further below. 



The concepts of generalization and overfitting\index{overfitting} are
particularly acute in \emph{unsupervised} learning, where 
the likelihood maximization of equation \ref{ML:eq:max_likelihood},  
combined with universal approximator assumption, must converge onto  $\hat{p}(x) = \frac{1}{n} \sum_{i=1}^n \delta(x - x_i)$, the empirical distribution of the dataset $\{x_i\}_{i=1,\dots,n}$.
This distribution has the highest likelihood on the training dataset and the lowest likelihood on the 
test data where it gives $\hat{p}(x)=0$ 
as long as the test dataset\index{testing dataset} are not 
identical to the 
training dataset\index{training dataset}. So the empirical distribution of the training dataset has the worst possible generalization property, yet 
it is the solution we converge to for 
sufficiently expressive architectures 
in the absence of 
any regularization. In contrast, 
in supervised learning we often observe 
the phenomenon of benign overfitting, 
where even zero loss can generalize well. 

In addition to approaches to density estimation that involve learning in the sense of minimizing a loss or risk function, we note that there are also classical density estimation techniques such as histogramming and kernel density estimation~\cite{parzen1962estimation, davis2011remarks,Cranmer:2000du}. These techniques often fail in very high dimensions. 

\subsection{Generative models}
\label{ML:sec:deep_generative}

Deep generative models\index{generative model} are powerful machine learning models that can learn complex, high-dimensional distributions and generate samples from them. Because of their inherently probabilistic formulation, generative models are rapidly becoming an indispensable tool for scientific data analysis in a range of domains.  
%
The goal of generative 
models is to draw samples 
from $p(x)$. For some 
formulations of learned $p(x)$, such as 
normalizing flows, 
the samples can be 
drawn directly. For 
other explicit formulations of $p(x)$, 
such as Boltzmann machines, one can use 
sampling techniques such 
as Monte Carlo Markov chain sampling. There 
are however many other 
approaches to drawing 
samples from $p(x)$
that do not rely on its 
explicit form. 

Generative models can be contrasted against discriminative\index{discriminative model} models that are primarily used for supervised learning\index{supervised learning} tasks. Roughly, discriminative models are used for prediction and $f(x)$ provides a point estimate of the target $y$, and they are more closely connected to function approximation. In contrast, generative models describe the data distribution $p(x)$ (or the joint data distribution $p(x,y)$ in a supervised setting). An enlightening discussion of these two approaches can be found in Ref.~\cite{NgJ01}.

There are a number of different types of deep generative models  that have various pros and cons as they do not all have the same capabilities. We will focus on variational autoencoders (VAEs)~\cite{kingma2013auto,rezende2014stochastic} \index{VAE, variational autoencoder}, generative adversarial networks (GANs)~\cite{GANs, radford2015unsupervised}\index{GAN, generative adversarial network}, normalizing flows (NFs)~\cite{pmlr-v37-rezende15,2014arXiv1410.8516D,dinh2016density,kingma2018glow,kobyzev2020normalizing}\index{NF, normalizing flow}, 
and flow-matching\index{flow-matching model} and diffusion models\index{diffusion model}~\cite{lipman2023flowmatchinggenerativemodeling,SongE19,SDE,Albergo},
though other approaches have been explored in this quickly developing area of research. Consider these three distinct types of functionality:
\begin{itemize}
    \item \textbf{generation:} ability to sample or ``generate'' a data point $x_i \sim p(x)$.
    \item \textbf{likelihood for generated data:} ability to evaluate the probability density (likelihood) $p(x_i)$ for a data point $x_i$ sampled from the model $x_i \sim p(x)$.
    \item \textbf{likelihood for arbitrary data:} ability to evaluate the probability density $p(x_i)$ for an arbitrary data point $x_i \in \mathcal{X}$.
\end{itemize}
Each of the models above can be used for generation; however, only normalizing flows provide all three capabilities. For reasons that we will describe below, GANs and VAEs do not provide a tractable likelihood function, and they are sometimes referred to as \textit{implicit models}. This establishes a connection to simulation-based inference\index{simulation-based inference} where most scientific simulators are also implicit models with an intractable likelihood. Because normalizing flows have a tractable likelihood, they can be trained via maximum likelihood (Eq.~\ref{ML:eq:max_likelihood}) as described in Sec.~\ref{ML:sec:density_estimation}. GANs and VAEs, on the other hand, need to employ some other loss function to be trained. In the case of VAEs, training is based on the ELBO\index{ELBO, evidence lower bound objective} used in variational inference\index{variational inference} (see Sec.~\ref{ML:sec:classification} and the discussion around the reverse KL\index{KL, Kullback-Leibler divergence} divergence below Eq.~\ref{ML:eq:KL}). While GANs are also implicit models they data they can generate is typically restricted to a lower-dimensional manifold $\mathcal{M} \subset \mathcal{X}$, meaning that almost all real training dataset doesn't ``live on'' the subspace of possibilities that the model can produce. In this case, the likelihood is for almost all data is zero, and so even ELBO-based training will not work. The breakthrough idea introduced in Ref.~\cite{GANs} was to use adversarial training where a classifier would be used to quantify how different the data generated from the model is from the data from the target distribution. 



VAEs, GANs, and normalizing flows introduce a mapping $g(z,\theta)$ from a base random  variable $z$ to the space of the data $\mathcal{X}$. The map $g(z,\theta)$ is typically implemented with a neural network.  The random variable $z$ is sampled from some known base distribution $p(z)$ that is both easy to sample and has a density that is easy to evaluate. Typically, the base distribution is a multivariate normal. 

In the literature on GANs and normalizing flows, this base random variable is often referred to as a latent variable and $p(z)$ is often referred to as a prior distribution. 
In the case of VAEs, one additionally adds some normally-distributed (Gaussian) random noise $\epsilon$ to the output so that $x=g(z,\theta)+\epsilon$. In this case, $x$ and $z$ are not deterministically related and $z$ is a legitimate latent variable in the model and $p(z)$ can be interpreted as the prior on that latent variable. In this case, the model can populate the full space of the data. Unfortunately, the marginal likelihood $p(x) = \int p(x,z) dz$ involves an intractable integral, thus maximum likelihood training is infeasible. However, the likelihood term $p(x|z)$ is tractable (\ie the Gaussian noise), so training with the ELBO is possible. 

Note that the dimensionality of $z$ need not be the same as that of $x$. If $z\in \mathbb{R}^q$ and $\mathcal{X} =\mathbb{R}^d$ with $q<d$, then all points $g(z, \theta)$ will lie on a $d$-dimensional surface in $\mathbb{R}^d$. In the case of a VAE, the Gaussian noise $\epsilon$ means that the generated data $x$ will be distributed in a thin region around the surface defined by $g(z, \theta)$. 
The presence of a bottleneck\index{bottleneck} (\ie $q<d$) leads to advantages and disadvantages. The disadvantages for GANs is that the likelihood assigned to almost all real world data (\ie data not generated by the model) will be zero, so training is more difficult and many tasks in probabilistic inference won't be applicable. However, often real world data is also effectively described by a low-dimensional subspace in the full space of the data -- random images look like noise, while natural images are in some sense special. For this reason, images produced by GANs for instance often have better visual quality than those produced by other techniques.   This points to the ambiguity encountered in quantifying how close two distributions are, and also motivates the use of distance measures such as the Earth movers distance or Wasserstein distance~\cite{arjovsky2017towards, wiatrak2019stabilizing}\index{Wasserstein distance}. Conversely, the lack of a bottleneck (\ie $q=d$) leads to very large models and scalability issues when the data is high dimensional.

Recent work has also focused on combining ideas from VAEs, GANs, and normalizing flows so that the generative model does involve a bottleneck but can still provide tractable likelihoods for density estimation\index{density estimation} restricted to that manifold~\cite{pmlr-v37-rezende15,rezende2020normalizing, gemici2016normalizing, Brehmer:2020vwc,bohm2020probabilistic}. Some of these models can also be used in the context of anomaly detection and out of distribution detection by identifying data that is off the manifold. 


The parametrization of the mapping (the architecture of the neural network\index{neural network}) should match the structure of the data and be expressive enough. For problems 
with explicit symmetries it is  beneficial to include them into the 
architecture of the network explicitly, which restricts the 
allowed space of the models and matches
their inductive bias (implicit 
regularization inherently built into the choice 
of architecture of the network) to the data. 
Different architectures have been proposed~\cite{kingma2018glow, van2017neural,karras2017progressive, karras2019style}, 
and to achieve the best performance on a new dataset one needs extensive hyperparameter explorations~\cite{lucic2018gans}.


\subsubsection{Variational autoencoders}\label{ML:sec:vae}

The autoencoder was described in Sec.~\ref{ML:sec:representation} as model for compression and representation learning. The model is  $f = g\circ e:\mathcal{X}\to \mathcal{X}$, where $e:\mathcal{X}\to \mathcal{Z}$ is referred to as the \textit{encoder}\index{encoder} and  $g:\mathcal{Z}\to \mathcal{X}$ is referred to as the
\textit{generator} or \textit{decoder}\index{decoder}. 
The standard autoencoder is not a probabilistic model, but additional probabilitic structure can be added.

One approach is VAE\index{VAE, variational autoencoder} mentioned above~\cite{{kingma2013auto,rezende2014stochastic}}. 
By equipping the latent space with a prior distribution $p(z)$, the decoder of the autoencoder $g(z, \theta)$ implies a distribution on a manifold in the output space $\mathcal{X}$. VAEs additionally add some normally-distributed (Gaussian) random noise $\epsilon$ to the output so that $x=g(z,\theta)+\epsilon$. This implies that $p_\theta(x|z)$ is a tractable quantity, and it is interpreted as the likelihood in this context. 

In a VAE one also elevates the encoder to have a probabilistic form. Instead of encoding $z=e(x)$ in a deterministic way, one seeks a distribution over $z$ given $x$. A natural target for the probabilistic encoder would be to probabilistically invert the decoder. 
This inverse problem is solved by the posterior distribution $p(z|x)$ via Bayes theorem 
\begin{equation}\label{ML:eq:pxy}
    p(z|x) = \frac{p(x|z) p(z)}{p(x)} \;.
\end{equation}
While the likelihood and the prior may both be tractable, the normalizing constant  $p(x) = \int p(x,z) dz$ involves an intractable integral (the same intractable integral that makes maximum likelihood training of the VAE infeasible). 

One approach to Bayesian inference in these settings is variational inference\index{variational inference} (VI). In VI one approximates the posterior with some parametric family $q_\phi(z|x)$ in a parametric form, 
and then optimizes the ELBO with respect to its parameters $\phi$.
\begin{eqnarray} 
   {\rm ELBO}
    = \mathbb{E}_{q(z)} {\log p({x}|z)}
       - D_\text{KL}[q(z)||p(z)] \le \log \mathbb{E}_{q(z)}\left[ \frac{p(x,z)}{q(z)} \right] =  \log p({x})
       \; ,
        \label{ML:eq:nll71}
\end{eqnarray}
where we used Jensen's inequality for 
concave functions ($\log$) and the  reverse 
Kullback-Leibler (KL) divergence\index{KL, Kullback-Leibler divergence} term is 
\begin{equation}
D_\text{KL}[q(z)||p(z)]=\mathbb{E}_{q(z)}[\log q(z)-\log p(z)] \ge 0\,.    
\end{equation}

In a VAE, the variational model for the posterior $q_\phi(z|x)$  
is often  assumed to be an uncorrelated Gaussian (this is often called mean field approximation) defined by the mean $\mu$ and variance $\Sigma$.  Instead of optimizing the mean and variance independently for each $x$, VAEs use neural networks to predict the mean $\mu_\phi(x)$ and the variance $\Sigma_\phi(x)$. This is called \textit{amortized inference}, since after an up-front training cost  the approximate posterior $q_\phi(z|x)$ can be evaluated efficiently with a single forward pass of the neural network. Note the standard auto-encoder is recovered if one only used the mean $\mu_\phi(x)$ for the encoder and did not add noise $\epsilon$ to the decoder. 

Both the probabilistic encoder  $q_\phi(z|x)$  and the probabilistic decoder $p_\theta(x|z)$ are trained jointly by optimizing the ELBO. 
Unlike the standard autoencoder, which only minimizes the reconstruction error, ELBO optimization of Eq. \ref{ML:eq:nll71} has a tradeoff between minimizing the reconstruction error in the first term (averaged over the approximate posterior $q(z)$), which encourages high quality reconstructions, and minimizing the KL divergence term, which forces the posterior $q(z)$ to be as close to the chosen prior $p(z)$, and thus controls the sample quality by matching the aggregate posterior with a chosen prior distribution~\cite{FixElbo}. This term regularizes the 
VAE latent space, such that every sample 
drawn from the prior $p(z)$ correspond
to a valid sample. Successful VAE training requires to find a delicate balance between the two contributing terms to the ELBO. Whether the VAE training process succeeds in striking this balance depends on a number of factors, including the network architectures, the chosen prior and the class of allowed posterior distributions.
Once trained, the VAE can be used as a generative model by sampling from the prior $z_i \sim p(z)$ and then decoding according to $p_\theta(x|z) = g(z,\theta)+\epsilon$.

VAEs allow for expressive architectures, enjoy the benefits of regularization through data compression and have a firm theoretical foundation. Compared to GANs~\cite{GANs} \ref{ML:sec:gan}, VAEs are of particular interest to the scientific community as they provide a lower bound to the marginal likelihood (albeit potentially 
with a large gap) and a posterior distribution for the latent variables. 

It is also interesting to consider a special case of the autoencoder and VAE where the encoder and decoder are restricted to be linear transformations, which is effectively PCA\index{PCA, principle component analysis}. In PCA the (linear) decoder can be written $g(z)=Oz$, where $O$ is a matrix. As in the case of the autoencoder, PCA is not a probabilistic model, but probabilistic structure can be added. 
Probabilistic PCA~\cite{TippingBishop1999} assumes that the latent variables follow a Gaussian distribution with mean zero and covariance $\Lambda$,  where $ \Lambda$ is a diagonal matrix with the rank-ordered eigenvalues $\lambda_i$ along its diagonal. The
true distribution of the PCA components may be non-Gaussian, but a Gaussian is the maximum entropy approximation given their first two moments. Note that in probabilistic PCA these moments are measured on training dataset (when finding the principal components).

One can generalize probabilistic PCA to use nonlinear encoder and decoder as in an autoencoder. A Gaussian prior is a poor ansatz for the latent space distribution of data proceed by an autoencoder. Instead one can learn the density of the training samples in latent space using a normalizing flow. 
This  model was introduced in $\mathcal{M}$-flows~\cite{Brehmer:2020vwc} and 
in probabilistic autoencoder (PAE)~\cite{bohm2020probabilistic}\index{PAE, probabilistic autoencoder}, which achieves similar performance to a VAE in 
terms of sample quality without explicit ELBO optimization. 
In all these cases the dimensionality of the latent space is a hyperparameter to be chosen or optimized by the user. 
Unlike a standard VAE, these models do not add noise to the decoded output, thus the data is strictly restricted to the manifold defined by the decoder $g(z,\theta)$. However, unlike a GAN there is a well defined way to take an arbitrary data point $x$, project it onto the manifold, and calculate the density of the data point projected onto the manifold. Thus these models can also be used in the context of anomaly detection and out of distribution detection by identifying data that is off the manifold. 

\subsubsection{Generative adversarial networks}\label{ML:sec:gan}

GANs~\cite{GANs}\index{GAN, generative adversarial network} also 
typically choose a low dimensional latent space 
$z$ with a known prior distribution 
$p(z)$, typically a normal (Gaussian) distribution 
with zero mean and unit variance. 
GANs do not add noise to the output $g(z,\theta)$, so the likelihood $p(x|z)$ (and marginal likelihood $p(x)$) for almost all of the data space is 0, which precludes training by maximum likelihood and the ELBO\index{ELBO, evidence lower bound objective}. 
Instead of  training on ELBO, GANs train on a dissimilarity measure defined implicitly by a discriminator $D(x)$ (also referred to as the critic). Calculating the dissimilarity often involves it's own learning problem (\ie, adversarial training of the discriminator). 

The training is usually framed as a mini-max game 
\begin{equation}
    \min_g \max_D\mathcal{L}_\text{GAN}=\min_g \max_D \{\mathbb{E}_{x\sim p(x)}\log D(x)+\mathbb{E}_{z\sim p(z)}\log[1-D(g(z))]\}.
    \label{ML:eq:JS}
\end{equation}
The goal of the discriminator is to distinguish between true and 
generated data, hence we want to maximize
this loss with respect to $D$, assigning 1 to true data and 0 to generated data. The goal 
of generator is to fool the discriminator
such that it cannot distinguish between 
true and generated data, hence we want to 
minimize this loss with respect to $g$ at fixed $D$. This can be viewed as a game theoretical setup in a zero sum 
game between generator and discriminator. 

Instead of this game theory interpretation we 
can view the internal objective $\max_D \mathcal{L}_\text{GAN}$ as an implicit loss function that measures the 
dissimilarity between the target and 
generated distributions. 
The loss of Eq. \ref{ML:eq:JS} corresponds to 
the Jensen-Shannon (JS) divergence\index{JS, Jensen-Shannon divergence}, which  is a symmetrized form of KL divergence\index{KL, Kullback-Leibler divergence}. However, JS divergence is hard to directly work with, and the adversarial training could bring many problems such as vanishing gradient, mode collapse (tendency of generator to cluster the 
samples around the training samples, with holes between them) and non-convergence \cite{arjovsky2017towards, wiatrak2019stabilizing}. One of the core issues is that the distribution generated by the GAN is not guaranteed
to cover the entire space. 
To address these issues Wasserstein GANs\index{Wasserstein GAN} train on 
\begin{equation}
 \min_g \max_D\mathcal{L}_\text{WGAN}=\min_g \max_D \{\mathbb{E}_{x\sim p(x)}D(x)-\mathbb{E}_{z\sim p(z)}D(g(z))\}.
    \label{WGAN}   
\end{equation}
Here again the goal of discriminator is to 
make the loss as large as possible
between the true data and the generated 
data, while the goal of generator is 
to make it as small as possible, so that the discriminator cannot distinguish between the two. 
There is no requirement for $D(x)$ to 
be between 0 and 1, which helps with 
the above mentioned problems of JS
divergence. Instead, this is 
replaced with a requirement that $D(x)$
is 1-Lipshitz, i.e. 
the absolute value of the norm of the gradient of the discriminator output with respect to the input has to 
be less or equal to 1. 

Eq.~\ref{WGAN} can be interpreted as the 
dual form of the 1-Wasserstein distance\index{Wasserstein distance} 
between the true and generated distribution~\cite{arjovsky2017wasserstein}. Wasserstein 
distances are a measure of dissimilarity 
between two distributions used 
in the context of optimal transport, a
mathematical theory of how to define a notion of distance between probability distributions. Since the transport distance 
increases with the separation between 
the two distributions when they 
are non-overlapping, there is no 
gradient collapse that plagues other 
measures. 
In its primal form 
$p$-Wasserstein distance, $p\in [1,\infty)$, between two probability distributions $p_1$ and $p_2$, is defined as
  $  W_p(p_1, p_2) = \inf_{\gamma \in \Pi(p_1, p_2)} \left(\mathbb{E}_{(x,y)\sim\gamma} \left[ |x-y|^p  \right]\right)^{\frac{1}{p}}$,
where $\Pi(p_1, p_2)$ is the set of all possible joint distributions $\gamma(x,y)$ with marginalized distributions $p_1$ and $p_2$. In 1D the Wasserstein distance has a closed form solution via cumulative distribution functions (CDFs), but this evaluation is intractable in high dimensions. 

In the dual form of 1-Wasserstein distance, one instead maximizes Eq.~\ref{WGAN}
over all possible functions $D(x)$
that are 1-Lipschitz. One way to 
implement this is through weight clipping of the parameters of discriminator network, but a simpler solution is to add a 
gradient norm penalty term explicitly to the 
loss function~\cite{NIPS2017_4588e674}. 

Because of the discriminative nature of the dissimilarity measure
defined in data space, GANs and Wasserstein GANs often generate more realistic
samples than VAE or normalizing flows in high dimensions such as 
natural images (although flow matching and diffusion models can outperform GANs). However, GANs do not provide an encoder
from data to latent space nor a tractable likelihood
$p(x)$.

\subsubsection{Normalizing flows and autoregressive models}\label{ML:sec:flows}

Normalizing flows (NFs)\index{NF, normalizing flow} provide a powerful framework for density estimation and sampling~\cite{pmlr-v37-rezende15,2014arXiv1410.8516D,dinh2016density,papamakarios2017masked,kingma2018glow,kobyzev2020normalizing}. These models map the data $x$ to latent variables $z$ through a sequence of invertible transformations $f = f_1 \circ f_2 \circ \dots \circ f_n$, such that $z = f(x)$ or $x = g(z) = f^{-1}(x)$. As in the VAE and GAN, $z$ is modeled as a random number with a simple base distribution $p_Z(z)$, which is typically chosen to be a standard normal (Gaussian) distribution. Since NFs are invertible
the dimensionality of the latent 
space equals the dimensionality of the data space, in contrast to VAE and GANs where the
latent space dimensionality is 
often lower. 
The probability density of the model be evaluated using the change of variables formula:
\begin{equation}
    \label{ML:eq:flow}
    p_{X}(x) = p_{Z}(f(x))\left| \det \left(\frac{\partial f(x)}{\partial x}\right)\right| 
    = p_{Z}(f(x)) \prod_{l=1}^n \left|\det \left(\frac{\partial f_l(x)}{\partial x}\right)\right| ,
\end{equation}
where we have added subscripts to $p_X(x)$ and $p_Z(z)$ for clarity.
The Jacobian determinant $\det (\frac{\partial f_l(x)}{\partial x})$ must be efficient to compute for density estimation to be practical, and the transformation $f_l$ should be easy to invert for sampling.
In contrast to VAE and GANs, standard normalizing flows preserve the dimensionality of the data space as they are invertible (though there are normalizing flows that are defined on lower dimensional manifolds embedded in the data space~\cite{pmlr-v37-rezende15,rezende2020normalizing, gemici2016normalizing, Brehmer:2020vwc,bohm2020probabilistic}). As such, unlike GANs and VAEs, they can be trained via maximum likelihood (Eq.~\ref{ML:eq:max_likelihood}) as described in Sec.~\ref{ML:sec:density_estimation}.

There are several popular architectures
of NFs. A method used by NICE, 
RealNVP and Glow \cite{2014arXiv1410.8516D,dinh2016density,kingma2018glow} is to split the space
into two disjoint sets $z_1$ and $z_2$, 
and then use an identity forward map  $z\rightarrow x$ for $x_1$, $x_1=z_1$, 
and an affine transformation for $x_2$ of the form
\begin{equation}
    x_2=\exp(s(z_1))\odot z_2+m(z_1),
\end{equation}
where $\odot$ is elementwise product 
and $m(z_1)$, $s(z_1)$ are neural networks. 
The Jacobian of this map is lower 
triangular, and its determinant is simply  the product of elements along the diagonal, which is 
tractable, as is the inverse of the transformation. At the 
next layer one then performs a 
different split of dimensions into 
$z_1$ and $z_2$. The affine transformation 
can be further generalized to a nonlinear form using 
rational splines \cite{durkan2019neural}.

One can interpret the sequence of invertible transformations  $f_1 \circ f_2 \circ \dots \circ f_n$ as $n$ discrete time steps in a continuous flow. In particular, one can think of a continuous-time flow described by an ordinary differential equation (ODE) and then interpret the discrete time steps as the result of a numerical integration of that ODE. This is the approach taken by the Ffjord algorithm \cite{grathwohl2018ffjord} and other variants. A residual 
flow has an update $f_i(x)=x_i+\delta u_i(x)$, which for $\delta_i=n^{-1}$
and taking $n \rightarrow \infty$
limit gives rise to an ordinary differential equation (ODE)
\begin{equation}
    dx_t=u_t(x_t)dt.
    \label{flow}
\end{equation}
Here $u_t$ is the velocity field 
that defines the flow and is a vector field. One can build the density estimator for all intermediate times $t$ $p_t(x)$ using its divergence,
\begin{equation}
    \ln p_t(x_t)=\ln p_0(x_0)-\int_0^t \nabla \cdot u_s(x_s)ds,
\end{equation}
where $p_0$ at $t_0$ is the initial base distribution and $p_1$ at $t=1$ is the target distribution.  
Continuous normalizing flows parametrize $u_t$ as a neural network. They are very expressive, but expensive to train using 
maximum likelihood.

A different approach creating a deep generative model with a tractable likelihood is to
model $p(x)$ autoregressively as
\begin{equation}
    p(x)=\prod_{i=1}^np(x_i|x_1,x_2, \dots, x_{i-1}) \; .
\end{equation}
This form describes each 
new dimension conditionally on all 
previous dimensions. It can model a general 
likelihood $p(x)$ as a sequence of 
conditional 1d distributions, whose 
conditional dependence on the
parameters $x_1,x_2, \dots, x_{i-1}$ can be modeled with 
neural networks. If $x$ is a 
time series this form imposes a causal 
structure where $x_i$ depends on 
all previous times $x_j$, $j<i$. 
WaveNet~\cite{2016arXiv160903499V}\index{WaveNet} and PixelCNN~\cite{PixelCNN})\index{PixelCNN} are two well known examples. 
Sampling from an autoregressive model 
is sequential, and can be slow in 
high dimensions. Inverse autoregressive 
flow reverses this process and makes 
sampling fast, but the likelihood evaluation is
slow. Some normalizing flows have
autoregressive coupling layers, such as masked
autoregressive flow (MAF)~\cite{papamakarios2017masked}\index{MAF, masked autoregressive flow}.

All of the methods above use maximum likelihood training of likelihood $p(x)$ against network parameters, so 
the training is to minimize KL divergence\index{KL, Kullback-Leibler divergence} between the data 
distribution and a Gaussian in latent space. This can be overly sensitive 
to small variance directions that 
dominate the likelihood, without 
being sensitive to the global 
structure of the data. 
An alternative is to 
use Optimal Transport Wasserstein distance\index{Wasserstein distance} between 
the density of the generated samples and the data, which can be evaluated either
in data space or in latent space.
As Wasserstein distance is difficult to evaluate 
in high dimensions, one can instead use slices, 1d  projections of the 
data along different directions in 
high dimensional space, to 
build the flow~\cite{SINF}. 
Because this training is less sensitive to 
small variance directions than maximum likelihood
training it achieves 
better results on anomaly 
detection tasks~\cite{SINF}.

We end by noting that normalizing flows, autoregressive models, and other deep generative models that provide a tractable likelihood are powerful tools for simulation-based inference\index{simulation-based inference}. They can provide surrogate models trained from large simulated datasets when the simulators have intractable likelihood functions, which is usually the case. As described in Sec.~\ref{ML:sec:SBI}, one would like to work with models that can provide conditional density estimation in order to model either the likelihood $p(x|\theta)$ or the posterior $p(\theta|x)$~\cite{Cranmer:2016lzt,NIPS2016_6084}. These techniques are being actively explored and applied to a number of scientific problems. 

\subsubsection{Flow-matching and diffusion models}\label{ML:sec:flowsdiffusion}

In flow-matching models\index{flow-matching model},
we start from a base distribution such as a Gaussian $p_0=\mathcal N(0,I)$, and use ODEs to generate 
samples with a flow 
vector field $u_t^\theta(x)$ as in Eq.~\ref{flow}. 
As discussed above, continuous 
normalizing flows are expensive to train via maximum likelihood.
Instead, one can learn directly 
the velocity field parametrized as a neural network $u_t^\theta(x_t)$ with parameters $\theta$ via the flow-matching loss
\begin{equation}  \mathcal{L}=\mathbb{E}_{t\sim U(0,1),~x \sim p_t} \left[u_t(x_t)-u_t^{\theta}(x_t)\right]^2,
\end{equation}
where the expectation is uniform over time $t$ and over all intermediate 
distributions $p_t$. This equation 
is however not practical since we 
do not know the target $u_t(x)$. Instead, 
one can take advantage of the target
conditional velocity field $u_t(x_t|z)$, where $z\sim \hat p$ is a training data sample
\begin{equation}
\mathcal{L}=\mathbb{E}_{t\sim U(0,1),~x \sim p_t,~z\sim \hat p} \left[u_t(x|z)-u_t^\theta(x_t)\right]^2.    
\end{equation}
It has been shown that this 
conditional target velocity field training
also leads to the correct 
distribution in the flow models~\cite{lipman2023flowmatchinggenerativemodeling,Albergo}. Figure~\ref{ML:fig:flow_diffusion}, taken from Ref.~\cite{holderrieth2025introductionflowmatchingdiffusion}, illustrates
the main idea, which is that 
training a conditional flow, and 
averaging over all the training data, 
is the same as training on unconditional flow. 

The advantage of this formulation is that conditional velocity fields are a lot simpler to construct. A typical 
case is a flow from the initial 
Gaussian $p_0(x|z)=\mathcal N(0,I)$ to a delta function at z $p_1(x|z)=\delta_z(x)$. A 
very simple linear flow that 
achieves this is $p_t(x|z)=\mathcal N(tz,(1-t)^2I)$. The flow itself moves from 
a random Gaussian variable $\epsilon \sim \mathcal N(0,I)$ to the 
data point $z$, so $x_t=\epsilon(1-t)+tz$. Finally, the conditional velocity field is given by $u(x_t|z)=z-\epsilon$, so the training loss is 
\begin{equation}
\mathcal{L}=\mathbb{E}_{t\sim U(0,1),~\epsilon \sim \mathcal N(0,I),~z\sim \hat p} \left [z-\epsilon-u_t^{\theta}(\epsilon(1-t)+tz)\right ]^2.  
\label{flowloss}
\end{equation}
This leads to a simple 
training algorithm where one 
randomly chooses a minibatch of 
data $z$, random Gaussian variables 
$\epsilon$, and a time $t$ to 
update the parameters $\theta$
based on stochastic gradient descent using the loss of Eq.~\ref{flowloss}.
The simplicity and efficiency of this training 
procedure has led flow matching 
to become one of the leading generative 
models for large image based data. 
Sampling from flow-matching 
models requires randomly choosing an initial condition $x_0 \sim N(0,I)$ and discretizing Eq.~\ref{flow}. Note that this is
a deterministic ODE and all the 
randomness is in the initial 
conditions. 

Diffusion models\index{diffusion model} also start from a Gaussian base distribution, but also continuously add noise during the evolution in time, \ie, they are based on a stochastic differential equation (SDE)
\begin{equation}
x_0 \sim p_0,\,\,\,dx_t=u_t^{\theta}(x_t)dt+\frac{\sigma^2_t}{2}s_t(x_t)+ \sigma_tdW_t,
\label{diffusion}
\end{equation}
where we define score\index{score} $s_t(x_t)=\nabla \ln p_t(x_t)$. 
Here, $dW_t$ is the stochastic term, which adds Brownian motion (also called a Wiener process) in the form of
uncorrelated Gaussian random noise. Note that with $\sigma_t=0$, a diffusion model becomes a flow model. 
The noise variance $\sigma_t$ is a 
free parameter that can be tuned for optimal performance. If our target is a static distribution so that $p_t=p$
then $u_t=0$ and we obtain the Langevin 
equation. 

In diffusion, we also need to learn 
the gradient field via the score 
function, and as before we can 
replace the marginal score with 
conditional score during the 
training to obtain score matching 
training procedure
\begin{equation}
\mathcal{L}=\mathbb{E}_{t\sim U(0,1),~x \sim p_t,~z\sim \hat p} \left [\nabla \ln p_t(x|z)-s_t^\theta(x_t)\right ]^2.
\end{equation}
In 
the simple Gaussian example with 
$p_t(x_t) = \mathcal N(\alpha_tz, \beta_t^2I)$, 
where $\alpha_0=\beta_1=0$ and $\alpha_1=\beta_0=1$, we have 
a trajectory $x_t=\alpha_tz+\beta_t \epsilon$, and the score loss
\begin{equation}
\mathcal{L}=\mathbb{E}_{t\sim U(0,1),~\epsilon \sim \mathcal N(0,I),~z\sim \hat p} \left[\frac{\epsilon}{\beta_t}+s_t^{\theta}(\alpha_tz+\beta_t \epsilon)\right]^2.  
\label{diffusionloss}
\end{equation}

It would appear that in diffusion, 
one must train both the flow and 
the score, but for simple linear 
models the two can be related to 
one another, and one can 
choose the flow-matching or score-matching training procedure. For 
the example in Eq.~\ref{diffusion}, the corresponding score-matching training is
\begin{equation}
 x_0 \sim p_0,\,\,\,dx_t=\left[\left(\beta_t^2\frac{\dot{\alpha}}{\alpha}-\beta_t\dot{\beta}_t+\frac{\sigma^2_t}{2}\right)s_t(x_t)+ \frac{\dot{\alpha}}{\alpha}x_t\right]dt+\sigma_tdW_t.  
\end{equation}

One of the advantages of 
score- and flow-based methods is 
that 
one can reduce the architectural restrictions 
imposed by normalizing flows or autoregressive 
models. Score- and flow-based training avoid the normalization requirement. 
Score-based models learn gradients of log probability density functions on a large number of noise-perturbed data distributions, and then generate samples by Langevin-type sampling. 

The generative models 
described in this subsection are called flow-based models~\cite{lipman2023flowmatchinggenerativemodeling}, score-based generative models~\cite{SongE19}, diffusion probabilistic models~\cite{SDE}, 
or stochastic interpolants~\cite{Albergo}. They
have several advantages over other model families. They often outperform GAN-level sample quality without adversarial training, and enable exact log-likelihood computation
through their connection to continuous-time flows, which can be represented as a 
probability flow ordinary differential equation~\cite{SDE}. 
The main advantage is that the 
distribution $p(x)$ can be specified solely by its score or flow.
 This in turn  
enables more flexible model 
architectures than what can be 
used in normalizing flows or autoregressive models. 

\begin{pdgxfigure}
    \centering
    \includegraphics[width=0.45\linewidth,clip=true,viewport=459 0 1459 452]{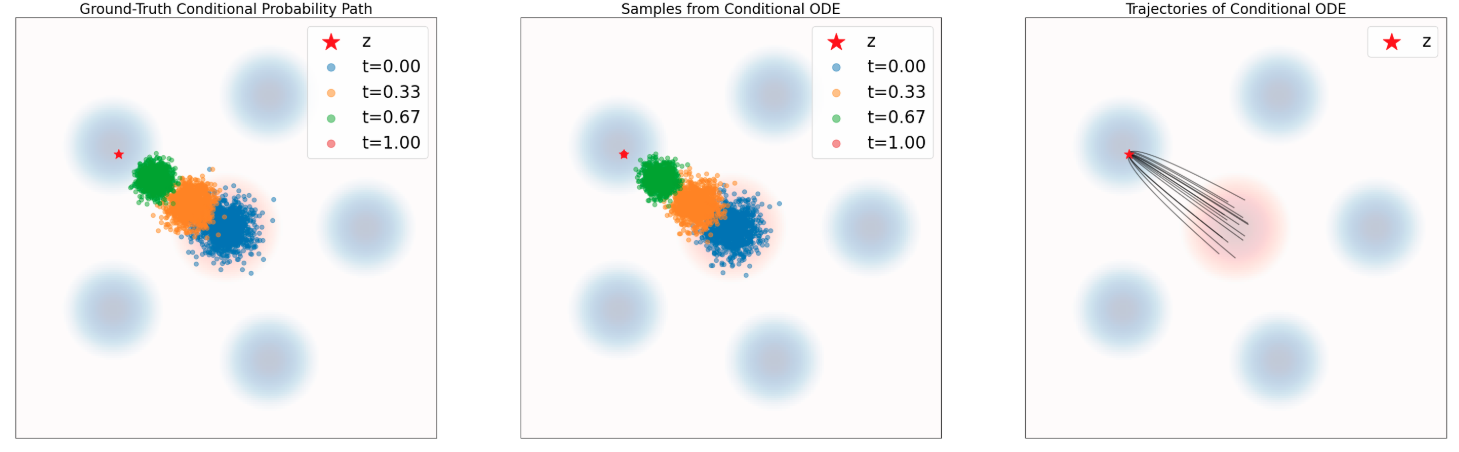}
    \includegraphics[width=0.45\linewidth,clip=true,viewport=459 0 1459 452]{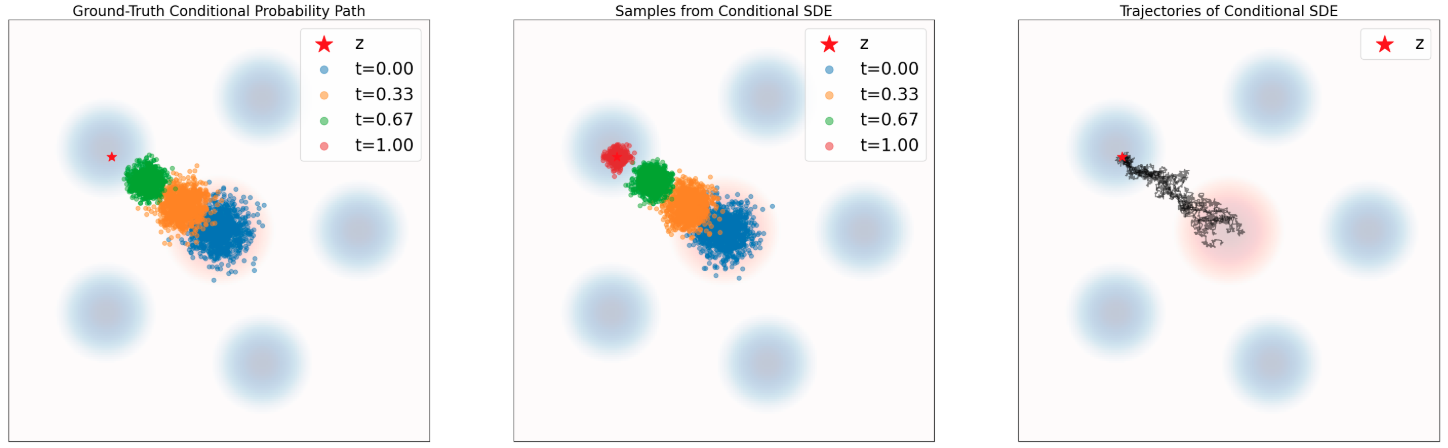}\\
    \includegraphics[width=0.45\linewidth,clip=true,viewport=459 0 1459 452]{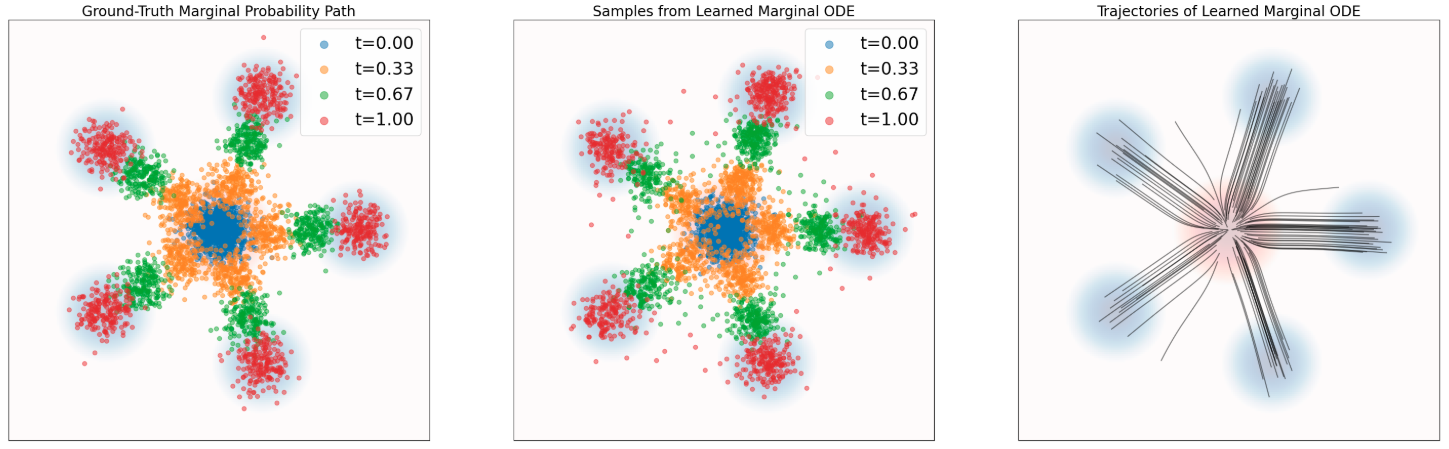}
    \includegraphics[width=0.45\linewidth,clip=true,viewport=459 0 1459 452]{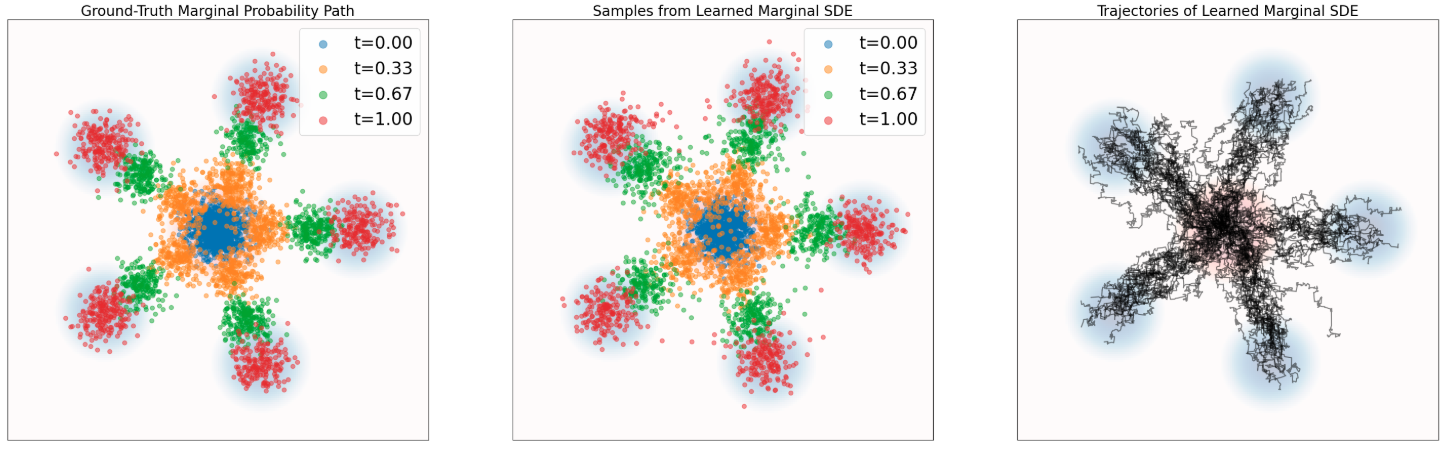}
    \caption{Illustration of the marginalization trick for flow-based models (left) and diffusion models (right), which simulate a probability path with ODEs or SDEs, respectively (Holderrieth and Erives, 2025).
    The data distribution $\hat p$ is the blue background, while the initial Gaussian distribution is the red background.
    The top graphs represent conditional probability paths, while the bottom graphs represent marginal probability paths.
    Both samples and trajectories are shown.}
    \label{ML:fig:flow_diffusion}
\end{pdgxfigure}
 
\subsection{Anomaly detection and out-of-distribution detection}\label{ML:sec:anomaly}
Unsupervised anomaly detection\index{anomaly detection}\index{unsupervised learning} techniques detect anomalies in an unlabeled test data set under the assumption that the majority of the in-distribution data are normal under some measure, while out-of-distribution (OOD)\index{OOD, out-of-distribution} data are not.
In the context of autoencoders\index{autoencoder}
a popular technique is to use the 
reconstruction error of Eq.~\ref{ML:eq:reconstruction_error} to 
identify an outlier as one with a 
large reconstruction error \cite{PhysRevD.101.076015,Farina_2020,Heimel_2019}. One issue with this 
method is that for higher dimensional 
latent space and flexible neural network architectures the
encoder-decoder map become identity for 
any input data, $f(x)=x$, regardless of whether input $x$ is from the in-distribution training dataset or from the out-of-distribution data. The choice of 
autoencoder latent space dimensionality is thus an 
important hyperparameter that must be 
tuned. 

Another set of anomaly detection techniques construct a model representing normal behavior from a given in-distribution training dataset, and then evaluate the likelihood of a test instance to be generated by the utilized model. For instance, one can use density 
estimation methods such as normalizing 
flows\index{NF, normalizing flow} (section \ref{ML:sec:flows}) to learn the density (likelihood) of the in-distribution training dataset $p(x)$, and apply it to 
the test data. The expectation is that 
out-of-distribution data will have a 
lower density (likelihood) under the in-distribution
density model. 
This expectation is however not always met in high dimensions and the 
method suffers because 
likelihood-based training\index{training} is sensitive to the smallest variance directions~\cite{ren2019likelihood}.
Low-variance directions may contain little or no information on the global structure of the image, so there is a mismatch 
between the training objective and outlier detection objective. 
Lower dimensional autoencoders with NF in the latent space
deal better with this issue \cite{bohm2020probabilistic}.

A related issue is that of typicality: an in-distribution data sample likelihood 
will typically be lower
than the maximum value, so an out-of-distribution data sample that 
is closer to the peak would have a higher likelihood. If 
this happens in low-variance
directions that dominate the likelihood, normalizing flows\index{NF, normalizing flow} can assign higher likelihoods to out-of-distribution data than to in-distribution training data~\cite{nalisnick2018deep}. 
A number of techniques have been proposed to 
circumvent these limitations, such as 
likelihood regret~\cite{xiao2020likelihood}, likelihood-ratio~\cite{ren2019likelihood},
likelihood in autoencoder latent space~\cite{bohm2020probabilistic}, and Wasserstein distance\index{Wasserstein distance} training of the likelihood
$p(x)$~\cite{SINF,CMS:2025lmn}. These methods can achieve better anomaly detection performance
than the autoencoder reconstruction error~\cite{Brehmer:2020vwc,bohm2020probabilistic,CMS:2025lmn}.
However, even perfect density estimation cannot guarantee good anomaly detection performance~\cite{Le_Lan_2021,Kasieczka:2022naq}.

Supervised anomaly detection\index{anomaly detection}\index{supervised learning} techniques require a data set that has been labeled as in-distribution and out-of-distribution and involves training a classifier (the key difference to many other statistical classification\index{classification} problems is the inherent unbalanced nature of outlier detection). These methods 
assume some form for what out-of-distribution data may look like, and their success relies
on whether the assumed form is a realistic
representation of actual out-of-distribution data. When this assumption is 
valid these methods
can be more powerful than unsupervised
methods, but the reverse is also true. 
A hybrid between the two approaches is to 
train a classifier without labels \cite{Collins:2018epr}. All 
these approaches are largely 
complementary to each other \cite{Collins_2021}. Examples of different 
anomaly detection methods applied to 
HEP are the LHC Olympics 2020 and Dark Machines challenges \cite{Kasieczka:2021xcg,Aarrestad:2021oeb}. 

\section{Self-supervised learning}
\label{ML:sec:ssl}

Self-supervised learning (SSL)\index{self-supervised learning} also aims to distill useful features in the data without requiring supervision labels for every sample in the input data.
Self-supervised methods make use of large unlabeled datasets to build meaningful representations.
They can generally be categorized as \textit{autoassociative}, where the model is trained to reproduce or reconstruct its own (masked) input or 
\textit{contrastive}, where the model is trained to learn a mapping that is insensitive to different ``views'' of the data.
These methods are often used to build ``foundation models'' (FMs)\index{FM, foundation model} discussed in Sec.~\ref{ML:sec:foundation_models}, which are pre-trained using self-supervised learning and fine-tuned using supervised learning for different downstream tasks.
However, FMs are not the only possible use case.

A classic autoassociative task is masked language modeling popularized by the bidirectional encoder representations from transformers (BERT) model~\cite{bert}.
In this task, BERT ingests a sequence of words, a fraction of which are randomly masked, and tries to predict the original words that have been masked.
For example, in the sentence ``The Milky Way is a [MASK] galaxy,'' BERT would need to predict ``spiral.''
This helps BERT learn bidirectional context.
A variant of this approach is next token prediction, popularized by the generative pretrained transformer (GPT)~\cite{gpt3}.
A common theme in these methods is \emph{tokenization}, in which elements of the input data are mapped to discrete vectors, known as tokens.
These approaches have been applied in the context of particle jets~\cite{mpmv1,mpmv2,omnijetalpha}, enabling the construction of backbone models that can be fine-tuned for different tasks and provide improvements for small training samples.

Sensory data (\eg, 1D waveforms, 2D images, or 3D scenes) pose a significant challenge for autoassociative tasks compared to symbolic data such as language, math, and high-level physical concepts like jets and particles. 
For symbolic data, the masking unit is naturally defined (\eg, a word for language) and associated with a strong semantic meaning, which yields a well-defined learning objective for mask-based self-supervision. 
On the contrary, sensory data captures raw information and a unit of data (\eg a single pixel in an image) does not carry meaningful information alone.
This challenge has resulted in in-depth R\&D for self-supervision techniques in computer vision.
The masked autoencoder (MAE) laid the initial ground work~\cite{MaskedAutoencoders2021}: the authors discovered that a large fraction of masking (\ie, 75\%) is crucial for successful training using an asymmetric encoder-decoder architecture.
Distillation with no labels (DINO) made another breakthrough by introducing a self-distillation technique where a student and teacher model pair---the teacher model typically being an exponential moving average of the student model---are forced to agree across different augmentations (\eg, cropping, adding jitter, and rotating) of the same data instance~\cite{caron2021emerging,oquab2023dinov2}.
For effective representation learning of 3D geometrical shapes, multi-view projection matching techniques~\cite{dust3r_cvpr24,duisterhof2025mastrsfm,wang2025vggt} are promising and a strong promise and relevant to time projection chamber (TPC) image data in high energy physics.
Exploration of these specialized techniques in computer vision has impacted HEP applications~\cite{young2025particletrajectoryrepresentationlearning,Hao:2025abk}.

In contrastive learning, portions of the input data are paired together and the model is tasked to find matching pairs.
Pairs can be constructed based on different data modalities, such as text and images, or based on data augmentations, that may be generic, such as adding noise, or domain-specific, like symmetry transformations.
For example, the contrastive language-image pre-training (CLIP)~\cite{clip} allows joint pretraining of a text encoder and an image encoder, such that a matching image-text pair have image encoding vector $\mathbf z_i$ and text encoding vector $\mathbf z_j$ that span a small angle, \ie, have a large cosine similarity
\begin{align}
c(\mathbf z_i,  \mathbf z_j)=\frac{\mathbf z_i\cdot \mathbf z_j}{|\mathbf z_i||\mathbf z_j|} = \cos\theta_{ij} \; ,
\label{ML:eq:cosine}
\end{align}
with $\theta_{ij}$ being the angle between the encoding vectors.
This approach has been applied in astrophysics~\cite{astroclip}.

Positive pairs may also be constructed by applying data augmentations.
For example, in the case of galaxy images, one may augment the data by performing image rotations, adding noise, size scaling, or adding point spread function smoothing, all of which are realistic transformations expected in a real galaxy image survey~\cite{Hayat2021,ChenK0H20}.
For particle jets, tailored augmentations may include rotations about the jet axis, translations in the $(\eta, \phi)$ plane, smearing the positions of the soft jet constituents, and collinear splitting of the jet constituents~\cite{Dillon:2021gag,PhysRevD.106.056005,10.21468/SciPostPhysCore.7.3.056,largescale}.
Another augmentation strategy is based on re-simulating the stochastic shower and detector interactions, thus generating multiple physical realizations of a primary particle's evolution~\cite{resimulation,ssljetphysics}.

A well-known approach for contrastive learning with augmentations is SimCLR~\cite{simclr}.
In this approach, the contrastive loss for a positive pair of an input and its augmentation $(\mathbf z_i, \mathbf z_i^\prime)$ is defined in terms of the cosine similarity of Eq.~\ref{ML:eq:cosine} as
\begin{equation}
\mathcal L(\mathbf z, \mathbf z_i') = -\ln \frac{ \exp[c(\mathbf z_i,\mathbf z_i^\prime)/\tau] }{ \displaystyle\sum_{j \neq i \in \text{batch}} \left[ \exp[c(\mathbf z_i,\mathbf z_j)/\tau] + \exp[c(\mathbf z_i,\mathbf z_j^\prime)/\tau] \right]} \; ,
\label{ML:eq:clloss}
\end{equation}
and the total loss is given by the sum over all positive pairs in the batch, $\sum_{i\in\text{batch}} \mathcal L(\mathbf z_i, \mathbf z_i^\prime)$.
The loss decreases when the distance between positive pairs decreases or when the distance between negative pairs increases.
The hyperparameter $\tau$ is known as temperature and controls the relative influence of positive and negative pairs.
SimCLR has been applied in radio astronomy~\cite{BaronPerez:2025}, neutrino physics~\cite{Wilkinson:2025nxv}, and collider physics~\cite{Dillon:2021gag}.
Another application of contrastive regularization is self-distillation introduced in DINO discussed above.
Self-distillation is a powerful technique that can be applied regardless of the target task, and improves the quality of self-supervision for many computer vision models for both image and point cloud data.

Finally, an alternative paradigm is the joint-embedding predictive architecture (JEPA)~\cite{ijepa}, which learns meaningful representations by modeling missing or unseen embeddings directly in the latent space without a decoder or full input
reconstruction.
The advantages of this approach are no data augmentations are required and unnecessary details of the input can be ignored.
This approach has been applied to particle jets~\cite{jjepa,hepjepa} and Square Kilometer Array (SKA) light cones~\cite{Ore:2024jim}.
A comparison between the different self-supervised learning approaches can be found in Fig.~\ref{ML:fig:jepa}, reproduced from Ref.~\cite{ijepa}.

\begin{pdgxfigure}
    \centering
        \includegraphics[width=0.25\textwidth]{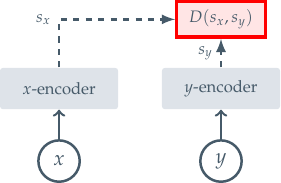}
        \includegraphics[width=0.3\textwidth]{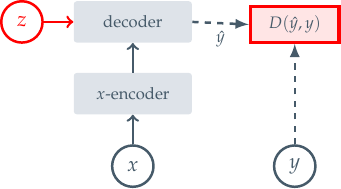}
        \includegraphics[width=0.3\textwidth]{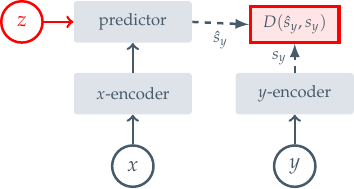}
    \caption{Common architectures for self-supervised learning, in which the system learns to assign a large scalar value to incompatible inputs, and a low scalar value to compatible inputs (M. Assran, et al. in ICCV, 2023).
    Joint-embedding architectures (left) learn to output similar embeddings for compatible inputs $x, y$ and dissimilar embeddings for incompatible inputs. 
    Generative architectures (center) learn to directly reconstruct a signal $y$ from a compatible signal $x$, using a decoder network that is conditioned on additional (possibly latent) variables $z$ to facilitate reconstruction. 
    Joint-embedding predictive architectures (right) learn to predict the embeddings of a signal $y$ from a compatible signal $x$, using a predictor network that is conditioned on additional (possibly latent) variables $z$ to facilitate prediction.}
    \label{ML:fig:jepa}
\end{pdgxfigure}

\section{Optimal control, reinforcement learning, and active learning}\label{ML:sec:RL}


Many problems in science and engineering can be cast as a control problem, which comprises a cost functional that is a function of state and some control variables that specify some underlying dynamical system. This is relevant for the control of accelerators where the dynamical system is physical. This formalism can also be used to  describe the design of experiments\index{experimental design optimization}, planning of an observational survey, and other decision making processes relevant to the scientific method. It is closely connected to planning, dynamic programming, and reinforcement learning\index{RL, reinforcement learning}. Optimal control\index{optimal control} generalizes the framing of learning presented in Sec.~\ref{ML:sec:loss_risk_Remp}.

\subsection{Optimal control}\label{ML:sec:optimal_control}
Optimal control theory deals with finding a control for a dynamical system over a period of time such that the objective function is optimized. The underlying system can be discrete or continuous and may be deterministic or stochastic.  The commonalities and differences between optimal control and reinforcement learning can be best understood through the formalism of a Markov decision process (MDP)\index{MDP, Markov decision process}, which is a discrete-time stochastic control process.  

A Markov decision process comprises four components often organized as a 4-tuple 
${\displaystyle (S,A,P_{a},R_{a})}$, where:
$S$ is a set of states called the state space, $A$ is a set of actions called the action space, ${\displaystyle P_{a}(s,s')=\Pr(s_{t+1}=s'\mid s_{t}=s,a_{t}=a)}$ is the probability that action 
$a$ in state $s$ at time $t$ will lead to state 
$s'$ at time $t+1$, ${\displaystyle R_{a}(s,s')}$ is the immediate reward (or expected immediate reward) received after transitioning from state 
$s$ to state 
$s'$, due to action $a$. 

The policy function 
$\pi$  is a mapping from state space to action space that can be either deterministic or probabilistic. For examples, the policy that drives a computer chess playing system, decides which move to make given the current state of the board. Similarly, policies dictate which experiment should be built next, which field of the sky should be observed, or how to adjust the operational parameters of an accelerator. 
%
The dynamics of the resulting system are then fixed by combining the policy with the underlying MDP. The evolution of the resulting dynamical system behaves like a Markov chain since the action chosen in state 
$s$ is completely determined by 
$\pi(s)$ and 
${\displaystyle \Pr(s_{t+1}=s'\mid s_{t}=s,a_{t}=a)}$ implies the Markov transition matrix 
${\displaystyle \Pr(s_{t+1}=s'\mid s_{t}=s)}$. 

The objective optimal control is to choose a policy 
$\pi$  that will maximize a cumulative function of the instantaneous rewards $R_a$. A common choice is the expected discounted sum:
\begin{equation}
{\mathbb{E}\left[\sum _{t=0}^{\infty }{\gamma ^{t}R_{a_{t}}(s_{t},s_{t+1})}\right]} \; ,
\end{equation}
where $a_t \sim \pi(s_t)$ are the actions given by the policy, the expectation computed with respect to the distribution 
${\displaystyle s_{t+1}\sim P_{a_{t}}(s_{t},s_{t+1})}$, and
$\gamma$ is the discount factor satisfying 
${\displaystyle 0\leq \ \gamma \ \leq \ 1}$. The discount factor is usually close to 1 and sometimes reparameterized as $\gamma = 1/(1+r)$, where $r$ is called the discount rate. A lower discount factor motivates the decision maker to favor taking actions early, rather than postpone them indefinitely.

A policy that maximizes the objective function is called an optimal policy and denoted $\pi^*$, though the optimal policy need not be unique. Importantly, the Markov property implies that the optimal policy is only a function of the current state.
Dynamic programming can be used to find the optimal policy for MDPs with finite state and action spaces. For instance, in value iteration (a.k.a. backward induction) can be used to solve the ``Bellman equation''~\cite{bellman1957}. For continuous-time systems, the optimal policy is defined by the Hamilton–Jacobi–Bellman equation~\cite{kirk2004optimal}.


In many settings, it is assumed that the state 
$s$ is fully known when action is to be taken and there are no latent variables. When this assumption is not true, the problem is called a partially observable MDP.
These problems are generally more difficult and the dynamic programming algorithms do not directly apply~\cite{aastrom1965optimal}. 




\subsection{Reinforcement learning}\label{ML:sec:reinforcement_learning}
The main difference between the classical dynamic programming methods and reinforcement learning (RL)\index{RL, reinforcement learning} algorithms is that the latter do not assume knowledge of an exact mathematical model of the MDP and they target large MDPs where exact methods become infeasible. For example, RL was used in the context of jet physics to search for the most likely jet clustering when the number of constituents was too large for the exact dynamic programming algorithm to be used~\cite{Brehmer:2020brs}.
In addition, RL can be used when the probabilities or rewards are unknown. Instead, the transition probabilities are often accessed indirectly through interaction with a real or simulated environment. 

Numerous variations to RL exist, which include so-called model-based and model-free approaches (referring to models of the instantaneous rewards and the state transitions) and on-policy and off-policy (which describes how the actions taken during learning are related to the current policy).  See Ref.~\cite{sutton1998reinforcement} for an introduction and Ref.~\cite{arulkumaran2017deep} for a recent review. Some examples of RL use in particle physics are in Refs.\cite{Carrazza:2019efs,Mendizabal:2025sbf,Wojcik:2024lfy}. 




\subsection{Multi-arm bandits}\label{ML:sec:multi-arm_bandits}

Multi-arm bandit\index{multi-arm bandits} problems are a classic reinforcement learning problem where one tries to maximize the expected gain by allocating a limited set of resources to various alternatives. The name is a reference to a gambler with a fixed amount of money that must choose between multiple slot machines (or ``one-armed'' bandits) when the payoff for the individual machines is unknown. 
A hallmark of multi-arm bandit problems is that they involve a tradeoff between exploration (playing machines to estimate their payoff) and exploitation (playing machines with the highest estimated payoff). 
Multi-armed bandits are used to manage large projects, organizations, and scheduling problems. The theory has a long history going back to Robbins in 1952 that used it to study the sequential design of experiments~\cite{robbins1952some} and Gittins who derived an optimal policy under some conditions~\cite{gittins1979bandit}. 

\subsection{Bayesian optimization}\label{ML:sec:BO}

A closely related set of techniques involve optimizing some expensive black box function $f(x)$. 
For instance, the function may be computationally expensive to evaluate or low-latency, \eg it may involve manually re-configuring a system. This is particularly relevant for analysis optimization in particle physics where evaluating $f(x)$ involves processing large numbers of simulated collisions. Another common use case involves optimizing the hyperparameters of a learning algorithm. 

Without any assumptions about the function $f(x)$ this is hopeless; however, if one assumes something about the functions (\eg some notion of smoothness) then one can leverage function evaluations evaluations $\{f(x_t)\}_{t=1,\dots,T}$ to say something about what value the function might take on at other values of $x$. This is usually cast in Bayesian terms, and Gaussian processes\index{GP, Gaussian process} (Section \ref{ML:sec:GP}) are often used to model the distribution over $f(x)$. The optimization techniques that use this framing are generically referred to as Bayesian optimization\index{Bayesian optimization}~\cite{mockus2012bayesian}.

Optimization in this context is usually characterized by an \textit{exploration-exploitation} tradeoff, similar to what is found in multi-arm bandits. Here, exploration refers to function evaluations that characterize the function in regions that haven't been evaluated, while exploitation refers to evaluations near what is predicted to be its maximum. This setting is similar to reinforcement learning in that it involves sequential decisions (\ie, where to evaluate the function next), but usually the target function $f(x)$ is assumed to be static. In that sense, the state referred to in the language of an MDP is the state of knowledge about the function after sequential evaluations $\{f(x_t)\}_{t=1,\dots,T}$. The reward at time $t$ is not the value of the function $f(x_t)$, but some quantity that characterizes what was learned about the function's maximum. In this literature, one often refers to the \textit{acquisition function}, which plays a similar role as the expected value of the reward in RL. Common acquisition functions include the probability of improvement, the expected improvement, and an upper-confidence bound~\cite{brochu2010tutorial}.

\subsection{Active learning}\label{ML:sec:active_learning4}

Active learning\index{active learning} is closely related to Bayesian optimization, described above. In Bayesian optimization one estimates the function $f(x)$ from some set of evaluations $\{y_t = f(x_t)\}_{t=1,\dots,T}$; however, the goal is to find the maximum $x^* = \arg\max_x f(x)$. In active learning, the goal is not to find the maximum of $f(x)$, but to approximate the function as one does in supervised learning. The main difference compared to vanilla supervised learning is that the labeled training dataset isn't provided a priori in a passive way, but the learning algorithm actively decides where to generate $(x_t, y_t=f(x_t))$  pairs. The function $f(x)$ is sometimes referred to as an \textit{oracle}. Active learning is particularly attractive when obtaining labeled data is a costly process. 

More broadly, a challenge of many machine learning applications is obtaining labeled data\index{label}, which can be a costly process. If a system could learn from small amounts of data, and choose by itself what data it would like the user to label via an external process called oracle, it would make machine learning more powerful. Such frameworks are also called experiment design or active learning.  In active learning, a model is trained on a small amount of data (the initial training dataset), and an acquisition function (often based on the model’s uncertainty) decides on which data points to ask for a label. The acquisition function selects one or more points from a pool of unlabeled data points, with the pool points lying outside of the training dataset. Once we  label the selected data points, these are added to the training dataset, and a new model is trained on the updated training dataset. This process is then repeated, with the training dataset increasing in size over time. The advantage of such systems is that they often result in dramatic reductions in the amount of labeling required to train an ML system (and therefore cost and time).



\section{Simulation-based inference}\label{ML:sec:SBI}
The goal of simulation-based inference\index{simulation-based inference} (related to, but distinct from, likelihood-free inference) is to extend the statistical procedures described in the Chapter on Statistics (\eg parameter estimation, hypothesis tests, confidence intervals, and Bayesian posterior distributions) to the situation where one does not know the explicit likelihood $p(x|\theta)$, the probability of the data given the parameters $\theta$, but has access to a simulator that defines the likelihood $p(x|\theta)$ implicitly~\cite{Cranmer:2019eaq,Brehmer:2020cvb}. In a typical setup we 
would like to solve the so 
called inverse problem\index{inverse problem} 
of getting the posterior of the 
parameters given the data, $p(\theta|x)$, but we cannot 
use Bayes theorem directly because 
we do not have explicit $p(x|\theta)$. 

In particle physics and cosmology, the simulators usually use Monte Carlo event generators (see Sec. \crossref{mcgen}) to sample unobserved latent variables $z$, such as the $z_p$ phase space of the hard scattering (see Sec. \crossref{kinema:sec:pardecay}), $z_s$ associated to showering and hadronization, $z_d$ associated to the interaction of particles with the detector
(see Sec. \crossref{passage}), or 
initial Gaussian modes of the universe realization. As such, the full simulation chain can be expressed approximately as
\begin{equation}
    \label{ML:eq:simulation_chain}
    p(x|\theta) = \int dz p(x,z|\theta)=\int \dif z_d \int \dif z_s \int \dif z_p \, p(x | z_d) p(z_d | z_s) p(z_s | z_p) p(z_p | \theta) \;, 
\end{equation}
where $\theta$ are the Lagrangian parameters that dictate the hard scattering. Evaluating the marginal likelihood $p(x|\theta)$ is intractable as it would require evaluating the integral above for each event.  

While the marginal likelihood is intractable, simulators provide the ability to generate synthetic data $x_i \overset{\text{i.i.d.}}{\sim} p(x|\theta)$ for any value of the parameters $\theta$. One can use a suitable proposal distribution $\tilde{p}(\theta)$, sample $\theta_i \overset{\text{i.i.d.}}{\sim} \tilde{p}(\theta)$, generate synthetic data $x_i \sim p(x|\theta_i)$, and then assemble a training dataset $\{x_i, \theta_i\}_{i=1,\dots,n}$ that can be used to train various machine learning models. 

There is thus a close analogy between 
simulation-based inference and 
data driven machine learning tasks discussed so far, replacing 
$\theta$ with $y$. One difference
is that in simulation-based inference we can always generate new 
samples by running additional simulations, 
while we typically view training dataset in machine 
learning as fixed. This property of 
simulation-based inference enables active learning\index{active learning}, 
where the additional simulations are 
chosen such as to minimize the error 
on the desired statistical inference task. 
Another difference is that we often
have access to the joint likelihood 
$p(x,z|\theta)$, where $z$ are unobserved
latent variables\footnote{For this reason we prefer to use simulation-based inference instead of likelihood-free inference: joint likelihood $p(x,z|\theta)$ is often available, it is the 
marginal integral over latent space $z$ that is assumed to be intractable.}.

Typically in particle physics, one uses histograms or kernel density estimation to model the distribution of observables (low-dimensional summary statistics such as the invariant mass) of simulated data~\cite{Diggle1984MonteCM}. 
Alternatively, one can use an explicit parametric family (such as a falling exponential or a Gaussian distribution) to model $\hat{f}(x |\theta) \approx p(x | \theta)$.
That model is then used as as a surrogate for the unknown density implicitly defined by the simulator.
A related approach is known as approximate Bayesian computation (ABC)\index{ABC, approximate Bayesian computation}, which approximates the likelihood through an acceptance probability that synthetic data is sufficiently close to the observed data~\cite{rubin1984, beaumont2002approximate}.
In practice, these techniques are limited to low-dimensional representations of the data.
Thus the potential of recent machine learning approaches to simulation-based inference is to extend this approach to higher-dimensional data, while maintaining the already well-established statistical procedures. 

For instance, one can use normalizing flows\index{NF, normalizing flow} (see Sec.~\ref{ML:sec:flows}) and the loss functions for density estimation\index{density estimation} (see Sec.~\ref{ML:sec:density_estimation}) to learn a surrogate model for the likelihood $\hat{f}(x |\theta) \approx p(x | \theta)$~\cite{Cranmer:2016lzt}.
Similarly, one can use conditional density estimation to learn a surrogate model for the posterior $\hat{f}(\theta | x) \approx p(\theta |x)$, which may involve including the prior-to-proposal ratio $\tilde{p}(\theta)/p(\theta)$~\cite{NIPS2016_6084}.
In addition to the unsupervised learning techniques, one can also use supervised learning to learn the likelihood-ratio $r(x| \theta_0, \theta_1) = p(x|\theta_0)/p(x|\theta_1)$ by leveraging the \textit{likelihood-ratio trick}\index{likelihood-ratio trick} of Eq.~\ref{ML:eq:lrt}~\cite{Cranmer:2015bka,Brehmer:2018hga}.

In some cases one can also augment the training dataset to include the joint likelihood-ratio
\begin{equation}
    \label{ML:eq:joint_ratio}
    r(x_i, z_i | \theta_0, \theta_1) \coloneqq p(x_i,z_i|\theta_0)/p(x_i,z_i|\theta_1)  \;,
\end{equation}
which can be used to reduce the variance for the squared-error or cross-entropy losses~\cite{Brehmer:2018hga,Stoye:2018ovl}.
While the marginal likelihood $p(x|\theta)$ is intractable due to the high-dimensional integral over the latent space, the joint likelihood is often tractable since no integration is necessary.


In some cases performing the 
marginal integral of Eq. \ref{ML:eq:simulation_chain} is 
tractable even for high dimensional 
latent space $z$. 
One of the approaches to make 
it feasible in high dimensional 
latent space is to 
make simulations differentiable with respect 
to all of its parameters, global variables $\theta$ and latent variables $z$. 
While differentiable simulations have not traditionally been developed for scientific applications, the success of machine learning based on backpropagation\index{backpropagation} combined with gradient descent\index{gradient descent} (see Sec.~\ref{ML:sec:grad_opt}), has inspired a renewed interest.
One example is FlowPM cosmological $N$-body simulation, which takes advantage of Mesh-Tensorflow to achieve a GPU-accelerated, distributed, and differentiable simulation~\cite{modi2020flowpm}.
Availability of simulation gradients in turn
enables gradient based 
Monte Carlo Markov chain methods to 
perform high dimensional marginal integral over the latent space
$z$ and over parameter space $\theta$ \cite{Jasche:2012kq}. 



Often SBI uses predetermined summary statistics, such as binned histograms in particle physics, or power spectrum in 
cosmology, to avoid the curse of dimensionality. It is however possible to 
train on uncompressed high dimensional data in cosmology by exploiting the symmetries \cite{2024PNAS..12109624D}. Yet another alternative is to 
train the network to search for the 
best possible summary statistic.
 The summary statistic can then simply be $\hat{\theta}$, which is the estimate of 
the parameters $\theta$ that emerge
from a supervised training on simulations. In SBI these can often be biased even after training, and one possible solution is to form a 
pseudo-likelihood to model the bias as a function of the true value of $\theta$ \cite{Ribli_2019}. 

\subsection{Latent space reconstruction and unfolding}
\label{sec:ML:unfolding}

While much of the work on simulation-based inference described above is aimed at inferring the parameters $\theta$ of the simulator, there is work that aims to infer the latent variables $z$.
A common approach in particle physics is 
to think of the parameters $\theta$ as parameters of a theory, such as masses, coupling constants, or Lagrangian parameters, while $z$ might describe the kinematics of a collision before the detector response. 
Inferring the distribution $p(z|\{x_1, \dots, x_n\})$ from a dataset of multiple observations is commonly referred to as unfolding\index{unfolding} in particle physics, and deconvolution in other contexts. Unfolding is a classic inverse problem, and the collection of ideas being used for machine-learning based simulation-based inference are also being applied in this setting~\cite{2022JInst..17P1024A}. 
For example, the OmniFold method~\cite{Andreassen:2019cjw} iteratively reweights a dataset in an unbinned way using machine learning to produce a simultaneous measurement of many observables.
In this method, samples $\vec{x}_r$ from detector-level MC simulation are first corrected by a learned weighting function $\omega(\vec{x}_r)$ to match data.
Then, samples $\vec{x}_p$ from particle-level MC simulation are corrected by another learned weighting function $\nu(\vec{x}_p)$ to match the $\omega(\vec{x}_r)$-weighted MC simulation.
The method is iterated multiple times, to achieve $\nu(\vec{x}_p)$-weighted MC events whose event yields and kinematics match those observed in data.
The H1~\cite{H1:2021wkz} and ATLAS~\cite{ATLAS:2024xxl} Collaborations have used the OmniFold method in experimental measurements.
It has also been applied to T2K~\cite{Huang:2025ziq} and CMS~\cite{Komiske:2022vxg} open data.

In cosmology, a common task is to 
reconstruct initial density distribution of the
dark matter, or its final distribution, from data such as galaxy positions. This can then be used for 
various downstream tasks such as 
cosmological parameter inference or 
making maps of dark matter in our 
universe. High dimensional SBI can be used for this 
task \cite{2025JCAP...09..039P}. An alternative 
is Bayesian inverse problem inference
using the 
forward model $g(z,\theta)$, which can be an N-body simulation with some 
simple galaxy formation model added to it \cite{Jasche:2012kq,Wang2014a,Seljak2017a}.  
Standard 
Bayesian methodology using for example MCMC can be used to 
solve this task and find the posterior
$p(z,\theta|x)$, which specifies 
initial distribution of dark matter. To draw samples of final dark matter 
distribution, and of 
the reconstructed 
data, we can first draw samples from 
the posterior $p(z,\theta|x)$, and then evaluate forward model $g(z,\theta)$ for each sample. 

\section{Data representations, inductive bias, and example applications}
\label{ML:sec:structured_data}

In Sec.~\ref{ML:sec:supervised} we describe the input data as living in an abstract space $x_i \in \mathcal{X}$.
In this section, we briefly discuss some of the common types of structured data that are encountered in physics and refer to the corresponding models classes that have been developed to work with them. We elaborate on the model classes in more detail in the following section. 

The most basic and common type of data structure is when $\mathcal{X} = \mathbb{R}^d$. This is often referred to as \textit{tabular data}\index{tabular data} since the entire data set $\{x_i\}_{i=1,\dots,n}$ can be thought of as a table with $n$ rows and $d$ columns.
It is common to think of an individual entry $x_i$ as a vector in $d$-dimensional Euclidean space, where the coordinates correspond to the columns of this table.
In some cases individual components of $x_i$ might be integers or take on only discrete values, in which case describing the space of the data as real-valued is a slight abuse of notation and representation.
For many years this was the dominant type of data in high energy physics as it is a natural input type for shallow neural networks, multilayer perceptrons\index{MLP, multilayer perceptron}, support vector machines\index{SVM, support vector machine}, and tree-based methods\index{tree} found in popular tools such as \texttt{TMVA}~\cite{Hocker:2007ht}. 

For categorical data, one typically uses a numerical representation such as \textit{integer encoding}\index{integer encoding} where different categories are mapped to integers with a corresponding dictionary. Another common representation of categorical data is based on the so-called \textit{one-hot encoding}\index{one-hot encoding} (aka `one-of-K' or `dummy'), in which case the category is mapped to a $k$-dimensional binary vector where $k$ is the number of categories and each component of this vector corresponds to a particular category. 
In the one-hot encoding, only one of the components is non-zero.
Finally, there are approaches in one learns an \textit{embedding}\index{embedding} that maps discrete categories into $\mathbb{R}^d$; an example of this is \texttt{Word2Vec}~\cite{mikolov2013efficient}.
Interestingly, such embeddings can preserve various types of semantics; for instance, the vector \texttt{walking - walk} is similar to the vector \texttt{swimming - swam} as are the vectors connecting countries and their capital cities.
This allows for a loose sense of algebra on the word embeddings such as \texttt{walking - swimming + swam = walk}.
Similar types of embeddings have also been used in a number of scientific use-cases including biological sequences (\eg, DNA, RNA, and proteins) for bioinformatics applications~\cite{asgari2015continuous}.

Particle physics data often is represented with an extension of the simple tabular data structure where the number of columns is not fixed. For instance, if the rows correspond to data for individual collisions, the number of electrons (and positrons) reconstructed in the event is variable. Thus the number of columns needed to represent the energy, momentum, and charge of these particles is also variable. A common solution to this problem is to fix a maximum number of particles and then \textit{truncate} and \textit{zero-pad} to fit the data into a fixed tabular representation, though this is not the natural representation of the data and it leads to a loss of information. 

\textit{Sequential data}\index{sequential data} is also commonly encountered in physics (\eg in time series). Here an individual entry $x_i = (x_{i}^1, \dots, x_i^t, \dots x_i^{T_i})$ where $t$ is index for the ordered sequence, $T_i$ is the length of the sequence (which might be variable), and the data associated to each ``time'' $x_i^t \in \mathbb{R}^d$.
This is similar to the previous example where the energy, momentum, and charge of the $t$th electron in the $i$th event would be $x_i^t$ and the electrons might be sorted according to their energy or transverse momentum.
Sequential data is also encountered in natural language processing, where $x_i^t$ correspond to individual words in a sentence. Recurrent neural networks\index{RNN, recurrent neural network} (see Sec.~\ref{ML:sec:rnn}) are particularly well suited to sequential data.
Examples applications from the Living Review include Refs.~\cite{Guest:2016iqz,Nguyen:2018ugw,Bols:2020bkb,goto2021development,deLima:2021fwm,ATL-PHYS-PUB-2017-003}.

\textit{Image-like data}\index{image data} is one of the most dominant forms of data in industrial applications of deep learning, is very relevant for astronomy and cosmology, and also appears in particle physics in various forms. Image-like data typically involves $d$-dimensional features associated to a regular grid or lattice that does not vary across the individual instances $x_i$. The canonical example is a standard image from a camera with $W\times H$ pixels where the $p$th pixel has data $x_i^p \in \mathbb{R}^3$ corresponding to the three \textit{channels}\index{channels} in the RGB color model. It is important to recognize that the data corresponding to the 2-dimensional image is not 2-dimensional; instead, it is $(W\times H \times c)$-dimensional, where $c$ is the number of channels. In astronomy, an image may be grey scale ($c=1$) or there may be more \textit{channels} ($c>3$) corresponding to different color filters. In other applications, the grid or lattice might be 3- or 4-dimensional. For example, the data associated to a regularly segmented calorimeter can be thought of as a 3-dimensional image and the data associated to a lattice simulation of a classical or quantum system can be thought of as a 4-dimensional image. Convolutional neural networks\index{CNN, convolutional neural network}, described in Sec.~\ref{ML:sec:cnn}, are particularly well suited to image-like data.  Example applications from the Living Review include Refs.~\cite{Pumplin:1991kc,Cogan:2014oua,Almeida:2015jua,deOliveira:2015xxd,ATL-PHYS-PUB-2017-017,Lin:2018cin,Komiske:2018oaa,Barnard:2016qma,Komiske:2016rsd,Kasieczka:2017nvn,Macaluso:2018tck,li2020reconstructing,li2020attention,Lee:2019cad,collado2021learning,Du:2020pmp,Filipek:2021qbe,Nguyen:2018ugw,ATL-PHYS-PUB-2019-028,Andrews:2018nwy,Chung:2020ysf,Du:2019civ,Andrews:2021ejw,Pol:2021iqw}.

It is also possible that the data (or features) associated to one ``pixel'' or lattice site may itself be structured. For example, the single read-out plane of a liquid argon time projection chamber (LArTPC) may involve a 1-dimensional or 2-dimensional grid, but the data associated to each ``pixel'' is itself a sequence or waveform. Example applications in neutrino physics from the Living Review include Refs.~\cite{Aurisano:2016jvx,Acciarri:2016ryt,Hertel:DLPS2017,Adams:2018bvi,Domine:2019zhm,Aiello:2020orq,Adams:2020vlj,Domine:2020tlx,DeepLearnPhysics:2020hut,Koh:2020snv,Yu:2020wxu,Psihas:2020pby,Alonso-Monsalve:2020nde,Abratenko:2020pbp,Clerbaux:2020ttg,Liu:2020pzv,Abratenko:2020ocq,Chen:2020zkj,SBND:2020eho,Qian:2021vnh,abbasi2021convolutional,Drielsma:2021jdv,Rossi:2021tjf,Hewes:2021heg,Acciarri:2021oav,Belavin:2021bxb,Maksimovic:2021dmz,Gavrikov:2021ktt,Garcia-Mendez:2021vts,Carloni:2021zbc,MicroBooNE:2021nss}. Similarly, in lattice quantum chromodynamics, the data associate to a particular site (or link) would be group valued (\eg $x_i^p \in SU(3)$ as in Refs.~\cite{Boyda:2020hsi,Kanwar:2020xzo}). 

Both sequential and image-like data have a notion of temporal or spatial structure. While it is possible to unroll an image into a $(W\times H \times c)$-dimensional vector, that would erase the spatial structure and obfuscate the fact that nearby pixels are highly correlated. Similarly, one could permute the time index for sequential data, but that would destroy the temporal structure of the data. The complementary point of view is that the model class should also be aware of the structure of the data. Recurrent and convolutional neural networks are good examples of \textit{inductive bias}\index{inductive bias} as the models incorporate the structure of the data. In some cases this can be formalized in terms of symmetry. For example, if we train model to classify images of cats and dogs, we would like it's prediction to be invariant to where in the image the cat is. This type of translational invariance\index{translational invariance} can be enforced in the design of the model class. 

While permuting the elements of a sequence destroys the temporal structure of a time series, attaching a temporal index $t$ to a set of  objects with features $x_i^t$ can also be problematic. If the data corresponding to $x_i$ are really a set $\{x_i^1, \dots, x_i^{T_i}\}$ (\eg, a point cloud), then we would like the output of the model to be \textit{permutation invariant}\index{permutation invariance} or \textit{permutation equivariant}\index{permutation equivariance} depending on if the output is per-set or per-element, respectively. A standard sequential or convolutional model will not generally be permutation invariant, but models such as deep sets\index{deep set}, various types of graph neural networks\index{GNN, graph neural network}, and transformers\index{transformer} can be made to enforce permutation symmetry. 
Example applications from the Living Review include Refs.~\cite{Komiske:2018cqr,Qu:2019gqs,Mikuni:2020wpr,Shlomi:2020ufi,Dolan:2020qkr,Fenton:2020woz,Lee:2020qil,collado2021learning,Mikuni:2021pou,Shmakov:2021qdz,Shimmin:2021pkm,ATL-PHYS-PUB-2020-014,Qu:2022mxj}.

The temporal and spatial structure of sequences and image like data can also be generalized. For instance, a 1-dimensional sequence can be generalized to a tree structured data like one finds in the hierarchical clustering of jets or as in a directed-acyclic graph (DAG).  Generalizations of recurrent neural networks have been constructed that can operate over these more complex data structures~\cite{Louppe:2017ipp,Cheng:2017rdo}. More generally, one can considered graph-structured data composed of nodes and edges or multi-graphs that group together three nodes into faces or $k$ nodes into $k$-edges. Graph neural networks are a class of models that work with this type of data. The emerging subfield of geometric deep learning aims to unify the notation, terminology, and theory that connect these considerations of the structure of the data and the corresponding model architecture. Example applications in the Living Review include Refs.~\cite{Henrion:DLPS2017,Ju:2020xty,Abdughani:2018wrw,Martinez:2018fwc,Ren:2019xhp,Moreno:2019bmu,Qasim:2019otl,Chakraborty:2019imr,Chakraborty:2020yfc,1797439,1801423,1808887,Iiyama:2020wap,1811770,Choma:2020cry,Alonso-Monsalve:2020nde,guo2020boosted,Heintz:2020soy,Verma:2020gnq,Dreyer:2020brq,Qian:2021vnh,Pata:2021oez,Biscarat:2021dlj,Rossi:2021tjf,Hewes:2021heg,Thais:2021qcb,Dezoort:2021kfk,Verma:2021ceh,Hariri:2021clz,Belavin:2021bxb,Atkinson:2021nlt,Konar:2021zdg}.

If the data are expected to have a symmetry associated to them but one is working with a model class that does not enforce this symmetry, then \textit{data augmentation}\index{data augmentation} is a common procedure used to improve generalization performance. Here one starts with an initial dataset $\{x_i\}_{i=1,\dots,n}$ and produces an augmented dataset $\{x_i'\}_{i'=1,\dots,n'}$ through some data augmentation strategy. For example, one might apply a random rotation $R_{i'}$ to an image to produce $x_i' = R_{i'}(x_i)$ if one assumes rotational invariance\index{rotational invariance} in the underlying problem.

In some cases some of the individual features (components) of $x$ are functions of other features. For instance, one may include components of a four-vector $(E, p_x, p_y, p_z)$ as well as redundant information such as transverse momentum, azimuthal angles, rapidity, etc. In this case, the data is restricted to a lower-dimensional surface embedded in $\mathcal{X}$. Even if the features aren't redundant, statistically the data are often effectively restricted to a small subspace of statistically likely samples and those that are exceedingly unlikely. For instance, the space of natural images is a small and highly structured subspace of all possible images, which are dominated by what we would perceive visually as noise. The term \textit{data manifold}\index{data manifold} is used to describe this restricted subspace where the data are to be found, even though it does not necessarily satisfy the formal requirements of a manifold in the mathematical sense.

These considerations on the structure of the data not only apply not to the input data $x_i \in \mathcal{X}$, but also to the output data $y_i \in \mathcal{Y}$. For instance, one might want a sequence-to-sequence model as in machine translation of written text~\cite{bff0e6bd8f4a4f0d9735bf1728fb43ef} or to learn a function that takes sets as input and produces graphs as output as in the Set2Graph mode~\cite{NEURIPS2020_fb4ab556}. One might also want the input and output of the model to be different in representations of an underlying symmetry group and for the model to enforce group-equivariance~\cite{Boyda:2020hsi,Kanwar:2020xzo}. The development of the necessary modeling components to enable practitioners to compose and train these types of models is a significant development for the field of physics. 

\section{Flavors of ML models}
\subsection{Support vector machines}

Support vector machines (SVMs)\index{SVM, support vector machine} are a class of supervised learning\index{supervised learning} models used for classification and regression\index{classification}\index{regression}. The learning algorithm involves a convex optimization problem that has a unique solution and can be solved with quadratic programming techniques. In this sense, they are robust and easier to characterize than neural networks that involve non-convex optimization. 


Linear support vector machines are used for binary classification\index{binary classification}, where $\mathcal{X} = \mathbb{R}^d$ and the target labels are conventionally defined as $\mathcal{Y} = \{-1,1\}$. The classification is simply based  on which side of a hyperplane the data lie. Any hyperplane can be written as the set of points $x$ satisfying ${\displaystyle  {w}^{T} {x} -b=0}$, where $w,b \in \mathbb{R}^d$ are the parameters of the model. The vector $w$ is normal to the hyperplane, but not necessarily normalized. The quantity ${\tfrac {b}{\| {w} \|}}$ quantifies the offset of the hyperplane from the origin along the normal vector $w$.

If the training dataset is linearly separable, then there is a region bounded by two parallel hyperplanes, called the \textit{margin}, that separate the two classes of data. The maximum margin classifier is uniquely defined by making the distance between these two hyperplanes as large as possible. The boundaries of the margin can be defined by ${\displaystyle  {w} ^{T} {x} _{i}-b = \pm 1}$, and the width of the margin is given by  ${\tfrac {2}{\| {w} \|}}$. Figure~\ref{ML:fig:max_margin} illustrates this for $x\in \mathbb{R}^2$. 

\begin{pdgxfigure}
    \centering
    \includegraphics[width=.43\textwidth]{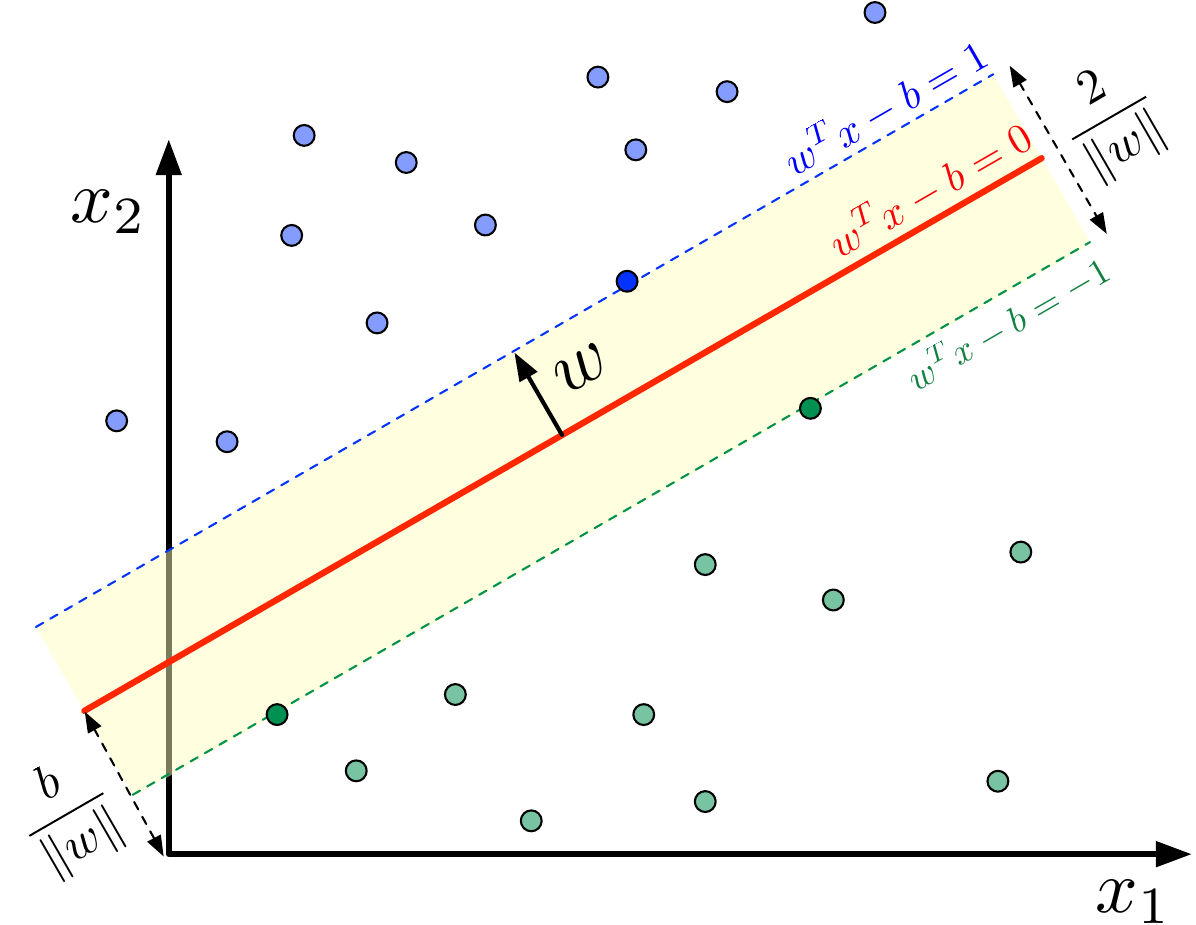}\hspace{.05\textwidth}
    \caption{Illustration of a maximum margin classifier for a linear support vector machine in the separable case.
    }
    \label{ML:fig:max_margin}
\end{pdgxfigure}


Since the width of the margin is maximized when $\|w\|$ is minimized, we can state the goal of the (hard) maximum-margin classifier in the linear separable case as the following constrained optimization problem:  
Minimize 
$\| {w} \|^2$ subject to the constraint
${\displaystyle y_{i}( {w} ^{T} {x} _{i}-b)\geq 1}$ for 
$i=1,\ldots ,n.$
The 
$w$  and 
$b$ that solve this problem uniquely determine the resulting classifier, 
$\hat{y}(x) = \operatorname {sgn}( {w}^{T} {x} -b)$.
This geometric description makes it clear that the maximum-margin hyperplane is completely determined by those 
${{x}}_{i}$ that lie nearest to it: the eponymous  \textit{support vectors}.

\subsection{From Bayesian linear regression to kernel regression and Gaussian processes}
\label{ML:sec:GP}
As discussed in Sec.~\ref{ML:sec:regression},
linear 
regression is a specific case of 
regression
where the solution is parameterized as
a linear combination of basis 
functions $\phi(x)$, 
\begin{equation}
    f_\phi(x)=\sum_k w_k \phi_k(x)=w^\intercal\phi,
\end{equation}
using a short-hand vector
notation. If we aggregate all the basis functions of the training data into $\Phi(x)$ and all $x$ into $X$, and assuming a
Gaussian noise model $\epsilon \sim \mathcal N(0,\sigma_n^2)$, we can write
the noise probability distribution as
\begin{equation}
    p(y|X,w)=\mathcal N(w^\intercal\Phi,\sigma_n^2I).
\end{equation}
In overparametrized models,
this needs to be regularized, 
with explicit L2 norm of the weights, as discussed in Sec.~\ref{ML:sec:regularization}.
If we view the process in the Bayesian context, we add a weight prior
$p(w)=\mathcal N(0,\Sigma_p)$ to the 
data likelihood. 
With this one can define the 
posterior of the weights as 
\begin{equation}
    p(w|X,y) \propto p(y|X,w)p(w)=\mathcal N(\sigma_n^{-2}A^{-1}\Phi y,A^{-1}),
\end{equation}
where $A=\sigma_n^{-2}\Phi \Phi^\intercal+\Sigma_p^{-1}$.

Our main task is not to predict the weights 
themselves, but to predict  
$f^*$ given some $x^*$. 
In the Bayesian view, 
one must model average over the weights, 
\begin{equation}
    p(f^*|x^*, X, y)=\int \dif w p(f^*|x^*,w)p(w|X,y)=\mathcal N(\sigma_n^{-2}\phi(x^*)^\intercal A^{-1}\Phi y,\phi(x^*)^TA^{-1}\phi(x^*)),
\end{equation}
where we used $p(f^*|x^*,w)=N(0,\sigma_n^2)$.
This can be rewritten as 
\begin{equation}
    p(f^*|x^*, X, y)=\mathcal N(\phi(x^*)^T\Sigma_p\Phi (K+\sigma_n^2I)^{-1}y,\phi(x^*)^\intercal\Sigma_p\phi(x^*))-\phi(x^*)\Sigma_p\Phi (K+\sigma_n^2I)^{-1}\Phi^T\Sigma_p\phi(x^*)),
\end{equation}
where $K=\Phi^T \Sigma_p \Phi$.
In general we call $k(x,x')=\phi^T(x) \Sigma_p \phi(x')$ a kernel or covariance function between $x$ and $x'$. 
The final expression for the regression mean and covariance has a
form 
\begin{equation}
    f^*=K(x^*,X)[K(X,X)+\sigma_nI]^{-1}y,
\end{equation} 
with covariance 
\begin{equation}
    \mathrm{cov}(f^*)=K(x^*,x^*)-K(x^*,X)[K(X,X)+\sigma_n^2I]^{-1}K(x^*,X).
\end{equation}
One can see that the prediction 
takes the observed values $y$ at the 
training data $X$, and predicts the value at $x^*$ by incorporating 
the strength of their correlations 
via $K(x^*,X)$. In addition, the 
prediction also suppresses the 
training points with a small inverse covariance, a generalization 
of inverse noise weighting. 

This expression simply rewrites the standard Bayesian 
regression, and the kernel is 
still defined as an inner product 
of the regression basis functions 
$\phi(x)$ with respect to $\Sigma_p$. However, the kernel
can be replaced by any kernel form that
describes the level of correlations 
between $x$ and $x'$, a process known as the \emph{kernel trick}\index{kernel trick}. A general 
condition for the kernel to be 
valid is that the covariance 
matrix is always invertible. 
In a Gaussian 
process the kernel is often stationary, defined as $k(x,x')=f(|x-x'|)$. This and many 
other kernels
cannot be related to a finite 
set of basis functions $\phi(x)$, 
which is why it is often stated that Gaussian process corresponds to an infinite 
basis function expansion. 
Yet another form of the kernels
are neural tangent kernels of neural networks in the infinitely wide network limit~\cite{JacotHG18}.

Another path to Gaussian processes 
is via kernel regression, where one makes kernel regression 
Bayesian~\cite{bishop2006pattern}. 
Standard kernel regression, also called 
Nadaraya-Watson regression, is of 
the form $f^*=\sum_x k(x,x^*)y(x)/\sum_x k(x,x^*)$, 
which can be interpreted as a soft
version of the $k$-nearest neighbors algorithm.
Bayesian kernel regression in the form of a Gaussian process replaces kernel sums with matrix operations, and 
is closer to linear regression than 
to the nearest neighbor methods. 


One advantage of Gaussian process is that one can work with a family of kernel functions parameterized by some hyperparameters $\eta$. One can then optimize the hyperparameters via gradient-ascent on the marginal likelihood function. In contrast, hyperparameter tuning in other models typically requires a grid search or some other black-box optimization procedure evaluated on held out data or some form of cross-validation.

While we only describe Gaussian process regression, there is a corresponding Gaussian process classification. 
Rasmussen and Williams provides an excellent review of Gaussian processes~\cite{williams2006gaussian}. Numerically, Gaussian process libraries are confronted with computing the inverse of the covariance kernel, which scales like $\mathcal{O}(n^3)$ in computational complexity. 
Gaussian 
processes are often used as 
emulators or surrogate models, specially in the context of low dimensional input $x$ and low  number of training data $n$ to avoid the steep $\mathcal{O}(n^3)$ scaling. They 
are used widely in cosmology, and there are a growing number of applications in (astro-)particle physics~\cite{Gandrakota:2022wyl}.
Recent works explore the design of physics-inspired kernels and use Gaussian processes to model the intensity for a Poisson point process like those found in experimental particle physics and $\gamma$-ray and X-ray astronomy~\cite{Frate:2017mai, Mishra-Sharma:2020kjb, Foster:2021ngm}.
Gaussian processes are also extensively 
used in Bayesian 
optimization\index{Bayesian optimization} (Section~\ref{ML:sec:RL}), because the 
uncertainty quantification that 
is automatically provided by the Gaussian process enables exploration-exploitation strategies where 
to evaluate the function next. 

\subsection{Decision trees}\label{ML:sec:decision_trees}


\paragraph{Tree-based models} 

Classification\index{classification} and regression\index{regression} trees (CART)\index{tree} typically partition the input space into $J$ disjoint regions $\mathcal{X} = \mathcal{X}^1 \cup \cdots \cup \mathcal{X}^J$ through a sequence of $J-1$ binary splits based on an individual components of $x \in \mathcal{X}$ (\eg $x_4 < 0.7$)~\cite{Breiman1984ClassificationAR}. The model is piecewise constant and assigns the value $b_j \in \mathcal{Y}$ to the $j$th terminal region $\mathcal{X}^j$.
The model can be written 
\begin{equation}\label{ML:eq:CART}
    \hat{y}(x) = f_\phi(x) = \sum_j b_j \mathbf{1}(x\in \mathcal{X}^j) \;.
\end{equation}
The parameters $\phi$ of the model comprise the components index and thresholds for the successive splittings and the coefficients $b_j$.

Tree learning refers to the algorithm used to choosing the tree structure and determining the predictions at leaf nodes. Optimization of the tree structure involves a difficult discrete optimization since the change in the loss with respect to the tree structure is non-differentiable and it is intractable to explore the combinatorially large space of possible trees with brute force. Therefore, the discrete optimization component of tree learning typically involves some approximate algorithm based on heuristics. In contrast, optimization of the $b_j$ for a given tree structure can exploit gradient-based optimization algorithms.

Common approaches to building the decision tree start with a root node and grow with splits based on individual attributes (components of $x$). These are referred to as top-down induction strategies. There are various impurity heuristics used for choosing the best attribute to split on such as the Gini index, cross-entropy and mis-classification error. Generally they aim to find a split that will refine the the terminal nodes such that they have higher purity than the parent node. 

Because most tree learning algorithms consider splits aligned with individual feature components, there are some failure modes for tree-based models. However, tree-based models work well with tabular data composed of a mix of continuous and discrete features. Tools such as \texttt{XGBoost}~\cite{xgboost} and LightGBM~\cite{lightgbm} are competitive on tabular data benchmarks like TabArena~\cite{tabarena} and are widely used in industry; the boosted decision trees (BDTs) implemented in~\\
\texttt{StatPatternRecognition}~\cite{Narsky:2005xpa} and \texttt{TMVA}~\cite{Hocker:2007ht} have been one of the most used techniques in particle physics~\cite{Radovic:2018dip}. 

Individual trees are often referred to as weak learners and they can be combined in various ways described below. Regularization is also an important consideration with tree-based models as one can always learn a tree that assigns exactly one training dataset point per terminal node and memorize the training dataset exactly. One approach to this is called \textit{pre-pruning}, which simply terminates the growing of the trees if the number of training samples reaching the terminal node drops below some threshold, the purity of a terminal nodes is below some threshold, or if the improvement in purity due to a proposed split is not above a threshold. Another regularization approach is called \textit{post-pruning}, which uses a validation data set that is disjoint from the training dataset to probe generalization performance. In this approach, after initially growing a tree with the training dataset, a sequence of pruned trees is considered where splits are removed based on some heuristic. The tree in this sequence of pruned trees that minimizes the generalization error on the validation set is chosen. Alternatively, in tools such as \texttt{XGBoost} there is an explicit regularization term included in the loss function (see Eq.~\ref{ML:eq:xgboost_regularization}). 

\paragraph{Ensemble methods}
\label{ML:sec:ensemble}
\index{ensemble methods}
The idea of ensemble methods is to combine multiple models into a more performant one by exploiting the bias-variance tradeoff~\cite{louppe2014understanding}.  This is most commonly achieved through averaging (\eg bagging and random forests), which primarily reduces variance, or boosting (\eg, AdaBoost and gradient boosting), which primarily reduces bias. Here, bias refers to the difference between the Bayes optimal model and the average model produced by the learning procedure with different training sets and variance quantifies how much the learned model varies from one training set to another.

The motivation of boosting\index{boosting} is to combine the outputs of many ``weak'' models to produce a more expressive model. Compared to averaging techniques like bagging and random forests, the model is built sequentially on modified versions of the data and the final predictions are combined through a weighted sum
\begin{equation}\label{ML:eq:boosted_tree}
    \hat{y}(x) = \sum_{t=1}^T \beta_t \hat{y}_t(x) \;,
\end{equation}
where $\beta_t$ expand the parameters of the model $\phi$.

\paragraph{Bagging}\index{bagging, bootstrap aggregating} The idea behind bagging (bootstrap aggregation) is to create $T$ bootstrap training datasets $B_1, \dots, B_T$ drawn from the training dataset $\{x_i, y_i\}_{i=1, \dots, n}$, then learn a model $\hat{y}_t$ for each, and finally construct an average model $\hat{y}(x) = (1/T) \sum_t \hat{y}_t(x)$. If one had $T$ independent training datasets each of size $n$, then the bias of the average model would be the same as the original model, but the variance would be reduced by a factor of $T$. By using bootstrap resampling, the bias may increase but the reduction in variance often dominates, which leads to improved performance. 

\paragraph{Random forests} Random forests\index{random forest} refers to a type of ``perturb and combine algorithm'' that combines bagging and random attribute subset selection. Again one builds trees $\hat{y}_t(x)$ from bootstrap training datasets $B_t$, but instead of choosing the best split among all attributes, one select the best split among a random subset of $k$ attributes. If $k$ includes all attributes, then it is equivalent to bagging. 

\paragraph{AdaBoost} In AdaBoost\index{AdaBoost, adaptive boost} (adaptive boost) the sequence of trees $\hat{y}_1, \dots, \hat{y}_T$ are trained with reweighted versions of the original training dataset such that the weight of individual training sample is based on the prediction error in the previous iteration~\cite{freund1997decision}. This requires working with a loss function that and learning procedure for the individual iterations that is amenable to weighted training dataset $\{x_i, y_i, w_i\}_{i=1, \dots, n}$. Incorporating the weights $w_i$ is straight forward when the risk is expressed as an expectation, since the emperical risk of Eq.~\ref{ml:eq:Remp} is just replaced with the weighted average. Similarly, the heuristic for many of the tree-based learning algorithms (\eg the Gini index) also have natural generalizations with weighted events. 

In the context of classification, the weighted error of the model $\hat{y}_t(x)$ is
\begin{equation}
    \textrm{err}_t = \frac{\sum_i w_i^{(t)} \mathbf{1}[y_i \ne \hat{y}_{t}(x_i)]}{\sum_i w_i^{(t)}} \;.
\end{equation}
Based on this weighted error, the coefficient $\beta_t$ of the component $\hat{y}_t(x)$ in Eq.~\ref{ML:eq:boosted_tree} is given by 
\begin{equation}
    \beta_t = \log\left( \frac{1-\textrm{err}_t}{\textrm{err}_t} \right) \;.
\end{equation}
Then for the next iteration the weights of the misclassified events are updated as $w^{(t+1)} = w^{(t)} \exp(\beta_t )$ and then renormalized so that the sum of all weights is 1. This reweighted dataset is then used to train the next model $\hat{y}_{t+1}(x)$ and the entire procedure is initialized with uniform weights $w_i^{t=0} = 1/n$.

There is an analogous procedure for regression with the squared loss function based on the residuals $r_i = y_i - \hat{y}_t(x_i)$ (see for example Ref.~\cite{Hocker:2007ht} for details).

\paragraph{Gradient boosting} 

One of the most powerful forms of tree based models, which is implemented in the tool \texttt{XGBoost} is referred to as \textit{gradient boosting}\index{gradient boosting}~\cite{friedman2001greedy}. In this setup, the model is purely additive as in the case of random forests, so the model is Eq.~\ref{ML:eq:boosted_tree} with all $\beta_t=1$. Note this is without loss of generality since the $\beta_t$ can be absorbed into the $b_j$ of Eq.~\ref{ML:eq:CART}. As with AdaBoost, the model is built sequentially through the sequence $\hat{y}_1, \dots, \hat{y}_T$. 

At each iteration, a new term $f_t$ will be added to the sum in Eq.~\ref{ML:eq:boosted_tree}. For a given decision tree defined by splits on attributes, one can approximate the objective function (the loss function $\mathcal{L}$ plus a regularization term $\Omega$) as a function of $b_j$ in a second order Taylor series:
\begin{equation}
    \text{obj}^{(t)} = \sum_{i=1}^n [\mathcal{L}(y_i, \hat{y}_i^{(t-1)}) + g_i f_t(x_i) + \frac{1}{2} 
    h_i f_t^2(x_i)] + \Omega(f_t) + \mathrm{constant} \;,
\end{equation}
where
\begin{equation}
    g_i = \partial_{\hat{y}_i^{(t-1)}} \mathcal{L}(y_i, \hat{y}_i^{(t-1)})
\end{equation}
and
\begin{equation}
h_i = \partial_{\hat{y}_i^{(t-1)}}^2 \mathcal{L}(y_i, \hat{y}_i^{(t-1)}) \;
\end{equation}
In \texttt{XGBoost}, the regularization term is taken to be
\begin{equation}\label{ML:eq:xgboost_regularization}
    \Omega(f) = \gamma J + \frac{1}{2}\lambda \sum_{j=1}^J b_j^2 \;,
\end{equation}
where $J$ is the number of terminal nodes in the tree. With the second-order approximation of the objective, one can directly solve for the optimal $b_j$ for the next tree and the corresponding value of the optimized objective function. The improvement in the objective function can then be used as a heuristic for choosing the best split. Specifically, define $G_j = \sum_{i\in I_j} g_i$ and $H_j = \sum_{i\in I_j} h_i$, where $I_j$ is the set of indices of data points assigned to the $j$th leaf. The heuristic used in \texttt{XGBoost} for splitting a node is
\begin{equation}
\textrm{Gain} = \frac{1}{2} \left[\frac{G_L^2}{H_L+\lambda}+\frac{G_R^2}{H_R+\lambda}-\frac{(G_L+G_R)^2}{H_L+H_R+\lambda}\right] - \gamma \;.
\end{equation}
This formula can be interpreted as the score on the new left leaf plus the score on the new right leaf minus the score on the original leaf minus a regularization penalty on the additional leaf. If the gain from splitting a leaf is smaller than 
$\gamma$, then the total Gain is negative and the split will not be added, which can be seen as implementing a form of pruning.

\subsection{Neural networks}\label{ML:sec:nn}
In this section we focus on the different types of components used in modern neural network\index{neural network} architectures. 
Gradient-based optimization\index{gradient descent} techniques are most commonly used for training\index{training} neural networks, and they are described in  Sec.~\ref{ML:sec:grad_opt}. Similarly, other important aspects to effectively training neural network models such as parameter initialization\index{initialization} and early stopping\index{early stopping} are discussed in Sec.~\ref{ML:sec:learning_algorithms}. 

The vanishing\index{vanishing gradient} and exploding\index{exploding gradient} gradient problem is a common challenge for gradient-based optimization of neural networks and is described in Sec.~\ref{ML:sec:vanishing_grad}.
That problem is referred to repeatedly in this section because it has motivated the development of numerous architectural components described below.



\subsubsection{Feed-forward multilayer perceptron}\label{ML:sec:mlp}

One of the core components in neural networks is the fully-connected, feedforward network or\textit{multilayer perceptron} (MLP)\index{MLP, multilayer perceptron}, which is composed of $L$ \textit{layers}: $f = f^{(L)} \circ \dots \circ f^{(1)}$. The $l$th layer defines a function that maps a $d_{l-1}$-dimensional input vector, called {\it features}\index{feature vector}, to an $d_{l}$-dimensional output $f^{(l)}: \mathbb{R}^{d_{l-1}} \to \mathbb{R}^{d_{l}}$. A unit producing an individual component of the $d_{l}$-dimensional output is called a {\it neuron}\index{neuron} or a {\it filter}\index{filter} interchangeably. For $l<L$, the functions $f_l$ are called hidden layers, and the number of neurons ($d_{l}$) is referred to as the width of the hidden layers. The layers in an MLP take on the form:
\begin{equation}
    \label{ML:eq:hiden_layer}
    f^{(l)}(u) = \sigma^{(l)}( W^{(l)} u + b^{(l)} ) \;,
\end{equation}
where $W^{(l)} \in \mathbb{R}^{d_{l} \times d_{l-1}}$ is called the \textit{weight matrix}\index{weight matrix}, the components of the vector $b^{(l)} \in \mathbb{R}^{d_{l}}$ are referred to as the \textit{biases}, $u \in \mathbb{R}^{d_{l-1}}$ is the input from the previous layer, $W^{(l)} u$ denotes a matrix-vector product, and $\sigma^{(l)}$ is a non-linear \textit{activation function}\index{activation function} that is usually applied element-wise. The parameters of the network comprise the full collection of weights and biases, $\phi = (W^{(1)}, \dots, W^{(L)}, b^{(1)}, \dots, b^{(L)})$.

\subsubsection{Activation functions}
\label{ML:sec:activation}

The activation functions\index{activation function} $\sigma$ in neural networks are nonlinear functions and key to the expressiveness of the resulting family of functions. Two traditionally used functions are the {\it logistic or sigmoid}\index{logistic activation} function $\sigma(x) = 1/(1+e^{-x})$ and {\it hyperbolic tangent}\index{hyperbolic tangent activation} function $\tanh(x)=(e^x-e^{-x})/(e^x+e^{-x})$. These functions 
are bounded to be $(0,1)$ and $(-1,1)$ respectively, and are symmetric about the input value of zero. On the other hand, away from the zero input value, a gradient of both functions quickly vanishes and this poses a challenge in using gradient-based optimization method (see Sec.~\ref{ML:sec:grad_opt}). This can be avoided, to some extent, by normalizing the input values and carefully initializing the values of $W^{(l)}$ and $b^{(l)}$. These are discussed in Sec.~\ref{ML:sec:nn_init}, \ref{ML:sec:input_norm} and \ref{ML:sec:batch_norm}. Yet, it becomes difficult to maintain a null input value for a {\it deep} neural network, a model with many layers. Instead, a popular choice for a deep neural network is the {\it rectified linear unit} (ReLU)\index{ReLU, rectified linear unit}:
\begin{equation}
    \label{ML:eq:relu}
    \text{ReLU}(x) =
    \begin{cases}
      x & \text{if $x>0$}\\
      0 & \text{otherwise}
    \end{cases} 
\end{equation}
whose computational cost is small and ensures that the gradient does not vanish for $x\in(0,+\infty)$~\cite{Fukushima1980,Nair2010RectifiedLU}. An alternative to preserve a non-zero gradient in negative input values are called {\it leaky ReLU}\index{leaky ReLU} and modifies the output to $0.01x$ for $x\in(-\infty,0)$~\cite{Maas13rectifiernonlinearities}. Another variant, called {\it parametric ReLU} (PReLU)\index{PReLU, parametric ReLU}, turns the coefficient $0.01$ into a variable that is optimized as a part of the model during optimization~\cite{He2015DelvingD}.

The choice of activation functions depends on the model architecture and applications. As described, while the use of ReLU types are a typical choice for a deep neural network, 
a logistic function is a popular choice at the final layer for classification tasks.
In the area of neural scene representation, 
sinusoidal activation functions have been found to be surprisingly effective~\cite{sitzmann2019siren}.

Recently, additional smooth loss functions have been found to work well with larger models, such as the Gaussian-error linear unit (GELU)~\cite{hendrycks2023gaussianerrorlinearunits},
\begin{equation}
    \text{GELU}(x) = \frac{1}{2}\left ( 1+ \text{erf}\left(\frac{x}{\sqrt{2}}\right)\right)
\end{equation} and 
swish function~\cite{ramachandran2017searchingactivationfunctions}
\begin{equation}
\text{swish}_\beta(x) = \frac{x}{1+e^{-\beta x}},
\end{equation}
which smoothly interpolates between a linear function ($\beta=0$) and ReLU ($\beta=\infty$).
In addition, the value of $\beta=1$ corresponds to the sigmoid-weighted linear unit (SiLU)~\cite{elfwing2017sigmoidweightedlinearunitsneural}.

\paragraph{Softmax}
\label{ML:sec:softmax}
A softmax\index{softmax} function is often used to normalize elements of a discrete vector $u$, or to interpret the output as a probability over a set of $n$ discrete categories. 
Given a real-valued input vector $u \in \mathbb{R}^n$, the softmax function computes the output vector $v\in \mathbb{R}^n$ the $i$th component is given by:
\begin{equation}
    v_i = \frac{\exp({u_i})}{\displaystyle\sum_{j=1}^n\exp({u_j})} \;.
\end{equation}
The result has the property that $v_i\in(0,1)$ and $\sum v_i=1$.  The components of the input vector $u$ are often referred to as \textit{logits}\index{logits} in reference to their connection to the logistic function used in logistic regression. The softmax function is commonly used as the last layer in multi-class classifier\index{multi-class classification}. The softmax is also used in the context of attention (see Sec.~\ref{ML:sec:attention})\index{attention}.




\subsubsection{Universal approximation and deep learning}\label{ML:sec:deep_learning}

There are a number of universal approximation theorems\index{universal approximation} in the theory of neural networks.
One of the first was that even with one hidden layer $(L=2)$, an MLP can approximate any continuous function if the nonlinear activation function $\sigma$ is not a polynomial and the width $d_1$ is large enough~\cite{cybenko1989approximation}.
However, it is often more efficient (in terms of the number of parameters) to increase the \textit{depth} of the network $L$~\cite{NIPS2011_8e6b42f1}. 

Training a deep network (\ie $L>2$) that generalizes well can be difficult, requiring large training datasets, many gradient updates, and suitable regularization.
The introduction of large labeled training sets, advances in computing (\eg graphic processing units or GPUs which enabled orders of magnitude acceleration in parallel computation including matrix multiplies~\cite{10.1145/1553374.1553486}), development of ReLU, research progress in initialization and optimization algorithms for model parameters, and regularization techniques like \textit{dropout}\index{dropout}~\cite{dropout} all played an important role in the rise of \textit{deep learning}\index{deep learning}~\cite{lecun_2018, schmidhuber2015deep}.
Though the name deep learning was originally a reference to the depth $L$ of such networks, modern deep learning is characterized more by the composition of various types of modules that are trained through gradient-based optimization.
Below we introduce some other common network architectures. 

\subsubsection{Convolutional neural networks}\label{ML:sec:cnn}


Convolutional neural networks (CNNs)\index{CNN, convolutional neural network} are widely used for image-like data\index{image data}.
They implement the convolution of the input image $u$ and a \textit{filter} $W$ (also referred to as a kernel\index{convolution kernel}).
The parameters of the filter are learnable and the convolution involves traversing over input and calculating the inner product of the filter $W$ with the part of the input in the {\it receptive field}, which has the same spatial shape as the filter and is centered at the target pixel.
At each location---indexed by $i$ and $j$ below---there is a pixel that may have a vector of features associated with it.
In the context of CNNs, these components of these features---indexed by $c$ and $c'$ below---are often referred to as \textit{channels}\index{channels} in reference to the red, green, and blue color channels in a traditional image.
The convolution operation is often denoted with a $\ast$, and the result can be expressed as
\begin{equation}\label{ML:eq:convolution}
v_c(j) = (W \ast u_c)(j) = \sum_{c'} \sum_{i} W_{c,c'}(i) u_{c'}(``j-i")  \; ,
\end{equation}
where $``j-i"$ is shorthand for the pixel index corresponding to the translation from pixel $j$ to $i$.
By repeating the operation over all pixels, the result of a kernel convolution is also an image as illustrated in Fig.~\ref{ML:fig:convolution}. 
Note that the the number of channels in the output $v$ does not need to be the same as in the input, and the collection of filters $W_{c,c'}$ is often referred to as a filter bank.
The entire image for a fixed channel index is often referred to as a \textit{feature map}\index{feature map}. 

A key feature of the CNN architecture is that it is \textit{equivariant} to translations\index{translational invariance}, meaning that if the input image is shifted (\eg, $u(i) \to u'(i) = u(i-k)$), then the output is also shifted by the same amount (\eg, $v(j) \to v'(j) = v(j-k)$).
This equivariance property is a natural consequence of using convolutions.
A fully connected MLP would not generally have this symmetry; however, it is enlightening to imagine transferring the computation performed by a CNN to the weights and biases of a fully connected MLP, which would result in duplicating the weights of the filters multiple times.
In this view, the CNN can be interpreted as a fully connected MLP with \textit{shared weights}\index{shared weights}, which would maintain the equivariance property. 
This view is helpful for gaining intuition about the inductive bias of models and makes clear that a CNN is a subset of the fully connected MLPs that satisfy the translation equivariance property. 

\begin{pdgxfigure}[width=1.0\textwidth]
	\includegraphics{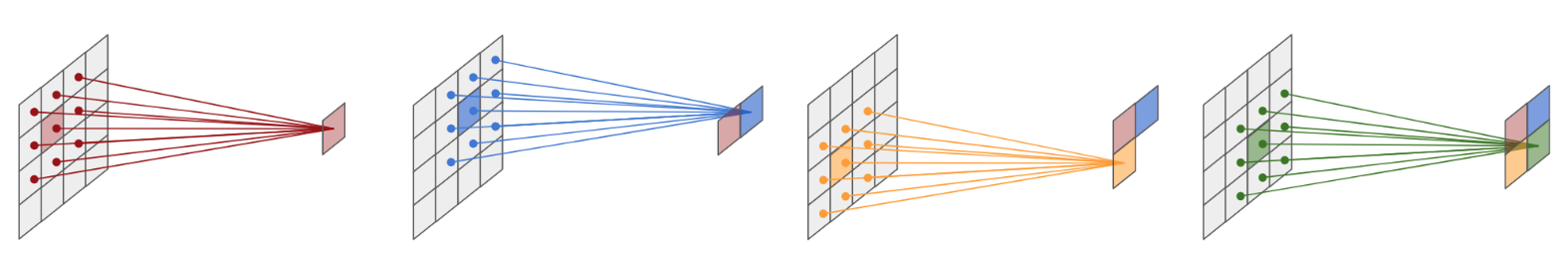} 
	\caption{A pictorial description of a kernel convolution over four input pixels. It takes a product of the weight matrix (kernel) and the local input matrix centered at a target pixel. The operation is repeated over the input image using the same kernel. The size of the output image depends on the size of the kernel, stride, and padding. In this figure, the kernel size of 3, stride of 1, and padding of 0 is used.}
	\label{ML:fig:convolution}
\end{pdgxfigure}

One may wonder how CNNs identify features with a spatial size larger than a typical kernel size.
One mechanism for this is by stacking multiple convolutional layers, \eg, the composition of two 3$\times$3 kernels will lead to an effective 5$\times$5 kernel in terms of the receptive field.
In addition, a typical CNN architecture uses pooling (described below), which effectively downsamples the image so that it can be processed at different  resolutions.
The effective receptive field\index{receptive field} in the input image may be much larger than the kernel size in this case.
An alternative approach is to use an {\it inception module}\index{inception}, which is designed to extract features simultaneously using kernels of different size~\cite{7298594}. 

\paragraph{Pooling}

\textit{Pooling}\index{pooling} plays an important role in convolutional neural networks both practically and in terms of their mathematical properties.
A pooling operation is a type of aggregation or downsampling that takes many pixels as input and produce one pixel for output.
Typically, the pooling operation is applied independently for each channel or feature component.
The most popular pooling operations are {\it max} and {\it average} pooling.
Max pooling\index{pooling, max} picks the highest activation pixel value within the specified receptive field, while the average pooling\index{pooling, average} computes the average pixel value in the receptive field.
The idea of pooling generalizes to other architectures, including graph neural networks where the receptive field includes the neighbors of a particular node in the graph (see Sec.~\ref{ML:sec:gnn}).
Pooling can make the model robust  to small, local deformations in the input, a property called \textit{geometric stability}~\cite{DBLP:journals/corr/CohenS16a,bietti2021sample,bronstein2021geometric}.
This type of local deformation is important and distinct from the equivariance to rigid translations provided by the convolutional structure. 
Repeated pooling operations that eventually lead to a single feature vector with no spatial index is what gives rise to the invariance of common CNN architectures to translations (\ie, an image with a dog will be labeled `dog' regardless of where the dog is in the image). 


\paragraph{CNN architectures for image analysis}
\label{ML:sec:CNNApps}
A typical CNN for extracting a 1-dimensional array of features is designed with repeating blocks of convolution layers and pooling operations~\cite{Simonyan2015VeryDC}. 
Figure~\ref{ML:fig:cnn_feature_extraction} shows an example evolution of a data tensor through the succession of convolution and pooling operations to extract a 1-dimensional array of features, which then can be fed into a block of MLP for an image classification\index{classification} (or a regression)\index{regression} task.
This type of architecture is referred to as an {\it encoder}\index{encoder} or {\it feature extractor}.
\begin{pdgxfigure}[width=1.0\textwidth]
	\includegraphics{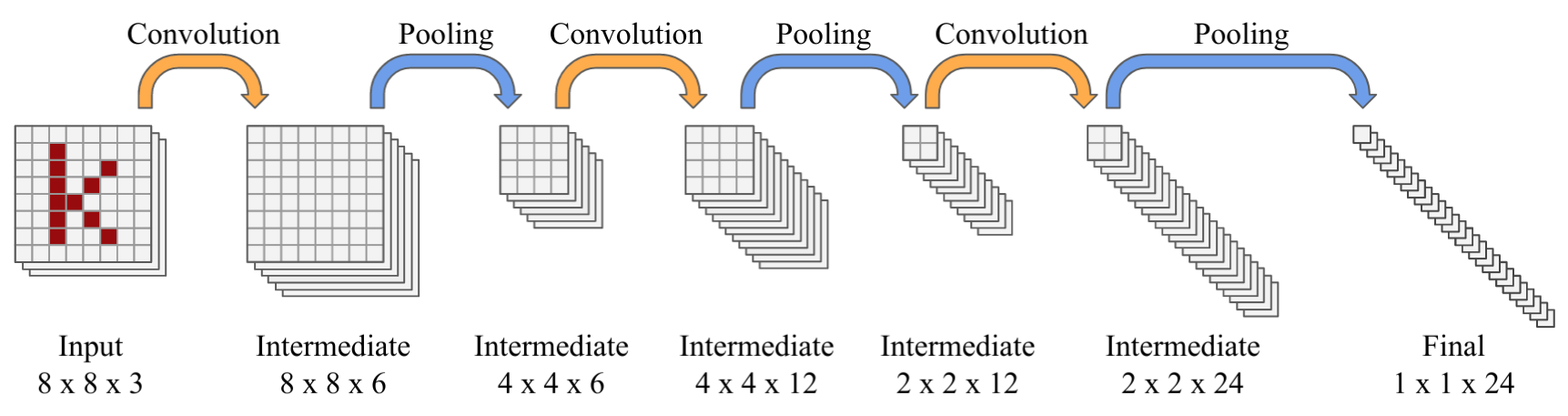} 
	\caption{An example CNN architecture to extract a 1-dimensional array of features from an image via succession of convolution layers and pooling operations. The (square) kernel, stride, and padding size of a convolution operation are 3, 1, and 1 in respective order. The pooling operation uses a square kernel size of 2. The number of filters at the first convolution layer is 6, and is increased by a factor of 2 at subsequent convolution layers.
	}
	\label{ML:fig:cnn_feature_extraction}
\end{pdgxfigure}

The reduction in the spatial size of an image is performed slowly, typically by a factor of 2, which is the minimum possible reduction factor. 
After the reduction of the spatial extent, the number of channels is typically increased (also by a factor of 2 in most cases), converging one set of feature maps into a larger number of downsampled feature maps.
There may be more than one convolution layer within each spatial resolution (\ie, between the grouping operations).
Following these design principles, CNN encoders typically become {\it deep}, consisting of dozens or sometimes hundreds of convolution layers, and face challenges of vanishing gradient problem (see Sec.~\ref{ML:sec:vanishing_grad}).
A standard practice to mitigate this issue is to explicitly {\it normalize} the input tensor input in each convolution layer using algorithms such as {\it batch normalization}\index{BN, batch normalization}.
This will be discussed in Sec.~\ref{ML:sec:batch_norm}.

There are three main categories of computer vision tasks where CNNs are often used:
\begin{itemize}
    \item {\it image classification or regression} requires a prediction of single value for the whole image (\ie, a category or target value),
    \item {\it object detection}\index{object detection} produces a list location information, typically as a rectangular shaped bounding box, for detecting arbitrary number of objects in the input image, and
    \item {\it semantic segmentation}\index{semantic segmentation} brings a classification task down at the pixel-level (or regression although less common) to identify the class or a feature of every pixel in an image.
\end{itemize}
As discussed previously, a CNN feature extractor followed by MLP is often used for image classification and regression tasks in wide range of applications including particle physics. Many successful CNN architectures for object detection and semantic segmentation applications share key designs which we briefly discuss below.

\begin{pdgxfigure}[width=0.6\textwidth]
	\includegraphics{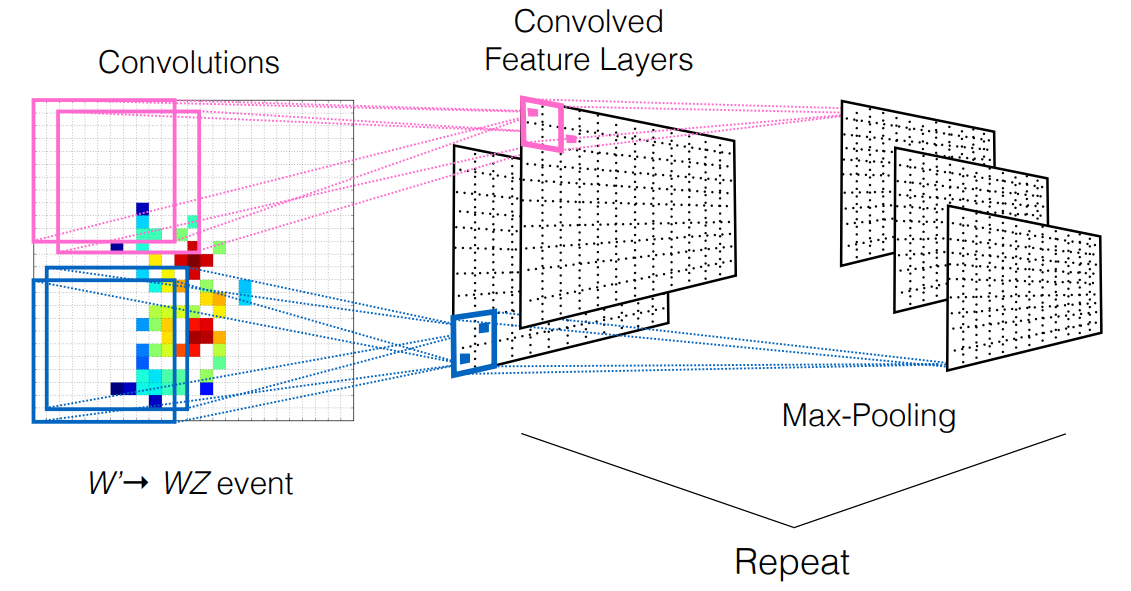} 
	\caption{CNN classifier for identifying highly boosted W bosons at ATLAS.
	}
	\label{ML:fig:cnn_collider}
\end{pdgxfigure}

\paragraph{Region convolutional neural network} (R-CNN)\index{R-CNN, region convolutional neural network} is one of the most successful design for object detection~\cite{NIPS2015_14bfa6bb}.
R-CNN has been explored in HEP experiments where the number and location of signal (\eg, neutrino interactions) are not known apriori in large image data such as neutrino detectors~\cite{Acciarri_2017,Radovic:2018dip,Domine:2020tlx}. 
R-CNN consists of multiple CNNs.
The first is a feature extractor which produces a spatially compressed feature tensor.
For every pixel in the compressed tensor, the second CNN applies $1{\times}1$ convolution to predict two information: an {\it object score} to inspect whether or not there is a target object in the (spatially compressed) pixel, and prediction of the location and size of a rectangular, axis-aligned bounding box that contains the object (if exists).
This second CNN is called the {\it region proposal network} (RPN)\index{RPN, region proposal network}, and the bounding box is called the {\it region of interest} (ROI)\index{ROI, region of interest}. 
For each ROI with an object score above threshold (hyperparameter), the third CNN operates in the corresponding sub-field of an already-compressed tensor (\ie by the first CNN) to perform a classification for an object inside the ROI.
This approach can produce multiple ROIs for the same object with a high overlap.
Those predictions are reduced using non-maximum suppression (NMS) algorithm\index{NMS, non-maximum suppression} which computes the intersection-over-union (IoU)\index{IoU, intersection over union} to combine overlapping ROIs that are likely detecting the same object.

\paragraph{Residual networks and skip connections}
\label{ML:sec:resnet}
\begin{pdgxfigure}[width=0.7\textwidth]
	\includegraphics{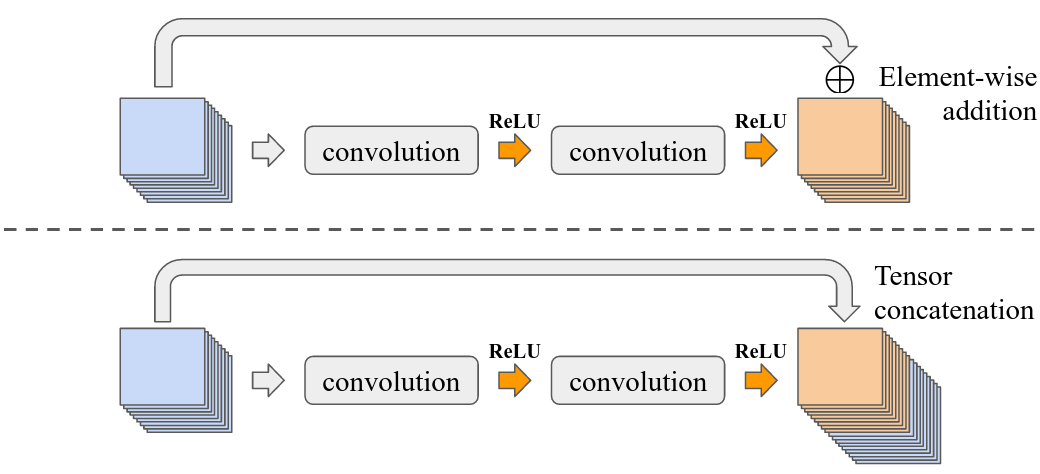} 
	\caption{Two types of skip connections: the top is from ResNet where the input is element-wise added to the output tensor of a block of convolution layers while the bottom shows a concatenation of the input to the output tensor as employed in other models including U-Net and DenseNet. 
	}
	\label{ML:fig:skip_connections}
\end{pdgxfigure}

The expressivity of a neural network increases as more hidden layers are added, but gradient-based optimization of deep models can be notoriously difficult due to vanishing gradients\index{vanishing gradient} (see Sec.~\ref{ML:sec:vanishing_grad}). 
One powerful technique to address this challenge is a {\it residual network} (ResNet)\index{ResNet, residual network}, which is a modular architecture design that can be applied to neural network models~\cite{7780459}.
Suppose a $f(x)$ as the target transformation to be learned by a few stacked layers where $x$ is the input to the first layer.
The authors of ResNet hypothesized that it may be easier for a model to learn a residual transformation $\tilde{f}(x)\coloneqq f(x)-x$, thus the objective to learn is $\tilde{f}(x)+x$ where $\tilde{f}(x)$ denotes the output of stacked layers.
This form assumes $\tilde{f}(x)$ and $x$ share the same tensor dimension and size.
If they differ in the feature dimension, equivalently the count of channels in an image tensor, one could use 1$\times$1 convolutions to transform and match the dimension. 
Adding the input tensor $x$ to the output of a convolution operation $\tilde{f}(x)$ in ResNet is a form of {\it skip connection}. For a residual block that outputs $y=\tilde{f}(x)+x$, the backpropagated gradient to the input $x$ becomes 
\begin{equation}
\frac{\partial\mathcal{L}}{\partial{x}}=\frac{\partial\mathcal{L}}{\partial y}\cdot\left(I+\frac{\partial\tilde{f}}{\partial{x}}\right),
\end{equation} meaning it sums an identity path with the gradient through the residual branch, helping prevent vanishing gradients.
ResNet authors demonstrated performance improvement at depths exceeding 1000 layers where the non-residual counterpart could not improve beyond a few dozen layers.
The ResNet design is widely used in many CNN architectures as it is modular, \ie, a {\it residual block}, and can be applied per a stack of convolution layers (\eg, U-ResNet introduced for LArTPC detectors uses ResNet modules within a U-Net architecture~\cite{Adams:2018bvi,Domine:2019zhm}).

\paragraph{U-Net}\index{U-Net} is one of the of most successful models used for semantic segmentation~\cite{10.1007/978-3-319-24574-4_28}.
Since the output of U-Net is also an image, it is more interpretable\index{interpretability} compared to models for image-level classification or regression.
The model is used widely in HEP experiments in both 2D and 3D image data~\cite{Adams:2018bvi,Domine:2019zhm,Koh:2020snv,SBND:2020eho,Abratenko:2020ocq}.
The architecture of U-Net consists of a CNN encoder and {\it decoder}\index{decoder}.
A decoder consists of convolution and {\it transposed convolution}\index{transposed convolution} layers (also called deconvolution).
The operation of a transposed convolution can be seen as the opposite of a convolution: for every input pixel, its value is multiplied by the kernel and copied to the output.
In contrast to a regular convolution layer that reduces input pixels via the kernel, a transposed convolution layer broadcasts input pixels via the kernel, producing an output that is larger than the input.
In the decoder of U-Net, transposed convolution layers are used to upsample spatially compressed feature tensors back to the original image resolution.
Standard convolution layers are placed between upsampling operations.
Features for every output pixel can be used for either a classification or a regression task.
The idea behind encoder-decoder\index{encoder-decoder} architecture is to extract features in the encoder, and the decoder interpolates those features back to the original spatial resolution.
The downsampling operation (\eg, max pooling) in the encoder is, however, a lossy process where spatial information is permanently lost. 
This is a major obstacle to achieve a high precision semantic segmentation task.
The U-Net architecture overcomes this challenge by concatenating intermediate tensors in the encoder block with the tensors of the corresponding spatial size in the decoder block.
This is a type of a {\it skip connection}\index{skip connection} discussed previously, which dramatically improves the performance of semantic segmentation.

\subsubsection{Recurrent neural networks}\label{ML:sec:rnn}

\begin{pdgxfigure}[width=.8\textwidth]
	\includegraphics{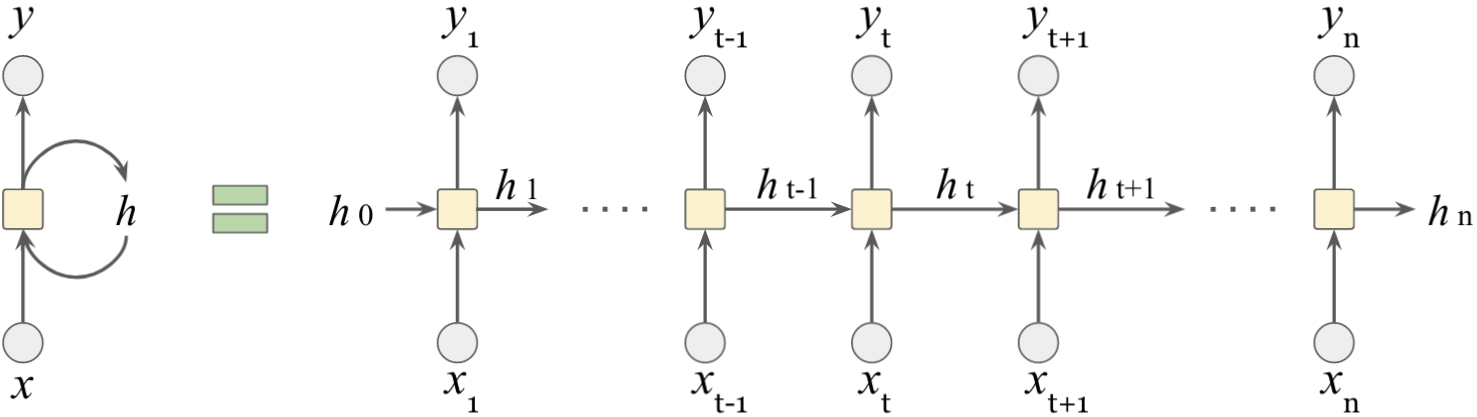} 
	\caption{Pictorial description of a RNN (on the left) which takes an input and produces an output at every step with a hidden-to-hidden connection. The right diagram is unrolled over discrete steps. The yellow box represents a cell: a set of operations unique to each architecture.}
	\label{ML:fig:rnn}
\end{pdgxfigure}
{\it Recurrent neural networks} (RNNs)~\cite{Rumelhart1986}\index{RNN, recurrent neural network} are a family of neural networks designed for sequential data\index{sequential data} (\eg, time series). Consider sequential data  where $x_t$ represents each step in a sequence with $t\in[1,n]$. A typical RNN takes the following form:
\begin{equation}
    \label{ML:eq:rnn_basic}
    h_t = g_h(h_{t-1},x_t,\theta)
\end{equation}
where $h_t$ and $\theta$ denote the {\it hidden state}\index{hidden state} of the system and parameters of $g_h$, the RNN model. The term {\it recurrent} refers the nature of the model operating on the previous state of the system (and hence the whole history). RNNs operate on three types of tasks:
\begin{itemize}
    \item {\it One-to-many} takes a single input and generates a sequence (e.g. generates a sequence data, such as a sentence or waveform, given a category).
    \item {\it Many-to-one} takes a sequence and generates an output (e.g. sequence-labeling).
    \item {\it Many-to-many} takes a sequence and generates a sequence where the length of input and output sequence may be same (e.g. classification of individual element in a sequence) or different (e.g. sequence to sequence mapping).  
\end{itemize}
Figure~\ref{ML:fig:rnn} shows an example for a many-to-many task, where $\{x_t\}_{t=1:n}$, $\{y_t\}_{t=1:n}$, and $\{h_t\}_{t=0:n}$ denote the inputs, outputs, and hidden states respectively. A set of operations at each time step is called a {\it cell}. A simple RNN cell may look like:
\begin{equation}
    \label{ML:eq:rnn_simple}
    \begin{aligned}
    h_t &= g_h(Wx_t + Vh_{t-1} + b)\\
    y_t &= g_o(Uh_t)
    \end{aligned}
\end{equation}
where $W\in\mathbb{R}^{d_h\times d_i}$, $V\in\mathbb{R}^{d_h\times d_h}$, $U\in\mathbb{R}^{d_o\times d_h}$ are matrices $g_h$ and $g_o$ represent functions. $d_i$, $d_h$, and $d_o$ are the dimension of input, hidden state, and output. $b\in\mathbb{R}^{d_h}$ is a bias term. An example application is sequence-labeling where the goal is for $y_t$ to classify each input $x_t$ in the sequence. In that case, one might use $g_h=\tanh$ and $g_o=\operatorname{softmax}$ and use a loss function that averages classification accuracy over the sequence.

\begin{pdgxfigure}[width=0.8\textwidth]
	\includegraphics{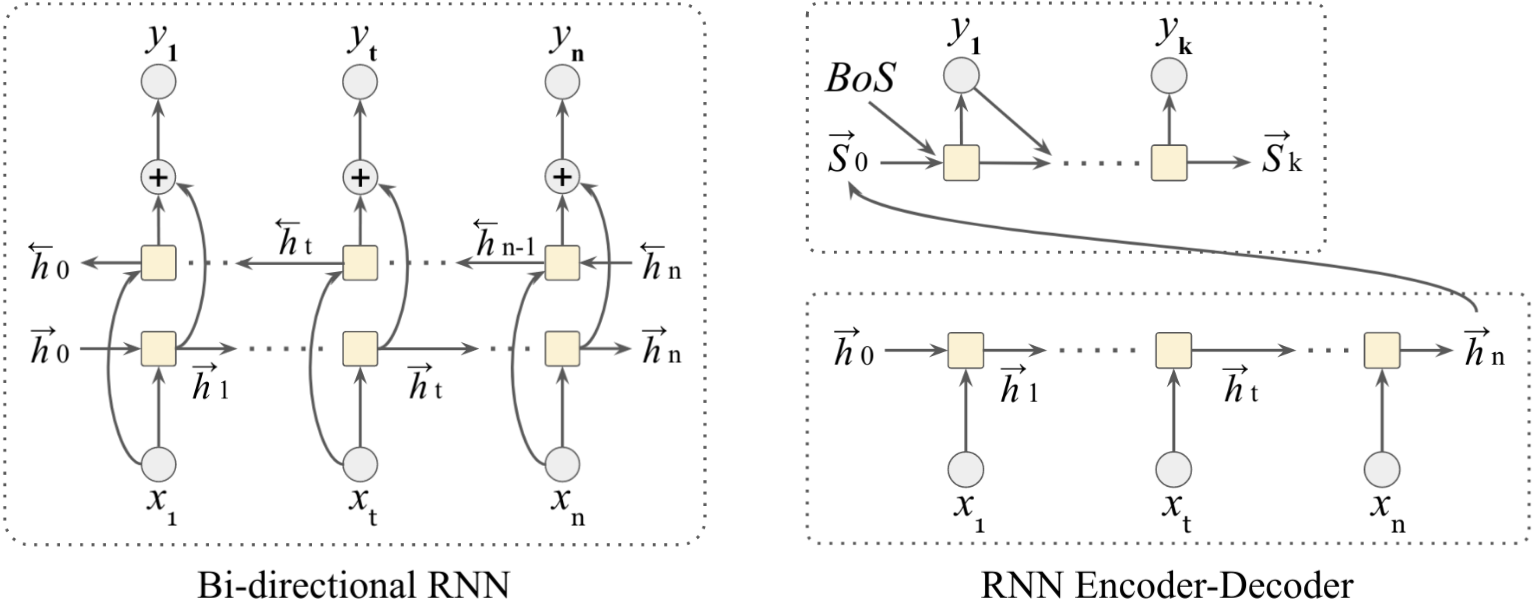} 
	\caption{Bidirectional RNN (left) provides contexts in the preceding and subsequent parts of the input sequence. RNN encoder-decoder (right) can generate an output with a different sequence length from an input. Each cell in the decoder may take a previously generated element, starting from a special marker that signals the beginning of the sequence (BoS) and ending when the end of sequence is generated.}
	\label{ML:fig:rnn_variants}
\end{pdgxfigure}

Variations in RNN architectures result from  the design of cells (described below) and flow of information across the cells. For instance, a bidirectional RNN (Figure~\ref{ML:fig:rnn_variants} left)\index{bidirectional RNN} employs two set of RNNs, one processing the sequence in the forward direction and the other in the backward direction, and the hidden states from both directions are then combined to capture the context from both parts of the sequence. An RNN encoder-decoder\index{encoder-decoder} (Figure~\ref{ML:fig:rnn_variants} right) use one RNN to generate a context vector that encodes the whole input sequence, and use a separate RNN to generate another sequence from the encoded context. This can be used for machine translation.

\paragraph{LSTM and GRU}
An RNN applies the same functions $g_h$ and $g_o$ in Eq.~\ref{ML:eq:rnn_simple} repeatedly for each element of the sequence. This repeated component is similar to the shared weights\index{shared weights} for a convolutional filter in a CNN. 

A hyperbolic tangent ($\tanh$)\index{hyperbolic tangent activation} is traditionally a popular choice for $g_h$ as it regulates the magnitude of the hidden states and prevents it from diverging. Yet, this simple model is challenging to train for a long sequence of data~\cite{bengio1994learning,pmlr-v28-pascanu13}. This is partially due to the fact that $\tanh$ contributes to the vanishing gradient\index{vanishing gradient} problem and because repeated multiplication of the same weight matrices (i.e. $V$ and $W$ in Eq.~\ref{ML:eq:rnn_simple}) can lead to gradients that can either explode\index{exploding gradient} or vanish (see Sec.~\ref{ML:sec:vanishing_grad}). Additionally, the way the signal accumulates means that changes early in the sequence have different impact from changes late in the sequence.


\begin{pdgxfigure}[width=0.85\textwidth]
	\includegraphics{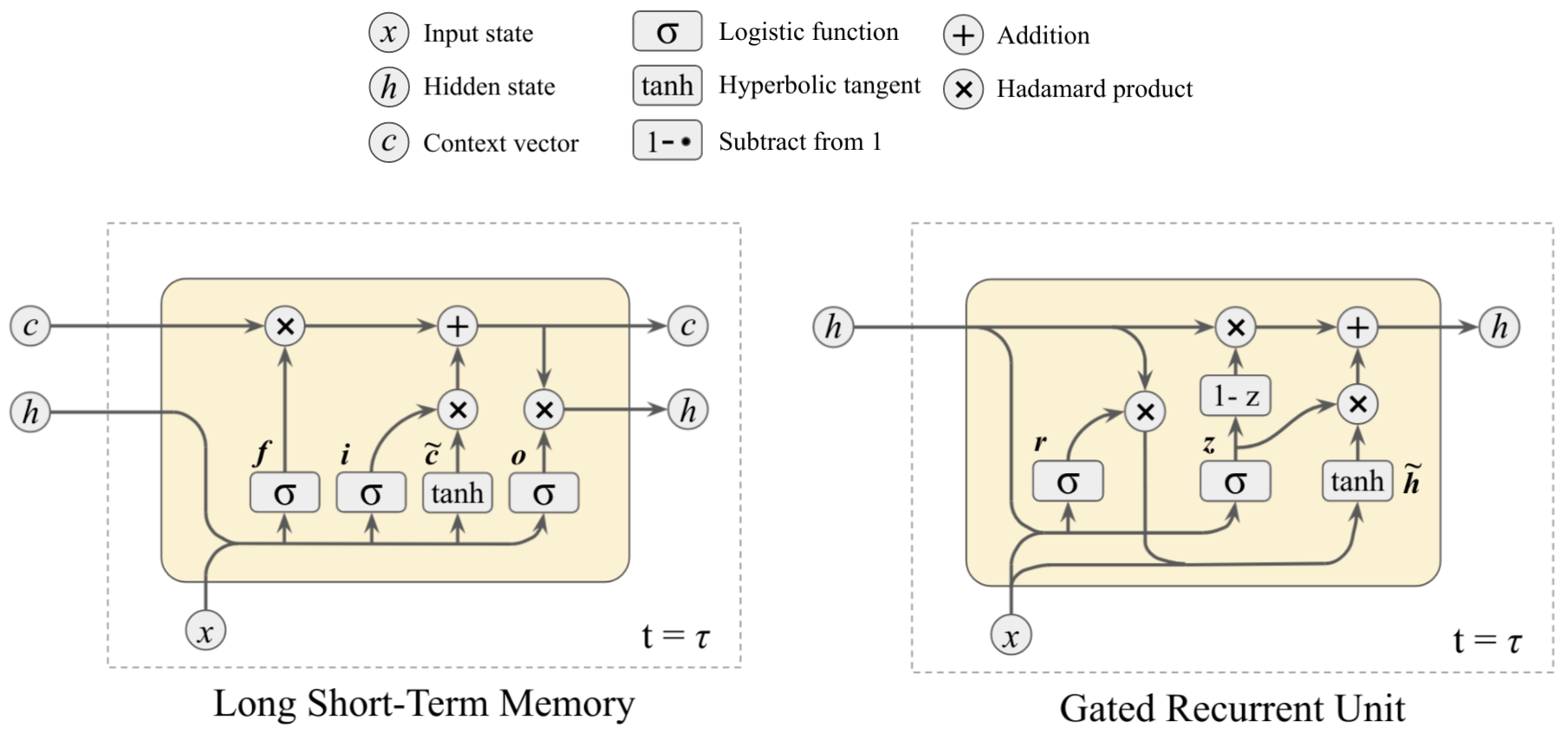} 
	\caption{LSTM (left) and GRU (right) are both gated neural network designed to address a vanishing gradient problem for RNNs.}
	\label{ML:fig:lstm_gru}
\end{pdgxfigure}
{\it Long short-term memory} (LSTM)~\cite{HochSchm97}\index{LSTM, long short-term memory} is a model designed to address the issue of vanishing gradient for RNNs. In this model, a {\it context} is introduced as a way to enable the model to hold long-term memory while the hidden states remain to hold short-term memory. The context $c_t$ and hidden state $h_t$ at step $t$ are computed as follows:
\begin{equation}
    \label{ML:eq:lstm}
    \begin{aligned}
        c_t &= f_t\bigodot c_{t-1} + i_t\bigodot \tilde{c}_{t} \\
        h_t &= o_t\bigodot c_t
    \end{aligned}
    \hspace{20pt}
    \text{where}
    \hspace{20pt}
    \begin{aligned}
        f_t &= \sigma\left(W^fx_t + V^fh_{t-1} + b^f\right) \\
        i_t &= \sigma\left(W^ix_t + V^ih_{t-1} + b^i\right) \\
        o_t &= \sigma\left(W^ox_t + V^oh_{t-1} + b^o\right) \\
        \tilde{c}_t &= \tanh\left(W^cx_t + V^ch_{t-1} + b^c\right)
    \end{aligned}
\end{equation}
where $\sigma$ and $\bigodot$ denote logistic function and an element-wise (\ie, Hadamard) product and $f_t$, $i_t$, and $o_t$ are referred to as \textit{gates}.
Each gate outputs a value between 0 and 1, and is associated with unique weights, $W$ and $V$, and a bias $b$.
One can see $c_t$ is a combination of the previous context vector $c_{t-1}$ and a new context vector $\tilde{c}_t$.
The \textit{forget} gate $f_t$ controls which and how much of the past context should be kept or forgotten.
The \textit{input} gate $i_t$ controls how much of the present context $\tilde{c}_t$ should propagate to the current state $c_t$.
The output gate $o_t$ controls which and how much of the context vector should represent the present hidden state $h_t$.
From Figure~\ref{ML:fig:lstm_gru}, one can see that the context vector $c_t$ evolves with a gated addition operation.
As such, it can be seen as an uninterrupted path for gradients to flow.
This is similar to a residual connection\index{skip connection}\index{ResNet, residual network} (see ResNet in Sec.~\ref{ML:sec:resnet}), which enabled training of CNNs with thousands of layers.

Another gated model to solve a vanishing gradient problem is the \textit{gated recurrent unit} (GRU)~\cite{bff0e6bd8f4a4f0d9735bf1728fb43ef}\index{GRU, gated recurrent unit}.
The GRU is similar to the LSTM with a few simplifications: the GRU merges the context vector and the hidden states and combines three gates into two.
As a result, it requires less computational resources while retaining a similar level of performance for long sequences.
The GRU operations are defined as follows:
\begin{equation}
    \label{ML:eq:gru}
    h_t = z_t\bigodot h_{t-1} + (1-z_t)\tilde{h}_t
    \hspace{20pt}
    \text{where}
    \hspace{20pt}
    \begin{aligned}
        r_t &= \sigma\left(W^rx_t + V^rh_{t-1} + b^r\right)\\
        z_t &= \sigma\left(W^zx_t + V^zh_{t-1} + b^z\right)\\
        \tilde{h}_t &= \tanh\left(W^hx_t + V^h\left(r_t\bigodot h_{t-1}\right) + b^h\right)
    \end{aligned}
\end{equation}
where $r_t$ and $z_t$ are referred to as {\it reset} and {\it update} gate.
As one can see in Figure~\ref{ML:fig:rnn}, the reset gate in GRU performs the same task as the forget gate in LSTM by removing or reducing the elements of its memory (\ie the hidden state).
The update gate $z_t$ determines the relative proportion of the previous hidden state $h_{t-1}$ and the new context $\tilde{h}_t$ to be mixed in producing the new hidden state.


In addition to sequential data, the LSTM and GRU units can be used for data that has a tree-like structure.
In this setting, the networks are often referred to as recursive neural networks or TreeRNN and they have found applications in natural language processing and jet physics~\cite{socher2011parsing,socher2011semi,chen2015sentence,Louppe:2017ipp,Cheng:2017rdo}. 

\subsubsection{Attention and transformers}
\label{ML:sec:attention}



The idea behind \textit{attention}\index{attention} is to form a representation for the input, but different parts of the input are weighted differently according to the task at hand.
By making the weights learnable, the network can learn to attend to the relevant parts of the input.
For the $i$th task, one can form a task-specific context $c_i$ by computing the weighted average of the hidden state representations $h_j$ for each component of the input.
A softmax\index{softmax} function is used to produce attention scores $\alpha_{ij}$ for the $j$th input and $i$th task because it assigns a positive value to each component of the input and sums to one, $\sum_j \alpha_{ij}=1$.
Putting these ingredients together, we have the \textit{additive attention mechanism}
\begin{equation}
    \label{ML:eq:attention}
    c_i = \sum_{j=1}^n\alpha_{ij}h_j 
    \hspace{20pt}\text{where}\hspace{20pt}
    \alpha_{ij}=\operatornamewithlimits{softmax}_{j}(\beta_{ij})\; ,
\end{equation}
where $\operatornamewithlimits{softmax}_{j}$ indicates that normalizing sum runs over the index $j$ and the logits\index{logits} $\beta_{ij}$ can be computed from a neural network.
In the case of a cell of an RNN encoder-decoder network (see Fig.~\ref{ML:fig:rnn_variants}) that is decoding element $i$ with an incoming input state $s_{i-1}$, the logits for the attention mechanism can be computed as
\begin{equation}\label{ML:eq:attention_logits}
\beta_{ij} = U\tanh\left(W s_{i-1} + \widetilde{W} h_j + b_i\right) \; ,
\end{equation}
where $U$, $W$, and $\widetilde{W}$ are the weights and $b$ is the bias term of the model.
Figure~\ref{ML:fig:attention} from Ref.~\cite{olah2016attention} illustrates the full attention mechanism. 
%
This idea was first implemented by a model called \textit{RNNSearch}\index{RNNSearch} that made a breakthrough in machine translation by combining a bidirectional RNN with an additive attention mechanism~\cite{38ed090f8de94fb3b0b46b86f9133623}.
\begin{pdgxfigure}[width=0.6\textwidth]
    \centering
    \includegraphics{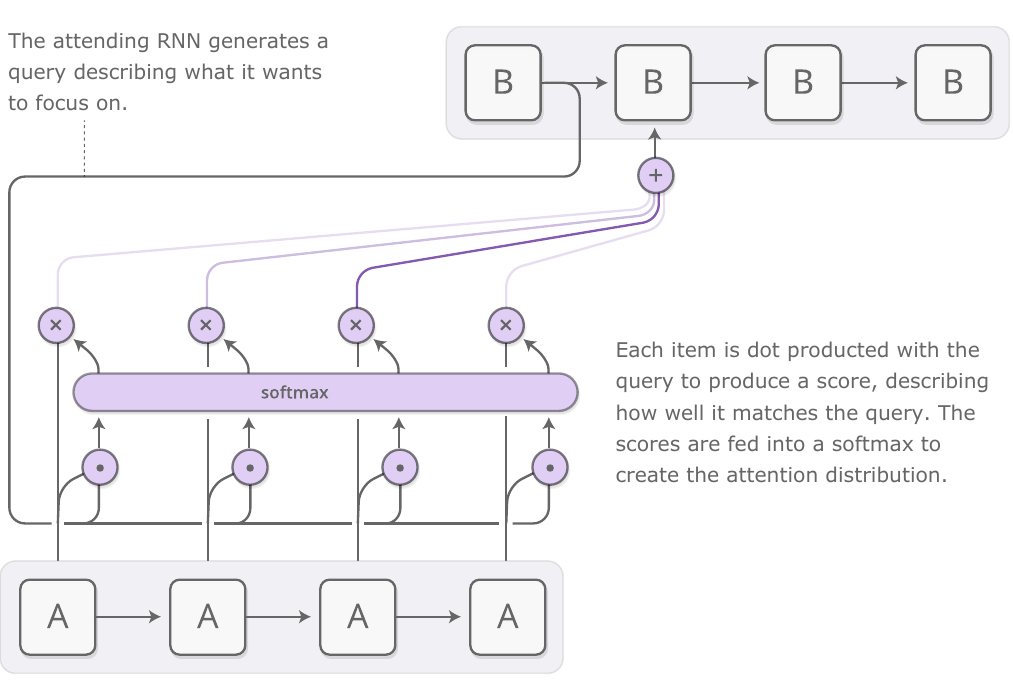}
    \caption{An illustration of the attention mechanism from Olah and Carter, ``Attention and Augmented Recurrent Neural Networks.''
    Lower boxes labeled \texttt{A} represent input elements in the sequence and upper boxes labeled \texttt{B} indicate output elements. The left-most line originating from the first \texttt{B} corresponds to the state $s_{i-1}$ in the text.}
    \label{ML:fig:attention}
\end{pdgxfigure}

The values $\alpha_{ij}$ can be used to visualize the influence of the $j$th input element on the $i$th output element, which improves interpretability\index{interpretability} of the model~\cite{olah2016attention} as shown in Fig.~\ref{ML:fig:attention_interp}.



In additive attention (Eq.~\ref{ML:eq:attention}), the hidden representations $h_j$, also called \textit{values}, are combined through a weighted average based on the coefficients $\alpha_{ij}$, resulting in a task-specific context vector $c_i$.
These values are often arranged in a matrix labeled $V\in\mathbb{R}^{m\times d_v}$, where the $m$ rows of the matrix correspond to individual hidden state vectors of length $d_v$.
The $\alpha_{ij}$ can also be represented as a $n \times m$ matrix $\alpha$ resulting from applying the softmax function to the $n \times m$ matrix $\beta$, normalized independently for each row.
With this notation, Eq.~\ref{ML:eq:attention} could be rewritten as $c = \operatorname{softmax}(\beta) V$, where the softmax is normalized per row. 

One powerful and widely used variant of the attention mechanism is \textit{scaled dot-product attention}\index{scaled dot-product attention}.
In scaled dot-product attention, instead of using an neural network to compute the logits $\beta$ as in Eq.~\ref{ML:eq:attention_logits}, the logits are computed by forming a dot product between an incoming \textit{query} and \textit{key}.
The set of $n$ query vectors can be arranged into the matrix $Q\in\mathbb{R}^{n\times d}$ and the set of $d$ key vectors can be arranged into the matrix (transpose) $K^\intercal\in\mathbb{R}^{d\times m}$.
One can interpret the keys as trying to detect certain types of queries and routing the attention to the relevant value.
Typically, the dot-product is scaled by a factor of $1/\sqrt{d}$.
The resulting task-dependent context is $c_i = \operatornamewithlimits{softmax}_{j}(q_i \cdot k_j / \sqrt{d}) v_j$.
A common, though confusing, notation is simply
\begin{equation}
    \label{ML:eq:sdp_attention}
    c = \operatorname{Attention}\left(Q,K,V\right) = \operatorname{softmax}\left(\frac{QK^\intercal}{\sqrt{d}}\right)V \;,
\end{equation}
where $c$ is a $n\times d_v$ matrix organizing the $n$ context vectors of length $d_v$ that are tailored summaries of the input vector for each of the $n$ tasks.

The \textit{transformer}\index{transformer} architecture is a powerful encoder-decoder model based on the scaled-dot product attention mechanism.
It was originally designed for sequential data and subsequently used in other areas of research including computer vision. 
One advantage of scaled-dot product attention is that computing the attention weights does not involve any sequential processing.
This allows the models to better leverage the parallelism of the hardware to train more expressive models faster than before.
In place of the gated units of an RNN that are key to avoiding the vanishing gradient problem, the transformer architecture employs a residual connections at every attention module (\ie, the input tensor is added to the output as in Fig.~\ref{ML:fig:skip_connections})\index{skip connection}. 


The second major ingredient in the transformer architecture is {\it multi-head attention}\index{multi-head attention}.
A multi-head attention module executes multiple scaled dot-product attention modules in parallel.
The query $Q$, key $K$, and value $V$ matrices in each scaled dot-product attention module are obtained by applying linear transformations (with learnable weights) to the common $Q$, $K$, and $V$ input matrices.
Each of them can be considered a different (albeit related) \textit{perspective} from which to derive attention. 

For a sequence-to-sequence mapping task, the output of encoder\index{encoder} is used to derive key $K$ and value $V$ matrices for the multi-head attention module in the decoder.
The decoder\index{decoder} is then responsible for mapping between the key-value features derived from the input (the encoder) and the queries from the decoder (which is still executed sequentially) in order to produce the final decoded output.

Finally, we note that the transformer architecture does not just employ an attention mechanism in the decoder.
By employing attention in the encoder as well the model has more capacity to ``interpret'' the input---a concept referred to as \textit{self-attention}\index{self-attention}. 
Transformer models have contributed to breakthroughs in many areas of scientific and industrial research~\cite{bommasani2021opportunities}. 
While transformers are very powerful, they also require a larger number of training samples due to the weaker inductive bias than other models.

\begin{pdgxfigure}
    \centering
    \includegraphics[width=\textwidth]{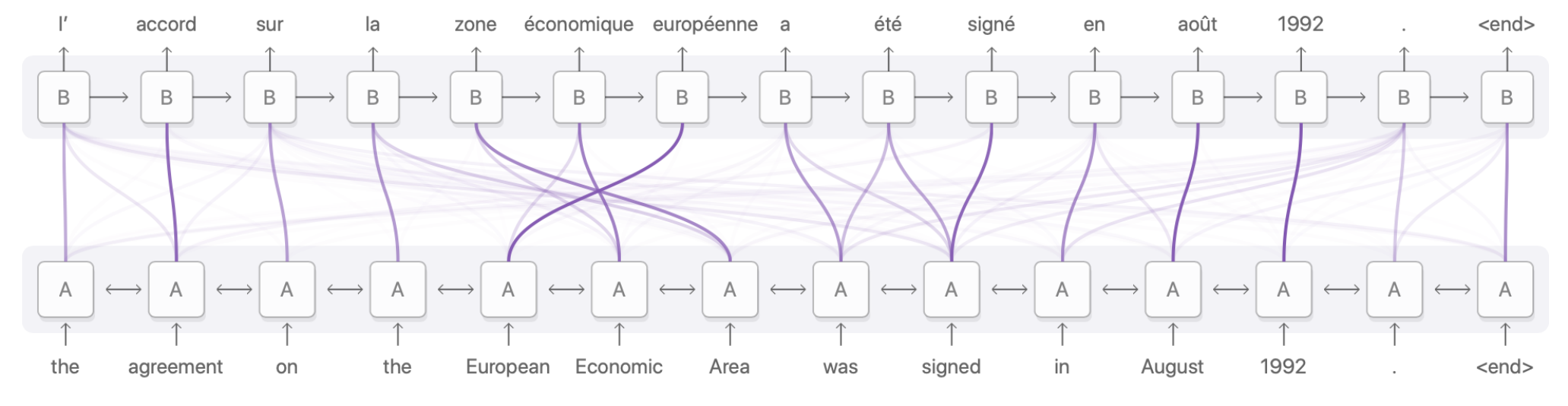}
    \caption{Visualization of the attention weights in a sequence-to-sequence problem from Olah and Carter, ``Attention and Augmented Recurrent Neural Networks.''
    The thickness of the lines is proportional to the attention weights $\alpha_{ij}$.}
    \label{ML:fig:attention_interp}
\end{pdgxfigure}

\subsubsection{Graph networks and geometric deep learning}\label{ML:sec:gnn} 

Graphs are a powerful archetype for representing structure data.
A graph consists of {\it nodes}\index{node} as elements and {\it edges}\index{edge} between between them.
Graphs are sufficiently flexible to describe many types of structured data including images and sequences.
Graph-based neural networks\index{GNN, graph neural network} can also be seen as a generalization of many common types of machine learning models such as recurrent and convolutional neural networks~\cite{bronstein2021geometric}.
The term \textit{geometric deep learning}\index{geometric deep learning} refers to this recent formulation that focuses largely on the symmetries of the data. 


An earlier attempt to organize the variations on different flavors of graph-based neural networks can be found in Ref.~\cite{47094}.
In their formalism, a graph network may be represented as $G(\mathbf{u},V,E)$ where $\mathbf{u}$ represents an array of global features, $V=\{\mathbf{v}_i\}_{i=1:N^v}$ represents a set of $N^v$ nodes with $\mathbf{v}_i$ as features for the $i$th node (e.g. such as RGB channels if a node represents a pixel in image data), and $E=\{(\mathbf{e}_k,r_k,s_k)\}_{k=1:N^e}$ represents a set of $N^e$ edges with $\mathbf{e}_k$ as features for the $k$th edge.
An edge may be (bi)directional where $r_k$ and $s_k$ denotes the destination and origin nodes respectively.
The features of a graph may evolve with three {\it update} functions $\phi$ and three {\it aggregate} functions $\rho$:
\begin{equation}
    \label{ML:eq:gn_functions}
    \begin{aligned}
        \mathbf{e}_k'&=\phi^e(\mathbf{e}_k,\mathbf{v}_{r_k},\mathbf{v}_{s_k},\mathbf{u})\\
        \mathbf{v}_i'&=\phi^v(\bar{\mathbf{e}}_i,\mathbf{v}_i,\mathbf{u})\\
        \mathbf{u}'&=\phi^u(\bar{\mathbf{e}}',\bar{\mathbf{v}}',\mathbf{u})\\
    \end{aligned}
    \hspace{20pt}
    \begin{aligned}
        \mathbf{e}_i'&=\rho^{e\rightarrow v}(E_i') & \hspace{5pt}\text{where}\hspace{10pt} & E'_i=\{(\mathbf{e}'_k,r_k,s_k)\}_{r_k=i,k=1:N^e}\\
        \bar{\mathbf{e}}'&=\rho^{e\rightarrow u}(E') & \hspace{5pt}\text{where}\hspace{10pt} & E'=\cup_iE_i'=\{(\mathbf{e}_k',r_k,s_k)\}_{k=1:N^e} \\
        \bar{\mathbf{v}}'&=\rho^{v\rightarrow u}(V') & \hspace{5pt}\text{where}\hspace{10pt} & V'=\{\mathbf{v}_i'\}_{i=1:N^v} \\
    \end{aligned}
\end{equation}
where $\mathbf{e}'$, $\mathbf{v}'$, and $\mathbf{u}'$ denote the updated node, edge, and graph features. In Graph Networks, three types of information are updated in the following order. The first step is $\phi^e$ to update every edge. The second step updates every node: for $i$th node, compute $\rho^{e\rightarrow v}$ to aggregate updated attributes from the edges with $r_k=i$ then compute $\phi^v$ to update the node attributes. 
The third step updates the graph attributes through $\phi^u$ which takes the original state $\mathbf{u}$, aggregated node and edge attributes by $\rho^{v\rightarrow u}$ and $\rho^{e\rightarrow u}$ respectively.

Graph neural networks~\cite{1555942} (GNNs)\index{GNN, graph neural network} are the class of neural networks that work on graph-structured data.
A related data format in computer vision and physics is the {\it point cloud}, which is an unordered set of points (\ie, a graph with no edges). Operations on point cloud need to be permutation invariant\index{permutation invariance} (\eg $\min$, $\max$, $+$, $\cdot$), and analysis of 3-dimensional physical object represented by point cloud need to be rotation and translation invariant as in the case for an image. PointNet~\cite{Qi_2017_CVPR,Qi2017PointNetDH}\index{Point Net}, a GNN that performs an object classification on point cloud of 3-dimensional positions, treats each point as a node, applies MLPs as $\phi^v$ to update node features, and global max-pooling operation as $\rho^{v\rightarrow u}$.
There is no explicit edge definition in PointNet (though the model applies affine transformation to all points using spatial transformer network~\cite{NIPS2015_33ceb07b}, which could be considered as a separate graph operation, to introduce rotation and translation invariance and to capture topological features).
Deep sets~\cite{NIPS2017_f22e4747}\index{deep set} follow the same manner except $\phi^v$ takes the global entities $\mathbf{u}$.
This is same for PointNet when performing point cloud segmentation: $\phi^v$ takes a step of simply concatenating $\mathbf{u}$ to node entities to combine a local and global features.
Dynamic graph CNN~\cite{10.1145.3326362}\index{dynamic graph CNN} is a variant that (re)define edges dynamically using attention mechanism: $\rho^{e\rightarrow v}$ aggregates $k$ neighbor nodes where the inter-node distance is defined as a Cartesian distance in the feature space. $\phi^v$ remains a MLP\index{MLP, multilayer perceptron} and, while edges are defined, there is no associated entity. A similar technique is used in nonlocal neural network~\cite{8578911} to efficiently propagate local feature information to points that may be far in the 3D cartesian coordinate.
Message-passing neural network~\cite{pmlr-v70-gilmer17a} (MPNN)\index{MPNN, message-passing neural network} explicitly defines a feature vector as edge entities.
In MPNN, $\rho^{e\rightarrow v}$ performs element-wise sum of features and feed into $\phi^e$, explicitly passing features across nodes as the name suggests.
While these are representative models that are frequently used in particle physics applications~\cite{IceCube:2018gms,Komiske:2018cqr,Qasim_2019,Moreno:2019neq,Moreno:2019neq, Qu:2019gqs,DeepLearnPhysics:2020hut,Alonso-Monsalve:2020nde,Ju:2020xty,NEURIPS2020_fb4ab556,Hewes:2021heg}, it is only a tiny fraction of GNN models developed over the past decade. 

Graph-based models are particularly interesting for science applications because they offer a natural way to organize the entities in the data and encode how those components interact each other.
This particular type of inductive bias is referred to as \textit{relational inductive bias}\index{inductive bias} in Ref.~\cite{1555942}.
Graph edges may be intrinsically defined in the data (\eg, when representing a social network) or not (\eg, a point cloud).
In the latter case, the graph structure must be chosen.
A naive approach may be defining a fully-connected graph.
However, for applications on hundreds of thousands of nodes (\eg, high resolution 3D point cloud), this may require a prohibitive amount of memory and computation.
On the other hand, if the graph is too sparse, it may negatively impact the performance.
One may need to compare the model performance among differently constructed graphs and balance against computational burden.
Ideally, the graph would be based on some knowledge of the interactions, but in the absence of such knowledge, popular graph construction methods include fully-connected, $k$-nearest neighbors, a Delaunay graph, minimum spanning tree, and locality-sensitive hashing~\cite{Pata:2023rhh}. 

Classification\index{classification} and regression\index{regression} tasks for graphs can be formulated such that the prediction is made for the entire graph or its individual nodes or edges. Graph-level prediction is like classifying an entire image, while node-level prediction is like semantic segmentation where individual pixels are classified. 
For clustering\index{clustering} of points, GNNs can approximate a transformation function for nodes into the latent space where an optimal clustering of points can be performed.
For instance, Ref.~\cite{Kieseler:2020wcq} proposes the \textit{object condensation} approach to extract particle information from a graph of detector measurements as well as grouping of the measurements.
The model predicts the properties of a smaller number of particles than there are measurements, in essence reducing the graph without explicit assumptions on the number of targeted particles.
Certain nodes are chosen to be the ``condensation'' point of a particle, to which the target properties are attached.
A special loss function mimics attractive and repulsive electromagnetic potentials to ensure nodes belonging to the same particle are close in the latent space.
Alternatively, one can formulate clustering as an edge classification task~\cite{Farrell:2018cjr,DeepLearnPhysics:2020hut,ExaTrkX:2021abe,Dezoort:2021kfk}. 
See Ref.~\cite{Shlomi_2021} for a comprehensive review on particle physics applications.

%
\subsection{Model design with physics inductive bias}
Designing neural networks that respect the structure of particle physics can materially improve sample efficiency, robustness to systematics, and physical interpretability.
Rather than relying on generic inductive biases (\eg, translation equivariance in image CNNs), particle-physics–informed models hard-wire symmetries, conservation laws, and kinematics into the architecture or loss.
This reduces the hypothesis space to functions that are a priori plausible, which is particularly valuable when training data is scarce, for extrapolation outside the training distribution, and for tasks where trust and uncertainty quantification matter as much as raw accuracy.

Exploitation of symmetry groups and hence {\it equivariance} is arguably one of the most important forms of physics-informed approaches.
At the constituent level, events and jets are sets of particles, so permutation symmetry is fundamental. Deep sets and graph neural networks implement this by aggregating over particles with symmetric operations~\cite{Komiske:2018cqr}. Beyond permutation symmetry, Lorentz symmetry is the natural arena for high energy physics experiments. 
Networks can be built from Lorentz scalars and tensors or by representing features as four-vectors and ensuring intermediate outputs transform equivariantly under boosts and rotations~\cite{pmlr-v119-bogatskiy20a,Gong:2022lye,Hao:2022zns,Bogatskiy:2022czk,Bogatskiy:2023nnw,Brehmer:2024yqw,Spinner:2024hjm}.
Gauge symmetry offers another avenue where symmetry-aware design pays off. In lattice gauge theory, gauge-equivariant networks ensures locality and exact invariance under gauge transformations~\cite{2019arXiv190204615C,Boyda:2020hsi,10.5555/3495724.3495890,PhysRevD.103.074504,Batzner_2022}.

While these models explicitly integrate laws of physics into mathematical operations and model architecture designs, it is also possible to implicitly enforce physics constraints through a loss definition and model optimization method.
For instance, physics-informed neural networks (PINNs) introduce regularization terms that force predicted physics quantities to follow laws of physics in the form of partial differentiable equations (\eg, acceleration as the time derivative of velocity)~\cite{RAISSI2019686,krishnapriyan2021characterizing}. 
Variants of PINNs incorporate differentiable physics models as a part of a model architectures~\cite{Newbury2024ARO}.
Finally, it is also possible to introduce physics constraints in an optimization process.
For example, for a machine learning model for data reconstruction, an output of the model may go through a forward physics simulator that infers the original input to the reconstruction model.
By minimizing the difference between the original input and the inferred one, the reconstruction model is forced to learn a solution that is consistent with the forward physics model~\cite{NEURIPS2020_a878dbeb}.

\section{Learning algorithms}
\label{ML:sec:learning_algorithms}
\subsection{Gradient-based optimization}\label{ML:sec:grad_opt}

Given a parameterized model $f(x,\theta)$ and a loss function $\mathcal{L}(x,\theta)$, where $x$ and $\theta$ denotes data and model parameters, one way to optimize $\theta$ is to first apply an appropriate initialization\index{initialization}, $\theta_{t=0}$ (e.g. Sec.~\ref{ML:sec:nn_init} for neural networks\index{neural network}), and perform an iterative update:
\begin{equation}
    \label{ML:eq:gd}
    \theta_{t} = \theta_{t-1}-\lambda\nabla_\theta \mathcal{L}(x,\theta), 
\end{equation}
where $\lambda$ is a small, real valued hyperparameter called {\it learning rate}\index{learning rate}. To see how this works, define $\delta\theta\equiv\theta_t -\theta_{t-1}$ and consider $\delta(\nabla_\theta \mathcal{L}(x,\theta))$:
\begin{equation}
    \delta\left(\nabla_\theta \mathcal{L}(x,\theta)\right) \approx \delta\theta\cdot\nabla_\theta \mathcal{L}(x,\theta) = -\lambda|\nabla_\theta \mathcal{L}(x,\theta)|^2
\end{equation}
which would monotonically decrease
the loss function, and locally 
move the parameter values in the 
desired direction of loss
function minimization.
This algorithm is called {\it gradient descent} (GD)\index{gradient descent}. We note that $\lambda$ needs to be sufficiently small for the approximation to hold. When $\lambda$ is too large, this can be a cause of a gradient explosion discussed in Sec.~\ref{ML:sec:nn_init}.

\subsection{Stochastic gradient descent}\label{ML:sec:sgd}

{\it Stochastic gradient descent} (SGD)\index{SGD, stochastic gradient descent} follows GD but replaces the exact gradient term $\nabla_\theta 
\mathcal{L}(x,\theta)$ with a stochastic approximation, where 
we subsample the data in the loss function using 
$N$ samples, where $N<n$, 
\begin{equation}
    \nabla_\theta \mathbb{E}_{\hat{p}(x)} \mathcal{L} \approx \displaystyle \frac{1}{N}\sum_{i}^N\nabla_\theta\mathcal{L}_i,
\end{equation}
where $\mathcal{L}_i$ is the loss function for
data sample $i$. 
It should be noted that $N$ needs to be randomly and independently sampled for the approximation to hold. Implementation of SGD follows three steps: take new samples of size $N$, approximate the gradient, then update the parameters $\theta$. 

In the case of optimizing the loss using a static database (\ie one cannot take new $N$ samples for every update), {\it mini-batch learning} is often employed. This replaces the first step with a randomly sampled {\it batch}\index{batch} of data, which is a subset of all the samples in the database. In this case, however, since a batch of data used for each parameter update is not entirely independent, a model may overfit\index{overfitting}. In practice, a part of the whole dataset is reserved as a {\it validation} sample, and the model performance is carefully monitored during the optimization process to avoid overfitting via an early stopping\index{early stopping} criterion (see Sec.~\ref{ML:sec:early_stopping} and Fig.~\ref{ML:fig:earlystop}).

SGD with slowly decreasing learning rate can be shown to converge to a local minimum 
almost surely under mild conditions, and to a 
global minimum for unimodal loss functions. SGD may 
also prevent getting stuck in 
shallow local minima of the loss function, thereby 
reaching a better local minimum for multi-modal 
loss functions. The noise in SGD with a 
constant learning rate can be 
viewed as a form of Langevin dynamics, which under proper conditions on the learning 
rate and mini-batch size 
converges to the stationary posterior 
distribution of the weights~\cite{MandtHB17}. Thus SGD at a constant 
learning rate can be viewed as a sampler
bouncing around and exploring the 
posterior surface for better solutions, 
descending onto the best found solution 
as the learning rate is decreased, a process
related to temperature annealing in 
global optimization.

Another advantage of SGD is simply the computational 
cost: rather than evaluating the loss over all 
the data samples at each update, we use a small 
subset of data instead at each update. Furthermore, 
mini-batching can take 
advantage of vectorization libraries and GPU 
architectures. Large batch training requires
specialized methods of training, such as 
layer-wise adaptive rate scaling (LARS)~\cite{you2017largebatchtrainingconvolutional}. 


\subsection{Optimization algorithms}\label{ML:sec:opt}

GD and SGD are the basic building blocks for
more advanced optimization algorithms. 
One can improve the convergence rate of gradient 
based optimization 
by considering the learning rate $\lambda$ to depend on 
individual $\theta_i$. Second order algorithms such 
as Newton's method
take into account second order derivatives (Hessian)\index{Hessian} to find the minimum, and give an exact solution 
in a single update when the loss is quadratic 
around the peak. However, 
this requires a matrix inversion of the Hessian, 
which is exceedingly expensive in ML applications, 
where the number of network parameters is very large. As a consequence, second order optimization 
is rarely used in ML. 

There are several improvements to the 
basic SGD even in the absence of 
Hessian information. Momentum based optimization 
takes a physics perspective of a viscous fluid
in an external potential, where one updates 
current velocity with the potential 
gradient (force), followed by an update in 
position based on velocity. This approach
therefore uses previous gradients in addition to the current one to 
compute a running average of the gradient, 
with a forgetting factor that controls how far 
back the averaging goes. This 
helps move faster towards the minimum in ravines, where 
gradient descent is usually inefficient due to 
the high condition number of the Hessian. 


Beyond SGD with momentum, modern optimizers adapt step sizes across parameters or layers using recent gradient statistics and decouple regularization from the update. RMSprop tracks a running RMS of gradients and scales steps inversely to damp oscillations. Adam adds momentum (first moment) and variance (second moment) estimates to provide per-parameter adaptive updates. AdamW~\cite{loshchilov2019decoupledweightdecayregularization} improves Adam~\cite{kingma2017adammethodstochasticoptimization} by decoupling weight decay from the adaptive step, applying true L2 regularization directly to weights, which typically yields better generalization and has become a common default in vision and language models. For very large-batch regimes, layer-wise trust-ratio methods such as LARS (Layer-wise Adaptive Rate Scaling)~\cite{you2017largebatchtrainingconvolutional} and LAMB (Layer-wise Adaptive Moments optimizer for Batch training)~\cite{you2020largebatchoptimizationdeep} scale each layer’s effective learning rate by the ratio of parameter norm to gradient norm, enabling stable training with batch sizes of tens of thousands. More recently, Lion~\cite{chen2023symbolicdiscoveryoptimizationalgorithms} uses the sign of the momentum update instead of a second-moment estimate, reducing memory and often improving speed and generalization; it can also be combined with decoupled weight decay. In practice, these optimizers are paired with warmup, cosine or exponential decay schedules, and sometimes gradient clipping; the best choice depends on model size, data regime, and whether you prioritize fast convergence, large-batch scalability, or final generalization.

\subsection{Automatic differentiation and backpropagation}\label{ML:sec:autodiff}

In practice, $f(x,\theta)$ might take a complex form and may include a large set of parameters. The term $\nabla_\theta \mathcal{L} = \nabla_\theta \mathcal{L} (f(x,\theta))$ requires computing partial derivatives with respect to individual parameter $\theta_i$. If $f$ is a composite model (i.e. $f=f_n(f_{n-1}(\cdots,\theta_{n-1}),\theta_n)$), and if all of $f_{i:1,n}$ are differentiable, a chain rule can be applied:
\begin{equation}
    \label{ML:eq:chainrule}
    \nabla_{\theta_i}\mathcal{L}  = \frac{\partial \mathcal{L} (f(x,\theta))}{\partial \theta_i} = \frac{\partial \mathcal{L} }{\partial x_n}\cdot\frac{\partial x_n}{\partial x_{n-1}}\cdots\frac{\partial x_i}{\partial\theta_i}
\end{equation}
where $x_n$ denotes the output of $n$th composite function $f_n$. In order to compute $\nabla_{\theta_i}\mathcal{L} $ for $f_i$, it needs computation of a gradient at all preceding (or subsequent if seen in the forward context) functions. As the gradients accumulate across differentiable functions in the reverse order of the model composition, this technique is called {\it backpropagation}~\cite{Rumelhart1986}\index{backpropagation}. An example of $f$ that satisfies conditions to apply backpropagation is a neural network, which consists of repeating blocks of a (differentiable) activation function and an affine transformation. 

 When the model $f(x,\theta)$ is implemented as a computer program in practice, {\it automatic differentiation} (AD)\index{AD, automatic differentiation}, also called {\it algorithmic differentiation}, is used to compute the derivatives. AD exploits the fact that any computer program consists of a sequence of elementary arithmetic operations (\ie, addition, subtraction, multiplication, and division) and functions (\eg, $\log$, $\exp$, $\sin$, and $\cos$) and apply chain rules to compute the target derivative. AD has advantages over traditional approaches including symbolic and numerical differentiation. The symbolic differentiation faces a serious difficulty of converting a program into a single expression, and the numerical differentiation  suffers from round-off errors. Finally, both methods scale poorly in speed of computation for calculating partial derivatives with a large number of inputs. AD delivers much faster speed and does not suffer from increasing errors for calculating higher derivatives.

There are two modes of AD: the {\it forward} and {\it backward} mode.  Consider a composite function $f(x,\theta)=f_n(f_{n-1}(\cdots f_1(x,\theta_1)\cdots),\theta_{n-1}),\theta_n)$. The forward mode applies the chain rule in the same order of the forward evaluation of $f$ by computing $\partial f_1/\partial x$ first, then $\partial f_2/\partial f_1$, and continue to $\partial f_n/\partial f_{n-1}$. The backward mode traverses the reverse direction: starting from the last (outer-most) function $\partial f_n/\partial f_{n-1}$, next $\partial f_{n-1}/\partial f_{n-2}$, and continue to $\partial f_1/\partial x$. Therefore, the backpropagation of gradients can be implemented using the backward AD, in which the target variable to be differentiated is fixed and the derivative is computed with respect to each sub-expression recursively as shown in Eq.~\ref{ML:eq:chainrule}.  The forward mode is simpler to implement as the order of gradient calculation follows the order of composite functions to be executed. The reverse mode typically requires less amount of computation than the forward mode, 
but more memory is required to store intermediate function output values to calculate derivatives efficiently. Another consideration is the mapping of dimensionality $f:\mathbb{R}^k\rightarrow\mathbb{R}^\ell$ as it concerns the number of variables to sweep from each end. The forward mode is efficient when $k\ll\ell$ while the reverse mode takes an advantage if $\ell\ll k$. For instance, in the case of an image classification where $(k,\ell)=(\text{pixel count},1)$, the reverse AD is more efficient.

Development of a differentiable physics simulator is an active area of research and AD-enabled programming frameworks are at the core of those research work. AD-enabled simulator can be used to solve an inverse problem of inferring the physics model parameters (e.g. calibration) or the input (i.e. reconstruction)~\cite{Gasiorowski_2024,Heinrich_2023}. A fully differentiable physics simulator often requires, however, a custom algorithm to approximate gradients to handle cases where gradient calculation is not straightforward (e.g. due to stochastic processes)~\cite{AEHLE2025109491}. Beyond AD, a specifically designed neural network that ensures accurate gradient calculation is also frequently used, sometimes in combination with AD-enabled framework~\cite{NEURIPS2020_a878dbeb,lei2022implicitneuralrepresentationdifferentiable}.

\subsection{The vanishing and exploding gradient problems}\label{ML:sec:vanishing_grad}
Gradient based optimization crucially depends on the size of gradient with respect to each model parameter. If the magnitude of gradient is too large with respect to the distance to an optimal parameter value, it may repeatedly overshoot the target and cause an oscillation preventing convergence. If the gradient is too small, it may take an impractically long time to converge. As shown in Eq.~\ref{ML:eq:chainrule}, the gradient of $i$th function $f_i$ is a product of gradients from the subsequent functions. If those gradients are too large or too small, the magnitude can can either increase or decrease exponentially in the number of layers. These are called {\it exploding} and {\it vanishing} gradient problem respectively\index{vanishing gradient}\index{exploding gradient}. 

Modern deep neural networks\index{neural network} consist with many composite functions (\ie, layers) and are particularly prone to this effect. Let us consider a simple RNN. From Eq.~\ref{ML:eq:rnn_simple}, we can write the backpropagating gradient:
\begin{equation}
    \frac{\partial h_t}{\partial h_{t-1}} = \text{diagonal}\left(f'\left(Wx_t+Vh_{t-1}+b1\right)\right)W
\end{equation}
where $f'$ denotes the derivative of an activation function. The gradient of the contribution to the loss $\mathcal{L}_i$ from the $i$th element in the sequence with respect to the $j$th hidden state $h_j$ is therefore:
\begin{equation}
    \label{ML:eq:vanishing_gradient}
    \frac{\partial\mathcal{L}_i}{\partial h_j}=\frac{\partial\mathcal{L}_i}{\partial h_i}V^{i-j}\prod_{j<t\leq i}\text{diagonal}\left(f'\left(Wx_t+Vh_{t-1}+b1\right)\right)
\end{equation}
where we can see that $V$ contributes multiplicatively with ${i-j}$ powers when $i-j>1$. This example is explored in depth for recurrent models~\cite{bengio1994learning,pmlr-v28-pascanu13} but is common for all types of deep neural networks. 

In practice, one may explicitly inspect the magnitude of gradients propagating across layers to ensure an effective optimization. One way to mitigate an exploding gradient is to set the maximum gradient value $\delta_{\text{max}}$ as a model hyperparameter and {\it clip} any larger gradients $\delta$ where it appears in the backpropagation:
\begin{eqnarray}
    \label{ML:eq:grad_clip}
    \delta = \frac{\delta_{\text{max}}}{\|\delta\|}\delta &\text{if}& \|\delta\|>\delta_{\text{max}}.
\end{eqnarray}
This is called {\it gradient clipping}~\cite{pmlr-v28-pascanu13}\index{gradient clipping}. 

Alternatively, there are many architecture designs that are motivated by the vanishing and exploding gradient problem or which aim to help  propagate gradients across many layers. These considerations drove the design of gated models like the LSTM and GRU for sequential data and also motivated the ReLU\index{ReLU, rectified linear unit} non-linearity. Other example architectural designs or components motivated by these considerations include identity mapping and skip connections\index{skip connection} used in ResNet, U-Net, and DenseNet, which allow gradients to flow across many layers.  

Other factors contributing to vanishing and exploding gradient include initialization\index{initialization} of model parameters and normalization of input data. These factors contribute in keeping the magnitude of activation, which also concerns the magnitude of gradient, within a reasonable range. A recommended practice for a gradient-based optimization of a neural network is to maintain the input values centered around zero and a similar level of covariance across the inputs (and the outputs that are the inputs to the next layer)~\cite{8c8eccbbe8a040118afa8f8423da1fe2}. These factors are discussed in the following.




\subsection{Early stopping}
\label{ML:sec:early_stopping}

\begin{pdgxfigure}
    \centering
    \includegraphics[width=0.5\textwidth]{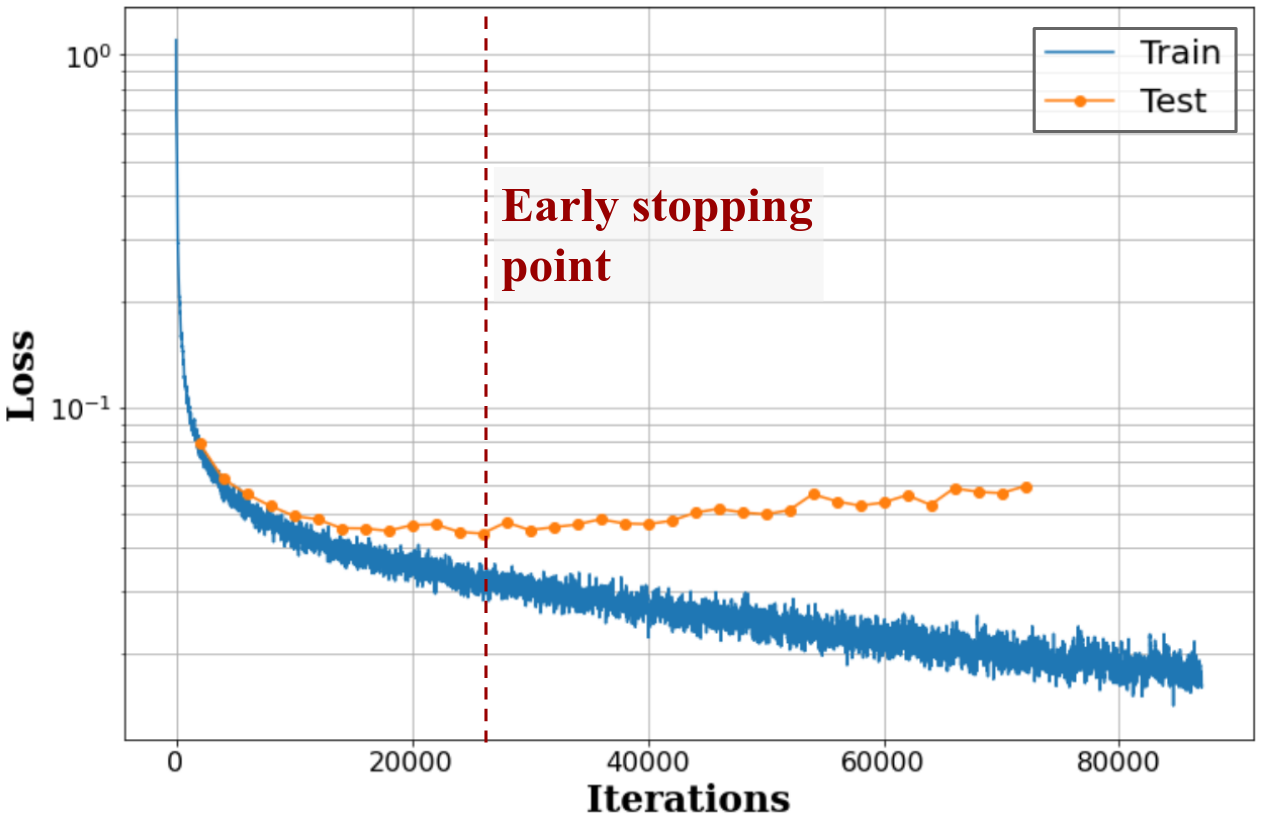}
    \caption{An example instance of overfitting. The training loss (vertical axis) shown in blue decreasing over iterations (horizontal axis) while the loss values evaluated on test samples shown in orange start to increase at around 26,000 iterations as indicated by the vertical line. }
    \label{ML:fig:earlystop}
\end{pdgxfigure}

Early stopping\index{early stopping} is a form of regularization\index{regularization} used to avoid overfitting\index{overfitting} when an iterative method, such as gradient descent, is used as a learning algorithm. Imagine a plot of the training loss and test loss as a function of iterations (\ie parameter updates). As learning proceeds, the training loss will generally decrease. However, the test loss will often decrease initially and then start to increase, which is the classic sign of overfitting as shown in Fig.~\ref{ML:fig:earlystop}. The basic idea of early stopping is simply to stop training before overfitting takes place. In some approaches to early stopping theoretical analysis of the learning problem provides a prescription for when to stop the training~\cite{yao2007early}; however, the most straight forward approaches use a held-out validation dataset to monitor the generalization performance~\cite{prechelt1998early}.

\subsection{Initialization of model parameters}
\label{ML:sec:nn_init}


 An improper initialization\index{initialization} can slow down the optimization process or even result in a loss of convergence.  While $b^{(l)}$ is typically initialized to zero, $W^{(l)}$ values need to be stochastic to avoid identical updates during optimization. One way is to sample $W^{(l)}$ from a zero-centered Gaussian distribution with a small variance (e.g. 0.01)~\cite{Krizhevsky2012ImageNetCW}. However, this method does not guarantee the same variance in the input to each layer, which depends on the size of the input layer, and makes it difficult to train a deep neural network~\cite{Simonyan2015VeryDC}. The \textit{Glorot} or {\it Xavier} initialization\index{Xavier initialization} takes this into account and sets the variance of a Gaussian distribution to be $\sigma^2=1/d^{(l-1)}$ assuming a symmetric activation function around zero, such as a logistic function or hyperbolic tangent~\cite{pmlr-v9-glorot10a}. The {\it He} initialization\index{He initialization} uses the variance $\sigma^2=0.5/d^{(l-1)}$, and is a simple extension of Xavier initialization for leaky, parametric, and standard ReLU activation~\cite{He2015DelvingD}.

\subsection{Input normalization}\label{ML:sec:input_norm}
\index{input normalization}
Input data to a neural network is often pre-processed for the same goals discussed previously: values are shifted to have the mean of zero and scaled to keep a similar covariance across features. Furthermore, a data may be transformed using techniques including PCA and whitening (sphering) to keep input features independent and uncorrelated from each other~\cite{8c8eccbbe8a040118afa8f8423da1fe2}. 

\subsection{Batch normalization}\label{ML:sec:batch_norm}
Even with careful normalization of the input data and initialization of model parameters, the mean and covariance of the data representations in hidden layers will evolve during training and may pose challenges for learning for downstream layers. This is called an {\it internal covariate shift}~\cite{DBLP:journals/corr/IoffeS15} and may cause negative effects to an optimization process. Accordingly, techniques to explicitly normalize features in between hidden layers are often employed for a deep neural network. One of them is {\it batch normalization (BN)}\index{BN, batch normalization}, which shifts and scales the input to a hidden layer:
\begin{equation}
    \label{ML:eq:batchnorm}
    \tilde{u}^{(l)} = \gamma\frac{u^{(l)}-\mu_B}{\sqrt{\sigma^2_B+\epsilon}}+\beta
\end{equation}
where $u^{(l)}$ and $\tilde{u}^{(l)}$ refer to the raw and normalized input to the $l$th layer, $\mu_B$ and $\sigma_B$ represent the mean and mean-squared-error of $u^{(l)}$ calculated using a {\it batch}\index{batch} of input data used to update the network parameters. $\gamma$ and $\beta$ are part of model parameters that are updated during the optimization. After optimization is complete, these parameters are fixed for model evaluation during production. $\epsilon$ is a small, fixed constant value to ensure numerical stability.  While it is popular (especially in computer vision), a downside of BN is its dependency on the batch size. In situations where the batch size is limited to be a small number (\eg, memory limitation for a large data or a model), the performance using BN could degrade since $\beta$ and $\gamma$ values may not be generalized for the dataset during training.

There are several variants to batch normalization with considerations on how to group a subset of values in $u^{l}$. For instance, an image naturally has three groupings: a set of pixels across spatial axis, features within one pixel (\ie, image {\it channels})\index{channels}, alongside with a grouping across multiple images (\ie, batch). Different groupings have been studied and found and some are found effective to particular type of applications: layer normalization groups values along the channel and spatial dimensions~\cite{ba2016layer}, instance normalization\index{instance normalization} groups along the spatial dimension but not along the batch nor channel~\cite{ulyanov2017instance}, and group normalization\index{group normalization} is similar to layer normalization\index{layer normalization} but forms multiple sub-groups of channels~\cite{wu2018group}. These variants do not apply normalization across samples within a batch, and thus are agnostic to the batch size.


\subsection{Transfer learning: pre-training and fine-tuning}

{\it Transfer learning}\index{transfer learning} is a technique to improve performance and accelerate optimization process by reusing a pre-trained machine learning model for a new task.
The two tasks and corresponding datasets may differ, but fundamental features, such as implicitly learned symmetries in the underlying data, may be reusable across tasks and datasets.
Transfer learning typically takes two steps: the first is to alter the model or data if necessary, then continue updating some or all of the model parameters on the new data or task, {\it fine-tuning}\index{fine-tuning} the model.
The first step is required, for example, when solving a different task that requires a different architecture (\eg, regression vs. classification), or when input data format requires a change (\eg, original model trained on three channel image, such as RGB images, while new data has a single channel).
Transfer learning has been widely practiced in the field of computer vision where large, labeled data sets are available~\cite{10.1007/978-3-319-10602-1_48,ILSVRC15,Cordts2016Cityscapes,Yi16}: a CNN\index{CNN, convolutional neural network} trained for classifying images of an animal can be largely reused for object detection, or even for analyzing image data in science (\eg, particle trajectories recorded by an imaging detector).
It is a critical aspect for the development of general AI as well as interdisciplinary sharing of models across research fields. 

While transformers were initially introduced for machine translation, later models such as GPT~\cite{gpt,gpt3} and BERT~\cite{bert} showed that these models can be generalized to multiple NLP tasks through transfer learning by \textit{pre-training}\index{pre-training} for several seemingly different tasks, including sentence classification, semantic similarity, question answering, and commonsense reasoning~\cite{gpt}.
These models are collectively referred to as large language models (LLMs) and some have been fine-tuned for physics domains~\cite{10.1063/5.0238090,PhysRevAccelBeams.28.044601}.

\subsection{Foundation models}
\label{ML:sec:foundation_models}
A key to successful transfer learning is an effective pre-training process through which general (and thus re-usable) representations are learned by a model.
When a large, comprehensive dataset within a certain domain is combined with an effective representation learning method, typically using self-supervision or multi-task supervision, which forces a model to learn foundational concepts, the resulting model may potentially be generalized for any task defined in the data domain.
Such AI models are referred to as \textit{foundation models} (FMs)\index{foundation models}, which are a central theme of general AI research today~\cite{tan2023promiseschallengesmultimodalfoundation}.
The LLMs such as GPTs are the first and the most successful FMs to date.
The core of LLM pre-training is based on self-supervised learning (see Sec.~\ref{ML:sec:ssl}), where, \eg, a model is tasked to predict the next or missing word in a sentence.
To solve this task, a model must learn not only grammar or parts of speech but also a concept of visual colors and the probability distribution over possible colors an apple can take.
Using the vast online corpus as a training data, LLMs are trained to learn foundational representations of the world interpreted and generated by humans in the form of texts.

The critical properties of FMs include the \textit{emergence}, \textit{homogenization}, and \textit{scalability}~\cite{tan2023promiseschallengesmultimodalfoundation}.
Emergence means the system behavior is implicitly induced rather than explicitly constructed (\eg, a model's capability to generalize through self-supervised pre-training).
Homogenization implies a single model or a system that can perform multiple tasks.
Scalability is improvement in model performance when increasing the computing resources, number of model parameters, and size of training dataset.
These properties are also used to evaluate the quality of FMs.
Following the initial success of LLMs, R\&D of FMs has expanded to computer vision and audio data domains. 
\begin{pdgxfigure}[place=h]
    \centering
    \includegraphics[width=0.7\linewidth]{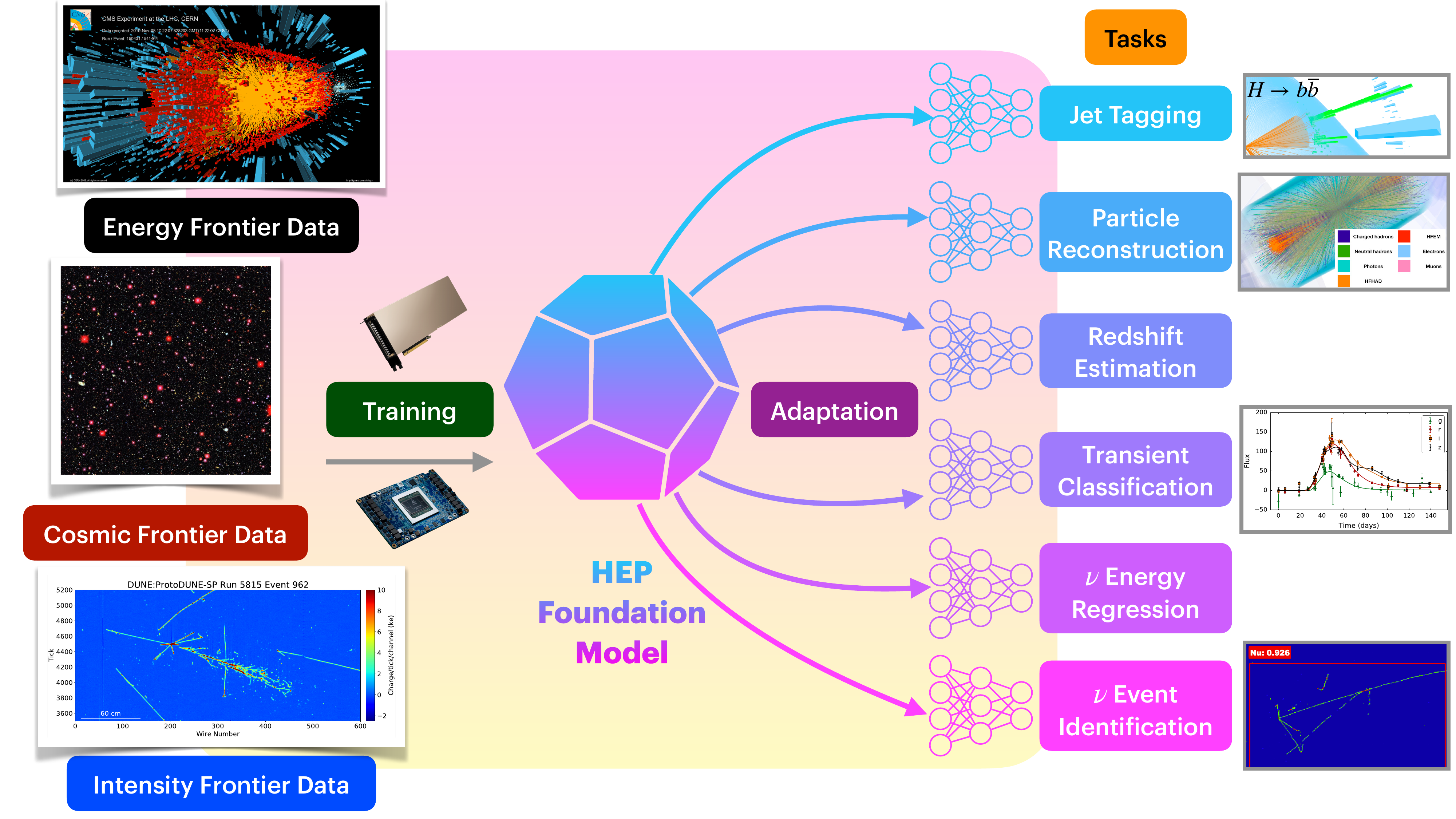}
    \caption{Schematic of a foundation model based on HEP data.
    Data is used to pre-train a large model, which can then be fine-tuned (or adapted) to various downstream tasks.}
    \label{fig:hepfm}
\end{pdgxfigure}

Many modern FMs combine multiple data modalities (\eg, image generation from text input, audio generation from text and image input)~\cite{DBLP:journals/corr/abs-2102-12092,clip,https://doi.org/10.48550/arxiv.2204.11824}.
Multi-modal FMs are trained to correlate features from different data modalities either during a pre-training or fine-tuning stage.
For example, the CLIP model achieves this by minimizing the distance between extracted features from an image and its corresponding caption (text) data~\cite{clip}.
It should be noted, however, pre-training FMs on sensory data (\eg, 1D waveforms, 2D images, or 3D scenes) is more difficult compared to symbolic data (\eg, language, math, or high-level physics data) as discussed in Sec.~\ref{ML:sec:ssl}.

HEP datasets present unique R\&D opportunities to advance understanding of FMs.
Particles in high energy collisions follow well-established physics models and can be learned by FMs similar to words in natural language~\cite{mpmv1,mpmv2}.
Large public datasets of galaxies recorded in multiple modalities (\eg, images and spectroscopy data) enable a contrastive learning approach based on CLIP~\cite{astroclip}. 
Similarly, a high-fidelity simulator in HEP can be used to formulate contrastive learning objectives across different scenarios in a stochastic process~\cite{resimulation}.
Finally, HEP detectors offer big, high-precision data sets with challenging tasks to extract complicated physics information~\cite{young2025particletrajectoryrepresentationlearning}.
A schematic of a foundation model based on HEP data is shown in Fig.~\ref{fig:hepfm}.

\section{Incorporating uncertainty}\label{ML:sec:uncert}



A fundamental aspect of data analysis is the quantification of uncertainty. This broad topic includes the traditional distinction between statistical and systematic uncertainty, procedures for propagation of errors, and the incorporation of uncertainty in to the statistical models (\eg with nuisance parameters) that are used in Bayesian or frequentist statistical procedures (see Sec. \crossref{stat}). Accounting for systematic uncertainty can be seen as a requirement, but ideally systematic uncertainties are also taken into account in the design of the analysis so as to mitigate their effect.  The introduction of machine learning into the analysis pipeline requires revisiting the techniques used for uncertainty quantification\index{uncertainty quantification} and exposes many fundamental issues that have nothing to do with the use of machine learning per se. See Ref.~\cite{Dorigo:2020ldg} for a recent review on this topic. 

 In machine learning research and industrial settings, the mismatch between the data distribution $p_\textrm{train}(x,y)$ used for training and the data distribution $p_\textrm{prod}(x,y)$ that the model will be applied to in production is referred to as \textit{covariate shift} or \textit{domain shift}\index{domain shift}. For example, one might train a classifier to identify cats and dogs with images from a well lit studio and then apply the classifier on images taken in doors with poor lighting conditions and a scratched lens. Not surprisingly, the mis-classification rate of the classifier will be different between the two settings. 

Physicists are keenly aware that the simulations that we use to describe the data are not perfect, and this mismodeling corresponds to a large fraction of the of systematic uncertainties accounted for in published works. Since simulated data is often used to train machine learning models (\ie $p_\textrm{train}(x,y)$), it is important to understand and account for how this mismodeling will influence results when applied to real data (\ie $p_\textrm{prod}(x,y)$). 

One of the primary approaches to incorporating this type of uncertainty is to introduce nuisance parameters $\nu$ corresponding to the  uncertain inputs to the simulation.
One then parametrizes various types of  perturbations (\eg, corrections to efficiencies or energy scales) in the hopes that the resulting family of distributions $p(x|y,\nu)$ is flexible enough to encompass the true data distribution for class $y$. In this approach one does not have just two ``domains'' for the data (\ie, $p_\textrm{train}$ and $p_\textrm{prod}$), but a continuous family of domains parameterized by the nuisance parameters $\nu$.

With this framing in mind, there are several approaches to incorporating uncertainty into an analysis that includes ML-based components:
\begin{itemize}
    \item \textbf{propagation of errors:}\index{error propagation} one works with a model $f(x)$ and simply characterizes how uncertainty in the data distribution propagate through the function to the down-stream task irrespective of how it was trained.
    \item \textbf{domain adaptation:}\index{domain adaptation}  one incorporates knowledge of the distribution for domains (or the parameterized family of distributions $p(x|y, \nu)$ ) into the training procedure so that the performance of $f(x)$ for the down-stream task is robust or insensitive to the uncertainty in $\nu$.
    \item \textbf{parameterized models:} instead of learning a single function of the data $f(x)$, one learns a family of functions $f(x; \nu)$ that is explicitly parameterized in terms of nuisance parameters and then accounts for the dependence on the nuisance parameters in the down-stream task. 
    \item \textbf{data augmentation:}\index{data augmentation} one trains a model $f(x)$ in the usual way using training dataset from multiple domains by sampling from some distribution over $\nu$. 
\end{itemize}

In this setting it is best to consider the trained model $f(x)$ or $f(x; \nu)$ to be a fixed function and decouple the variability associated to training or the choice of architecture. The fact that one could have chosen a different architecture or learning algorithm should be treated in the same way as other choices that are made in the data analysis pipeline. While it is reasonable to want downstream inference and decisions to be robust to these choices, they are of a different nature than the uncertainty in the modeling of the data distribution. We return to this point in Sections~\ref{ML:sec:aleatoric} and \ref{ML:sec:bayesian_nns}.

\subsection{Propagation of errors}\label{ML:sec:propagation_of_errors}

In this Section, we consider the common scenario in which one has used some machine learning technique to train a model $f(x)$ for classification or regression and wants to assess the sensitivity of the output of $f(x)$ to uncertainty in the input $x$. We regard the function $f(x)$ as fixed and we are not concerned with how the model was trained.

Propagating uncertainty through a ML-based model $f(x)$ is not fundamentally different than for any other function, and one can use the standard propagation of errors formula of Sec. \crossref{prob:sec:errprop}. As always, it is important to recognize the limitations of the propagation of errors formula, which is accurate when the uncertainty on $x$ is Gaussian and the function $f(x)$ is approximately linear within the region set by the uncertainty on $x$. 

Similarly, classifiers are often used for particle identification or event selection. In that case, one is primarily interested in the efficiency $\epsilon$ to satisfy a cut on the classifier output. The efficiency depends on the distribution $p(x|y)$  through the equation $\epsilon_y = P(f(x) > f_\textrm{cut} | y) = \int H(f(x) - f_\textrm{cut}) p(x|y) dx$, where $H$ is the Heaviside step function and $y$ is an index or label for the category of data that is being considered (\eg, signal vs. background or electron vs. jet). Thus, the question in this context is what is the uncertainty on the efficiency $\epsilon_y$ due to uncertainty in the distribution $p(x|y)$. In practice, the quantification of the uncertainty in the efficiency $\epsilon_y$ is usually based on either a calibration measurement on real data or estimated with simulated data. These procedures typically treat the classifier as a black-box, and thus nothing precludes  using those procedures on a ML-based classifier. An early example of this approach for b-tagging can be found in Ref.~\cite{ALEPH:1997umh}.

In the case where simulation is used to estimate the efficiency $\epsilon_y$ and its uncertainty, one usually varies nuisance parameters $\nu$ associated to the simulation. One then uses simulated samples to approximate $\epsilon_y(\nu) = P(f(x) > f_\textrm{cut} | y,\nu) = \int H(f(x) - f_\textrm{cut}) p(x|y,\nu) dx$. Again, the procedure for incorporating uncertainty isn't fundamentally different if the classifier $f(x)$ is based on machine learning or a hand-crafted observable. 


\subsection{Domain adaptation}\label{ML:sec:domain_adaptation}
\index{domain adaptation}
While estimating the uncertainty for a ML-based model is not fundamentally different than any other hand-crafted observable used for regression or classification, the worry of many physicists is that by working with a high-dimensional set of features $x$ that one is more susceptible to mismodeling of subtle correlations. This is a valid concern, and it should be appreciated that a great deal of prior knowledge and physical insight goes into the construction of hand-crafted observables so that they will be robust to the most uncertain aspects of data. However, much of  this craft is based on heuristics that are difficult to systematize. Furthermore, one can only validate that such an observable is robust if one can explicitly evaluate the performance for a perturbed distribution. Thus in the settings where one can validate the robustness to a perturbed scenario $\nu_0$, one must have access to $p(x|y,\nu_0)$.  

One approach to formalize this type of robustness is to consider the dependence on the distribution of the output of the model $f(x)$ to the nuisance parameters. In statistics, if the distribution of $f$ is independent of the nuisance parameters, then $f$ is referred to as a \textit{pivotal quantity}. This is a property that we can incorporate directly into the training procedure to target a particular notion of robustness. The authors of Ref.~\cite{Louppe:2016ylz} introduced an adversarial approach (similar to what is used in the generative adversarial network\index{GAN, generative adversarial network} of Sec.~\ref{ML:sec:gan}) to penalize a model during training if the distribution of the output varies with the nuisance parameters. To construct the training dataset $\{x_i, y_i, \nu_i\}_{i=1,\dots,n}$, one must sample $y$ and $\nu$ according to some proposal distribution (similar to a prior, but only used for the creation of training dataset, not necessarily for statistical inference\index{inference}), corresponding to a joint distribution $p(x,y,\nu)$. Instead of minimizing the target loss $\mathcal{L}_f$ (\eg cross-entropy or squared-error) with respect to the parameters $\phi_f$ that parameterize the model $f$, one trains with a minimax strategy that also includes an adversary $q$ with parameters ${\phi_r}$. The trained model is characterized by the saddle point
\begin{equation}\label{ML:eq:pivot}
    \hat\phi_f, \hat\phi_r = \arg\min_{\phi_f}\arg\max_{\phi_r} E_\lambda(\phi_f, \phi_r) \;,
\end{equation}
where the value function $E_\lambda$ includes the target loss as well as a regularization term associated to the adversary
\begin{equation}\label{ML:eq:pivot_E}
    E_\lambda(\phi_f, \phi_r) = \mathcal{L}_f(\phi_f) - \lambda \mathcal{L}_r(\phi_f, \phi_r) \;.
\end{equation}
The constant $\lambda$ is a hyperparameter, since generally there is a tradeoff between the two terms and only in special cases can the model that minimizes $\mathcal{L}_f$ also be a pivotal quantity. The regularization term 
\begin{equation}
  \mathcal{L}_r(\phi_f, \phi_r) =  \mathbb{E}_{p(x,y,\nu)}[ -\log q_{\phi_r}(\nu|f_{\phi_f}(x))]
\end{equation}
is an example of conditional density estimation\index{density estimation} (see Sec.~\ref{ML:sec:density_estimation}), where the model $q_{\phi_r}(\nu | f)$ is trying to predict the distribution of the nuisance parameter $\nu$ given the output of the model $f(x)$. This term is maximized when $f$ is independent of $\nu$. Earlier work had also used an adversarial technique for domain adaptation, but was limited to just two domains~\cite{edwards2015censoring, ganin2015unsupervised, 2014arXiv1412.4446A}, while here $\nu$ parametrizes a continuous family of distributions and can have multiple components corresponding to different sources of uncertainty. Furthermore, the previous work aimed to make the distribution for a high-dimensional, intermediate representation of the data be invariant to the domain shift as opposed to just the final output $f(x)$.

One way of interpreting Eq.~\ref{ML:eq:pivot} is that the goal is to minimize a regularized loss function $\tilde{\mathcal{L}}(\phi_f) = \argmax_{\phi_r} E_\lambda(\phi_f, \phi_r)$, where the optimization with respect to $\phi_r$ is not exposed. This motivates another approach in which the regularization is not achieved through a learned adversary, but by a measure of discrepancy between distributions that can be computed directly from samples. In particular, the authors of Ref.~\cite{Kasieczka:2020yyl} proposed the use of \textit{distance correlation} to avoid what can be a challenging min-max optimization problem.

In either case, the optimization of the hyperparameter $\lambda$ is based on the downstream task. For example, in Ref.~\cite{Louppe:2016ylz} considered the case where $f$ was a signal vs. background binary classifier where the nuisance parameter $\nu$ was associated to uncertainty in the background model. The hyperparameter $\lambda$ was then optimized to maximize the approximate median significance (AMS)\index{AMS, approximate median significance}. Similarly, the authors of Refs.~\cite{Shimmin:2017mfk} and \cite{Kasieczka:2020yyl} considered new physics searches in the context of boosted jet tagging, where the hyperparameter controls the sculpting of the side-bands used for background estimation.

While these strategies modify the training procedure so that the sensitivity to the nuisance parameters is reduced, it does not typically eliminate it. As a result, one still needs to propagate the uncertainty in the data distribution through the learned model as described in the preceding section. 
Furthermore, care must be taken in interpreting the loss of sensitivity to the nuisance parameter.
For example, for theoretical uncertainties estimated as the difference between two different calculations (\ie, two-point uncertainties), decorrelation methods may reduce the apparent uncertainty while the true uncertainty remains much larger~\cite{Ghosh:2021hrh}.

Note, this adversarial technique has also been employed in other settings where one would like to decorrelate the output of the classifier with an observed quantity so that it can be used for background estimation~\cite{Shimmin:2017mfk}, although other techniques like moment decomposition~\cite{Kitouni:2020xgb} may suffice without full decorrelation.
Widely used alternative approaches to decorrelation include uboost~\cite{Stevens:2013dya}, DDT~\cite{Dolen:2016kst}, and using dedicated training samples that vary the chosen quantity to be decorrelated~\cite{CMS-DP-2020-002}. Other examples of the domain adaptation and decorrelation use cases from the Living Review include Refs.~\cite{Louppe:2016ylz,Dolen:2016kst,Moult:2017okx,Stevens:2013dya,Shimmin:2017mfk,Bradshaw:2019ipy,ATL-PHYS-PUB-2018-014,Kasieczka:2020yyl,Xia:2018kgd,Englert:2018cfo,Wunsch:2019qbo,Rogozhnikov:2014zea,10.1088/2632-2153/ab9023,clavijo2020adversarial,Kasieczka:2020pil,Kitouni:2020xgb,Ghosh:2021hrh}.




\subsection{Parameterized models}\label{ML:sec:parameterized_models}

An alternative to learning a model $f(x)$ that is pivotal---\ie, whose distribution is independent of the nuisance parameter $\nu$---is to learn a family of models $f(x; \nu)$ that is parameterized in terms of the nuisance parameters.
In general, there is a tradeoff between the two terms of Eq.~\ref{ML:eq:pivot_E} for a single model $f(x)$.
In a parameterized model, $f(x; \nu)$ optimizes the performance of the model for every value of $\nu$.
Parameterized classifiers were first advocated in Ref.~\cite{Cranmer:2015bka} in the context of simulation-based inference (see Sec.~\ref{ML:sec:prob}) and in Ref.~\cite{Baldi:2016fzo} for new physics searches.
It has also been applied to simulation-based inference for effective field theory parameters in Ref.~\cite{Brehmer:2018eca} and Ref.~\cite{Ghosh:2021roe} provides additional pedagogical examples.

The training of a parameterized model is similar to the standard procedure. For example, if one originally wanted to minimize the squared loss function $\mathcal{L}(y,f(x)) = (y-f(x))^2$ with training dataset $\{x_i, y_i\}_{i=1,\dots,n}$, then the corresponding training procedure for the parameterized model would be as follows.
One would need to construct a training set $\{x_i, y_i, \nu_i\}_{i=1,\dots,n}$ as described in the preceding section, construct a parameterized model $f(x; \nu)$ that takes as input the original feature vector $x$ as well as the nuisance parameters $\nu$, and then train using the same loss  $\mathcal{L}(y,f(x;\nu)) = (y-f(x; \nu))^2$.

One complication of the parameterized approach is that it is no longer possible to evaluate the model on a dataset $\{x_i\}$ and pass on only $\{f_i\}$ for downstream analysis tasks since $f(x_i; \nu)$ still depends on $\nu$. Instead, one delay evaluating the model to some down-stream stage when the dependence on $\nu$ would accounted for.
For example, in the context of a likelihood based analysis where one is testing a hypothesis where the nuisance parameters take on a particular value $\nu_\textrm{test}$, then one will consider the data distribution $p(x|\nu_\textrm{test})$, and at that point one would evaluate the model at the corresponding nuisance parameter value, \ie $f(x; \nu_\textrm{test})$. Explicit examples are given in Refs.~\cite{Cranmer:2015bka, Brehmer:2018eca, Ghosh:2021roe,Dorigo:2020ldg}. While this may seem complicated, it actually corresponds to what is done in a typical likelihood-based fit when the statistical model has nuisance parameters; \ie the likelihood-ratio corresponds to the model $f(x; \nu)$ as in Eq.~\ref{ML:eq:lrt}. 

\subsection{Data augmentation}
\index{data augmentation}
An intuitive approach to building in robustness to systematic effects that can lead to domain shift, is simply to augment the training dataset so that it includes examples corresponding to several values of the nuisance parameter or systematic variations. As before one can construct a dataset $\{x_i, y_i, \nu_i\}_{i=1,\dots,n}$, but instead of leveraging the information about $\nu_i$, one simply discards this information. This corresponds to sampling from the marginal distribution $x_i, y_i \sim p(x,y) = \int d\nu p(x,y|\nu)p(\nu)$, and is often referred to as \textit{smearing}. One can then use this smeared dataset for supervised learning in the traditional way. While it is possible that this approach will lead to improved robustness to systematic variations (\ie generalization for $\nu$ other than the nominal value) than if systematic uncertainty weren't considered at all), this intuitive approach has several shortcomings. The approach does not yield a pivotal quantity as in the adversarial approach, so propagation of uncertainty through the network is still required. Moreover, there is no direct way to control the tradeoff between independence from the nuisance parameter and the original target loss as in the adversarial approach. Finally, it can lead to significant performance loss  compared to what is possible with the parameterized approach. These tradeoffs were studied in Refs.~\cite{Baldi:2016fzo,Ghosh:2021roe} with both pedagogical and physically-motivated examples.

\subsection{Aleatoric and epistemic uncertainty}\label{ML:sec:aleatoric}

In the machine learning and risk assessment literature, uncertainty is often characterized in terms of \textit{aleatoric}\index{aleatoric uncertainty} and \textit{epistemic}\index{epistemic uncertainty} uncertainty~\cite{OBERKAMPF200411, OHAGAN2004239,DBLP:journals/corr/abs-1910-09457,NIPS2017_2650d608}. Familiarity with these terms is useful, but the distinction between the two can be ambiguous, the terms are not always consistently used, and they do not clearly map onto the concepts used physics. 

For example, Ref.~\cite{DBLP:journals/corr/abs-1910-09457}, states that ``roughly speaking, aleatoric (\aka, statistical) uncertainty refers to the notion of randomness, that is, the variability in the outcome of an experiment which is due to inherently random effects'', while ``epistemic (\aka, systematic) uncertainty refers to uncertainty caused by a lack of knowledge (about the best model).'' This seems clear enough, but in that same reference (and in Ref.~\cite{KIUREGHIAN2009105}) the aleatoric uncertainty is considered irreducible, while the epistemic uncertainty could be reduced with additional information.
This may seem backwards for many physicists since often in particle physics, we think of how uncertainties scale as we collect more data but keep the experimental design fixed. In that case, the statistical uncertainty will be reduced with time while the systematic uncertainty will remain constant%
 \footnote{Further complicating the relationship between the terms is that many experimental uncertainties that are characterized as systematic are actually statistical in nature as auxiliary measurements and control regions are used to constrain the corresponding nuisance parameters.}. 
There is no paradox here, it is simply a different point of view. The emphasis of the risk\index{risk} assessment community is not on collecting more data with the same experimental design, but collecting different types of data  that will inform the models themselves. Clearly even for physicists, data from new experiments or calibration measurements could also reduce our systematic uncertainties. 
While there are exceptions in the literature, the bulk of it associates aleatoric uncertainty with the randomness of classical probability (\ie, the statistical uncertainty associated to repeating the same experiment many times) and epistemic uncertainty with our state of knowledge. 


Perhaps a more important distinction between the perspective of physicists and machine learning researchers has to do with the use of the term ``model'' and what exactly is uncertain. In physics, the systematic and epistemic uncertainty is typically associated to our understanding of the underlying physics and ``the model'' usually refers to the physics model, detector model encapsulated in a simulation. In contrast, for machine learning research, ``the model'' usually refers to the trained model $\hat{f}\in \mathcal{F}$ used as described in Section~\ref{ML:sec:loss_risk_Remp} (or the class of functions $\mathcal{F}$ itself). This makes sense if we recall that in the bulk of machine learning research, one has little insight into the process that generated the data (\eg, images of cats and dogs, or natural language). In that sense, the epistemic uncertainty in machine learning is usually associated to uncertainty in the model parameters $\phi$ after training, which would be reduced if one could collect more training dataset (see Ref.~\cite{NIPS2017_2650d608} for this point of view). 

In the literature on uncertainty quantification (UQ), which is more closely connected to physics given the role of computer simulations, the terminology is more fine grained and less ambiguous. That community uses the terms parameter uncertainty (\ie nuisance parameters), structural uncertainty (\ie, mismodelling), algorithmic uncertainty (\ie numerical uncertainty), experimental uncertainty (\ie, uncertainty from experimental resolution and statistical fluctuations), and interpolation uncertainty (\ie, uncertainty due to interpolating between different parameter values due to lack of computational resources).

\subsection{Model averaging and Bayesian machine learning}\label{ML:sec:bayesian_nns}

The core of  
Bayesian machine learning\index{Bayesian machine learning} is the 
model averaging view. Here 
one often takes a more ambitious view of learning than described in Sec.~\ref{ML:sec:loss_risk_Remp}, which is framed mainly as function approximation. While in Sec.~\ref{ML:sec:loss_risk_Remp}, the goal is to find a function that minimizes the risk, in Bayesian machine learning one explicitly builds a probability model $q_\phi(x,y)$ for the training dataset $\mathcal{D} = \{x_i, y_i\}_{i=1,\dots,n}$. It is the same change in perspective that one has when one views the squared error loss\index{MSE, mean-squared error} function $\mathcal{L}_\text{MSE} = (y-f_\phi(x))^2$ as the log-likelihood for a probability model $y \sim N(f(x), \sigma)$. In addition, one assumes some prior on the model parameters $p(\phi)$, which is often a Gaussian distribution, and is analogous to Tikhonov regularization\index{regularization} (see Sec.~\ref{ML:sec:regularization}). In this way, a single trained model $\hat{f}=f_{\hat{\phi}}$ is the MAP\index{MAP, maximum a posteriori} point estimate and the more complete Bayesian solution is the entire posterior distribution over the model parameters $p(\phi | \mathcal{D})$. With this view, it is  clear how increasing the number of training examples $n$ will lead to a reduction in uncertainty on $\phi$. However, this notion of epistemic uncertainty has little to do with the notion of systematic uncertainty as the term is used by particle physicists.

Bayesian methods can be applied to 
non-probabilistic regression problems, 
in which case they can provide uncertainty 
quantification. 
Consider the case of regression in traditional (non-Bayesian) machine learning. The trained model $f_{\hat\phi}(x)$ is used to predict the target label $y$. For a fixed $x$, the model does not provide any notion of uncertainty on the prediction. One could propagate uncertainty on $x$ through $f(x)$, but that is also not the desired quantity to characterize the intrinsic spread $p(y|x)$ in the data, which may exist even if $x$ has negligible uncertainty. In contrast, Gaussian process regression (a Bayesian method) does provide a natural way to communicate the uncertainty on the prediction, which is possible because one first had to specify a prior on the mean and covariance of the Gaussian process. 

In the context of Bayesian deep learning and Bayesian 
neural networks, one would place a prior on the weights and biases of the neural network $p(\phi)$ and then use one of the many emerging techniques to calculate the approximate posterior $p(\phi | \mathcal{D})$. However, we should recognize that we have little-to-no insight into the parameters of a deep neural network, so the prior on $\phi$ is hardly well-justified. Furthermore, just as in all Bayesian approaches, the prior is not invariant to reparametrizing the model: $\phi \to \eta(\phi)$. While it is difficult to justify the choice of the prior on the parameters (and, thus, the resulting posterior), the resulting model may perform well empirically. In such high-dimensional parameter spaces, the bias-variance tradeoff can be dramatic.  

Bayesian model averaging (BMA)\index{BMA, Bayesian model averaging} performs 
Bayesian average over the posterior 
$p(\phi | \mathcal{D})$. This can 
be applied to any quantity $f_\phi$, such as a
regression or classification 
prediction $y$. Suppose we can draw from 
the posterior $\phi \sim p(\phi | \mathcal{D})$. 
For each draw we can evaluate the predicted regression variable $y=f_\phi(x)+\epsilon$, where 
$\epsilon$ is some noise to account for uncertainty  in the predictions. We can denote this process
as a draw from $p(y|x,\phi)$, $y \sim p(y|x,\phi)=N(f_\phi(x),\sigma_{\epsilon}^2)$, where 
$\sigma_{\epsilon}^2$ is the noise variance. The BMA then performs
\begin{equation}
    p(y|x,\mathcal{D})= \int d\phi p(\phi | \mathcal{D}) p(y|x,\phi).
    \label{BMA}
\end{equation}
In practice $p(y|x,\phi)$ is evaluated by drawing 
samples of $y$ and $\phi$, so the posterior 
is defined implicitly by the samples. For 
example, the mean prediction is obtained 
by averaging $f_\phi(x)$ over the samples 
of $\phi$, and the covariance matrix is similarly 
evaluated by averaging the second moments over the samples 
of $\phi$. 
Ref.~\cite{Yao_2018} provides a different perspective on BMA analyzed in what are referred to as the $\mathcal{M}$-open and $\mathcal{M}$-closed settings~\cite{Yao_2018}. The $\mathcal{M}$-closed 
setting refers to the situation where the true data generating process is in the space of models, even if it is unknown to us. In contrast, the $\mathcal{M}$-open setting refers to when the true data generating process is not in model space (\ie the model is mis-specified). Interestingly, in the $\mathcal{M}$-open case one can potentially do better than any one model in the model class by considering an average over the models, since averaging can create a new model that is not in the  model class. BMA provides one such averaging, but other averages, which are not weighted by $p(\phi | \mathcal{D})$, can be a better choice. When the weights of each model  are optimized against appropriate loss the resulting procedure is called stacking, which has been shown to be superior to BMA in the $\mathcal{M}$-open setting~\cite{Yao_2018}. Ref.~\cite{SnoekOFLNSDRN19} performed experiments indicating that in some cases model averaging can also improve predictive uncertainty estimates under domain shifts. 
 

Neural network\index{neural network} model averaging beyond BMA comes in several different 
flavors. Two successful model averaging 
procedures  are Monte Carlo dropout~\cite{GalG16}\index{Monte Carlo dropout}, which uses dropout ensembling, and deep ensembles \cite{Lakshminarayanan17}, which use random initialization ensembling. These methods may not  only be superior to BMA, they are also  often significantly faster than BMA. 
Whether these model averages are an approximation to BMA, or an alternative to it,  remains a debated topic, and both views have been advocated.  
BMA itself can be accelerated using approximate methods, such as stochastic Variational Inference\index{variational inference}  with reparametrization  trick \cite{KingmaSW15}. 

Finally, in the context of Bayesian model uncertainty estimation, there are two practical ways to capture: repulsive ensembles and evidential regression. Repulsive ensembles are standard deep ensembles trained with an extra diversity penalty so the ensemble members make deliberately different but plausible predictions, improving coverage with fewer models. Evidential regression uses one network to predict the parameters of a simple probabilistic family plus an “evidence” term; when data are ambiguous it learns low evidence and returns wider intervals, and when data are plentiful it narrows them. The latter has been studied for uncertainty quantification for neutrino applications~\cite{Koh_2023}. In practice, both approaches can yield similarly well‑calibrated uncertainties. An open direction is to bring such calibrated epistemic uncertainty into the density estimators of generative models (\eg, flows, VAEs, or diffusion), for example via ensemble/Bayesian variants with diversity or evidential parameterizations, so we can represent and propagate uncertainty over whole distributions, not just point predictions.

\subsection{Connection to probabilistic machine learning}\label{ML:sec:prob}

We end this Section by reinforcing the connection between uncertainty quantification\index{uncertainty quantification} in traditional machine learning and the more probabilistic approaches to machine learning exemplified by simulation-based inference\index{simulation-based inference} (see Sec~\ref{ML:sec:SBI}) and deep generative models\index{generative model} (see Sec.~\ref{ML:sec:deep_generative}). In the standard approach to supervised learning (\eg classification and regression) the model $f(x)$ provides a point estimate for $y$. Estimating an uncertainty on $y$ goes a step further, but the complete picture would be to model the posterior distribution $p(y|x)$. Gaussian processes (see Sec.~\ref{ML:sec:GP}) are an example, but the form of the models is limited to Gaussian posteriors. In Sec~\ref{ML:sec:SBI} we discussed approaches to model $p(y|x)$ using conditional density estimation~\cite{Cranmer:2016lzt,NIPS2016_6084,Cranmer:2019eaq}. If we extend this task to include a family of distributions parameterized by some nuisance parameters $\nu$, then the task is to model $p(y|x,\nu)$, which is structurally similar. 

In the context of classification, the output is already probabilistic, and  the interpretation of the resulting classifier is $\hat{f}_\text{MSE}(x) \approx p(y=1|x)$ (see Eq.~\ref{ML:eq:fstar_MSE_binary}). Incorporating the dependence on the nuisance parameter, then connects to the likelihood-ratio trick\index{likelihood-ratio trick} (see Eq.~\ref{ML:eq:lrt}), approaches to simulation-based inference that involve learning the likelihood-ratio, and the parameterized approaches described in Sec.~\ref{ML:sec:parameterized_models}. 

If one pairs the training procedure for classification\index{classification}, regression\index{regression}, or density estimation\index{density estimation} used in the approaches above with model averaging techniques such as BMA, then it would be possible to incorporate both  uncertainty associated to finite training dataset and the uncertainty associated to systematic uncertainties. However, as described in Sec.~\ref{ML:sec:aleatoric} and Sec.~\ref{ML:sec:propagation_of_errors}, it is not clear that in physics applications it is desirable to account for the variability associated to training when the more common practice is to regard the trained model $\hat f(x)$ as fixed. 

While these probabilistic approaches to machine learning are attractive conceptually, it is known in the machine learning community that classifiers often are poorly calibrated and often overly confident in their predictions. This is a problem even if one regards the trained model $\hat f(x)$ as fixed. Various approaches, including model averaging, are being pursued to improve the calibration of trained models, but the problem is unlikely to be eliminated entirely. Miscalibration can be verified by evaluating the true positive and false positive rates on held out data. This is common practice in experimental particle physics, where the output of a binary classifier is rarely taken at face value. Instead, the true and false positive rates are estimated with simulated data or control samples as described in Sec.~\ref{ML:sec:propagation_of_errors}. Furthermore, the true and false positives can be characterized as a function of the nuisance parameters. These procedures can be used to help calibrate parameterized models based on the likelihood-ratio trick~(see Refs.~\cite{Cranmer:2015bka,Ghosh:2021roe}). 
Unfortunately, calibration in the context of density estimation is more challenging. This connects to topics and challenges in anomaly detection\index{anomaly detection} (see Sec.~\ref{ML:sec:anomaly}).

\section{Model compression and deployment in experiments}
\label{ML:sec:deployment}

The software and computing needs of training a machine learning model are different than those encountered when it is deployed for use.
The two stages are referred to as \textit{training}\index{training} and \textit{inference}\index{inference}, \ie making a prediction $\hat f(x)$ given an input $x$ and a trained model $\hat f$.
Sometimes this transition also involves using different programming languages for implementing the trained model from the ones used for training them.
Modern machine learning frameworks support various serialization formats to exchange trained models.
For instance, \texttt{ONNX}~\cite{onnx} provides an open source format for many types of models, is widely supported, and can be found in many frameworks, tools, and hardware.
This is important when integrating a trained model into the software frameworks used by the large experiments.

While hardware acceleration with GPUs is important for efficiently training modern machine learning techniques, there are also advantages of hardware acceleration at inference time.
This may include GPUs or field programmable gate arrays (FPGAs), and the Living Review includes many example works focusing on efficient inference for a given hardware architecture~\cite{Strong:2020mge,Gligorov:2012qt,Weitekamp:DLPS2017,Nguyen:2018ugw,Bourgeois:2018nvk,1792136,Balazs:2021uhg,Rehm:2021zow,Mahesh:2021iph,Amrouche:2021tio,Pol:2021iqw,Goncharov:2021wvd}.
Programming FPGAs requires the use of dedicated hardware description languages (HDLs) such as VHDL or Verilog as well as a design methodology that is aware of the limitations and nature of the relevant device.
Recently, high-level synthesis (HLS) tools~\cite{vitis,quartus,catapult}, which ingest algorithms written in C/C++ code, have lowered the barrier to entry for using FPGAs.
Several tools, including hls4ml~\cite{Duarte:2018ite}, FINN~\cite{FINN,blott2018finnr}, Conifer~\cite{Summers:2020xiy}, and fwXmachina~\cite{Hong:2021snb}, have been developed to automatically create firmware from ML algorithms.
These tools have been used for applications ranging from jet tagging~\cite{Khoda:2022dwz,Odagiu:2024bkp,CMS-DP-2025-032} to muon transverse momentum regression~\cite{CMSP2L1T}, on-detector data compression~\cite{DiGuglielmo:2021ide}, charged particle tracking~\cite{Elabd:2021lgo,Huang:2023bny}, calorimeter reconstruction~\cite{Iiyama:2020wap}, and anomaly detection~\cite{Govorkova:2021utb,CMS-DP-2024-059,CMS-DP-2024-121}.

    \begin{pdgxfigure}[place=!t]
    \centering    
    \includegraphics[width=0.17\textwidth]{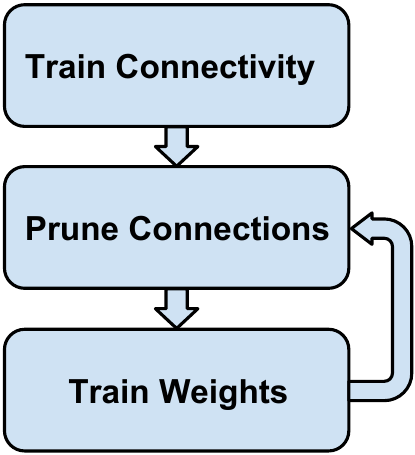}
    \hspace{0.05\textwidth}
    \includegraphics[width=0.4\textwidth]{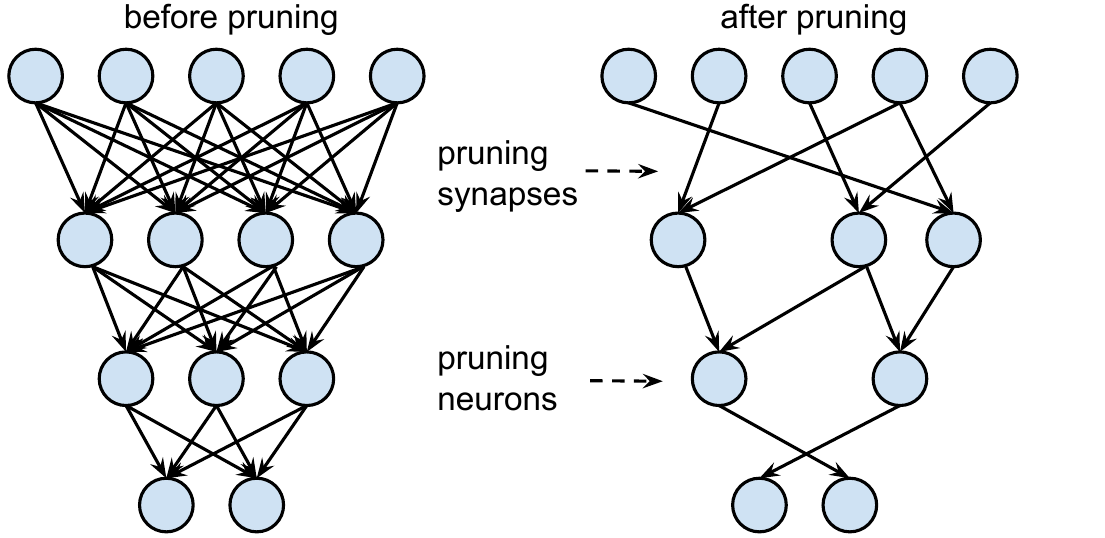}  
    \includegraphics[width=0.6\textwidth]{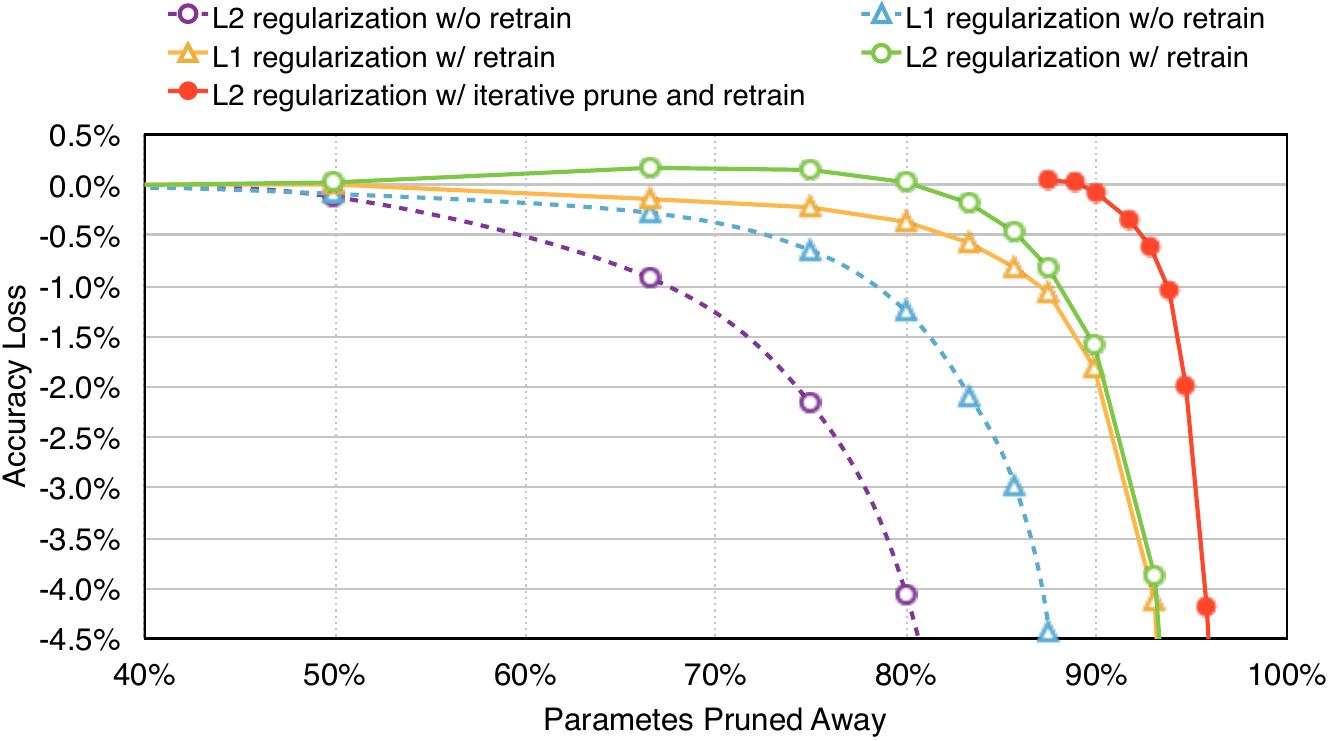}
    \caption{
        Illustration of the iterative magnitude-based parameter pruning and retraining with regularization procedure from Han et al. in NeurIPS, 2015.
        The top-5 accuracy loss is shown as a function of parameter reduction (sparsity) for VGG-16 on ImageNet following different pruning procedures.
        Without retraining, L1 regularization performs better than L2, but L2 performs better than L1 with retraining. 
        Iterative pruning gives the best result.
    }
    \label{ML:fig:pruning}
\end{pdgxfigure}

For applications where latency is a key concern (\eg, triggering at collider experiments), various accelerators have been investigated~\cite{Duarte:2018ite,DiGuglielmo:2020eqx,Summers:2020xiy,1808088,Iiyama:2020wap,Mohan:2020vvi,Carrazza:2020qwu,Rankin:2020usv,Heintz:2020soy,Rossi:2020sbh,Aarrestad:2021zos,Hawks:2021ruw,Teixeira:2021yhl,Hong:2021snb,DiGuglielmo:2021ide,Migliorini:2021fuj,Govorkova:2021utb}.
To enable the use of an ML model in resource-constrained or latency-sensitive experimental settings, reducing the size and computational complexity of the model through \textit{compression} is often essential.
Compression techniques aim to improve the computational efficiency of models, while keeping the performance as close as possible to the original.
The two most ubiquitous methods are \textit{quantization}~\cite{nagel2019datafree,han2016deep,meller2019same,zhao2019improving,banner2019posttraining,bertmoons,NIPS2015_5647,zhang2018lq, ternary-16,zhou2016dorefa,JMLR:v18:16-456,xnornet, micikevicius2017mixed,Zhuang_2018_CVPR, wang2018training,hawq,hawqv2}\index{quantization}, which modifies the number of bits used to calculate and store results in the model, and \textit{pruning}~\cite{optimalbraindamage,han2016deep,lotteryticket,learningraterewinding,supermask,stateofpruning}\index{pruning}, which removes model parameters.
However, symbolic regression\index{symbolic regression}~\cite{Tsoi:2023isc} and knowledge distillation\index{knowledge distillation}~\cite{Bal:2023bvt} have also been explored to learn compact algorithms.

While it is common to use 32-bit floating-point precision, for many applications, this may not be required to ensure adequate performance.
Reduced-precision formats, such as integer or fixed-point precision, may be used instead.
We can distinguish \textit{post-training quantization} (PTQ)\index{PTQ, post-training quantization}, in which model parameters are quantized after a traditional training is performed with 32-bit floating-point precision, and \textit{quantization-aware training} (QAT)\index{QAT, quantization-aware training}, in which the training procedure is modified to emulate reduced precision formats.
QAT results in better performance for a smaller bit width, but requires (re)training with a dedicated framework~\cite{Coelho:2020zfu,hawq,hawqv2,Sun:2024soe,brevitas}.
Serializing and exchanging quantized models is a challenge addressed by the \texttt{QONNX} format, which extends \texttt{ONNX} to represent arbitrary-precision quantized neural networks~\cite{Pappalardo:2022nxk}.

Pruning is the removal of unimportant weights, quantified in some way, from a neural network.
The two main categories are \textit{unstructured pruning}, where weights are removed without considering their location within a network, and \textit{structured pruning}, where weights connected to a particular node, channel, or layer are removed.
Pruning reduces the number of computations that must be performed to produce an inference result, thus reducing the hardware resources or algorithm latency.
The development of pruning algorithms and understanding their behavior is an active area of research~\cite{stateofpruning}.
One relatively simple method is iterative, magnitude-based pruning~\cite{learningeff,Duarte:2018ite}, as shown in Fig.~\ref{ML:fig:pruning}.
In this process, the model is trained with L1 or L2 regularization (discussed in Sec.~\ref{ML:sec:regularization}), resulting in a set of optimal parameters, where some are close to zero.
Those parameters with values below a certain threshold can be set to exactly zero (thereby removing them from the model), and training can be repeated.
Successive iterations of this procedure can remove more parameters until the desired reduction in parameters, or \textit{sparsity}, is achieved.
This process usually results in models that have slightly reduced performance, although the performance loss is typically negligible for sparsities $\lesssim$90\%~\cite{learningeff}.
Pruning and quantization can also be applied together~\cite{Hawks:2021ruw}.

Finally, some solutions for deployment of ML models involve using cloud resources~\cite{Kuznetsov:2020mcj,SunnebornGudnadottir:2021nhk} or using hardware coprocessors, like GPUs and FPGAs, \textit{as a service}~\cite{sonic,sonic_neutrinos,sonic_cms,sonic_tracking}.
In this approach, coprocessor resources are decoupled from CPUs, and CPU-based clients can send inference requests to coprocessor-based servers via network calls.
The advantages of this approach are coprocessors can accept inference requests from local or remote CPUs, certain types of coprocessors can be allocated for specific tasks, the coprocessor-to-CPU ratio can be optimized, the separation of software support for coprocessor and CPU workflows reduces the maintenance burden, and developments from industry can be more easily leveraged~\cite{sonic_cms}.
\IfFileExists{ml.bib}{\putbib[ml]}{}

\end{bibunit}
\fi

\ifdefined\isdraft
\clearpage
\renewcommand{\twocolumn}[1][]{
     \twocolumngrid
     #1
}
\printindex
\fi

\end{document}